\documentclass[12pt]{article}
\usepackage{amssymb}
\usepackage{epsfig}
\usepackage{float}
\usepackage{graphicx}
\begin{document}
\begin{center}
{\bf Neutrino. History of a unique particle }
\end{center}
\begin{center}
S. M. Bilenky
\end{center}
\begin{center}
{\em  Joint Institute for Nuclear Research, Dubna, R-141980,
Russia\\}
{\em TRIUMF
4004 Wesbrook Mall,
Vancouver BC, V6T 2A3
Canada\\}
\end{center}
\begin{abstract}
Neutrinos are the only fundamental fermions which have no electric
charges. Because of that neutrinos have no direct electromagnetic
interaction and at relatively small energies they can take part only
in weak processes with virtual $W^{\pm}$ and $Z^{0}$ bosons (like
$\beta$-decay of nuclei, inverse $\beta$ process $\bar\nu_{e}+p\to
e^{+}n$, etc.). Neutrino masses are many orders of magnitude smaller
than masses of charged leptons and quarks. These two circumstances
make neutrinos unique, special particles. The history of the neutrino
is very interesting, exciting and instructive. We try here to follow
the main stages of the neutrino history starting from the Pauli
proposal and finishing with the discovery and study of neutrino
oscillations.
\end{abstract}

\section{The idea of neutrino. Pauli}
\begin{center}
{\bf Introduction}
\end{center}
The history of the neutrino started with the famous Pauli letter. The
experimental data "forced" Pauli to assume the existence of a new
particle which later was called neutrino. The hypothesis of the
neutrino allowed Fermi to build the first theory of the $\beta$-decay
which  he considered as a process of a quantum transition of a
neutron into a proton with the creation of an electron-(anti)neutrino
pair. During many years this was the only experimentally studied
process in which the neutrino takes part. The main efforts were
devoted at that time to the search for a Hamiltonian of the
interaction which governs the decay.
\vskip0.5cm

The hypothesis of neutrino was proposed by W. Pauli in December 1930
in the famous letter addressed to  participants of a nuclear
conference in T\"ubingen. At that time protons and electrons were
considered as elementary particles and nuclei were considered as
bound states of protons and electrons. In the framework of this last
assumption there were two fundamental problems:
\begin{enumerate}
  \item The problem of continuous $\beta$ spectra.
  \item The problem of spin of some nuclei.
\end{enumerate}
From the point of view of the proton-electron model the $\beta$-decay
of a nucleus $(A,Z)$ is a process of emission of an electron in the
nuclear transition $(A,Z) \to (A, Z+1) + e^{-}$. From the
conservation of the energy and momentum it followed in this case that
the electron produced in the $\beta$-decay had a fixed kinetic
energy, approximately equal to the released energy
$Q=(M_{A,Z}-M_{A,Z+1})-m_{e}$. In experiments, however, continuous
$\beta$-spectra with an end-point energy equal to $Q$ were observed.

There was a belief that continuous $\beta$- spectra could be
explained by the loss of energy of electrons in the target. However,
in 1927 Ellis and Wooster performed a calorimetric $\beta$-decay
experiment \cite{EllisWooster}. They found that the energy per
$\beta$-decay of a nucleus was equal to the energy averaged over the
$\beta$-spectrum. Thus, it was proved that the energy detected in the
$\beta$-decay was smaller than the total released energy.

After the Ellis and Wooster experiment the situation with
the continuous $\beta$-spectrum became dramatic. Pauli was the
first who understood that under the condition of energy-momentum
conservation the only possibility to explain the continuous $\beta$-spectra
was to assume  that there existed a new, neutral particle
which was emitted in the $\beta$-decay together with the electron
and is not detected in an experiment. Pauli called the new particle the neutron.
Let us notice that there was at that time also an idea that energy in the
$\beta$-decay  is not conserved (A. Bohr).

If the $\beta$-decay is a three-body process $$(A,Z) \to (A, Z+1) + e^{-}+ "n",$$
the released energy is shared between the electron and the "neutron",
 and a continuous electron spectrum will be observed.
As in $\beta$-experiments only electrons were detected, Pauli assumed
that the absorption length of the "neutron" was "the same or probably
10 times larger than the absorption length of the
$\gamma$-quantum".\footnote{Pauli suggested in the letter that the
"neutron" had a magnetic moment $\mu$. He wrote: "The experiments
seem to require that the ionizing effect of such a neutron can not be
bigger than the one of a gamma-ray, and then $\mu $ is probably not
allowed to be larger than $10^{-13}$~e$\cdot$cm".}
Pauli further assumed that  "neutrons" had spin 1/2 and together with
electrons and protons were  constituents of nuclei.  This allowed him
to solve another problem which existed at that time, the problem of
the spin  of some nuclei.

Let us consider the $^{14}\mathrm{N}_{7}$ nucleus. From the point of
view of the electron-proton model this nucleus is a bound state of 14
protons and 7 electrons, i.e. the spin of  $^{14}\mathrm{N}_{7}$
nucleus has to be half-integer. However, from the investigation of
spectra of molecular nitrogen it followed that $^{14}\mathrm{N}_{7}$
nuclei satisfied the Bose-Einstein statistics. Thus, according to the
general theorem on the connection between spin and statistics the
spin of the $^{14}\mathrm{N}_{7}$ nucleus must be integer. If in
addition to   electrons and protons the spin 1/2 "neutrons"  are also
constituents of nuclei, the spin of $^{14}\mathrm{N}_{7}$ can  be
integer.

Pauli also assumed that the "neutron", a constituent of nuclei, must
have a mass different from zero. He wrote in the letter that "the
mass of the neutrons should be of the same order of magnitude as the
electron mass and in any event not larger than 0.01 of the proton
mass".\footnote{Pauli was the first who suggested the existence of a
new particle which was not directly observed in an experiment (but
was needed for the explanation of the experimental data). Nowadays it
is common practice but in Pauli's time it was a very courageous
proposal. It is interesting that in the framework of the wrong
electron-proton-"neutron" model of nuclei Pauli correctly predicted
that the new particle emitted in the $\beta$-decay was a particle
with spin 1/2 and nonzero mass.}

In 1932 the  neutron, a heavy particle with the mass practically
equal to the mass of the proton, was discovered by J. Chadwick
\cite{Chedwick}. Soon after this discovery  Heisenberg
\cite{Heisenberg}, Majorana \cite{Majorana} and Ivanenko
\cite{Ivanenko} put forward a hypothesis that nuclei are bound states
of protons and neutrons. This hypothesis could successfully  describe
all nuclear data. The problem of the spin of $^{14}\mathrm{N}_{7}$
and other nuclei disappeared ($^{14}\mathrm{N}_{7}$ nucleus is the
bound state of 7 protons and 7 neutrons and has an integer spin in
accordance with the theorem on the connection between spin and
statistics.)

\section{Neutrino and the first theory of the $\beta$-decay}

The next fundamental contribution to the development of the idea of
the neutrino was made
 by E. Fermi in 1934 \cite{Fermi}. Fermi built
the first theory of the $\beta$-decay of nuclei. The theory was based
on the Pauli assumption that in the $\beta$-decay together with the
electron a neutral, spin 1/2, light   particle was emitted.

After the discovery of the heavy neutron  Fermi proposed to call the
light Pauli particle {\em the neutrino} (from Italian neutral,
light). Fermi built the theory of the  $\beta$-decay assuming that
nuclei are bound states of protons and neutrons. There was a problem
to understand how an electron-neutrino pair was produced. By analogy
with the emission of a photon by an electron Fermi assumed that {\em
the electron-neutrino pair is produced in the quantum transition of a
neutron into a proton}\footnote{ We know today that in the
$\beta$-decay together with the electron an antineutrino $\bar\nu$ is
produced. Later we will explain  the difference between neutrino and
antineutrino.}
\begin{equation}\label{ndecay}
n\to p +e^{-}+\bar \nu
\end{equation}
The simplest electromagnetic Hamiltonian which induces the quantum transition
\begin{equation}\label{bdecay}
p\to p + \gamma
\end{equation}
has the form of the scalar product of the electromagnetic (vector) current $\bar p(x)\gamma_{\alpha}p(x)$ and vector electromagnetic field $A^{\alpha}(x)$
\begin{equation}\label{electromagnetic}
    \mathcal{H^{\mathrm{EM}}}(x)=e~\bar p(x)\gamma_{\alpha}p(x)~ A^{\alpha}(x)
\end{equation}
By analogy Fermi assumed that the Hamiltonian of the decay
(\ref{ndecay}) was the scalar product of the vector $\bar
p(x)\gamma_{\alpha}n(x)$ and the vector $\bar
e(x)\gamma_{\alpha}\nu(x)$ which could be built from electron and
neutrino fields\footnote{The current $\bar p\gamma_{\alpha}n$ induces
the transition $n\to p$. It changes the electric charge by one
($\Delta Q=1$) and is called the hadronic charged current (CC). The
current $\bar e\gamma_{\alpha}\nu$ provides the emission of the pair
$(e^{-}-\bar \nu)$. It is called the leptonic CC.}
\begin{equation}\label{bdecay}
    \mathcal{H^{\beta}}(x)=G_{F}~\bar p(x)\gamma_{\alpha}n(x)~
\bar e(x)\gamma_{\alpha}\nu(x)+\rm{ h.c.},
\end{equation}
where $G_{F}$ is a constant (which is called Fermi constant).

Let us stress the fundamental difference between the Hamiltonian of
the electromagnetic interaction (\ref{electromagnetic}) and the
Hamiltonian of the $\beta$-decay (\ref{bdecay}). The electromagnetic
Hamiltonian $\mathcal{H^{\mathrm{EM}}}$ is the Hamiltonian of the
interaction of two fermion fields and a boson field and
$\mathcal{H^{\beta}}$ is the Hamiltonian of the interaction of four
fermion fields. As a result of that {\em the constants $e$ and
$G_{F}$ have different dimensions.} In the system of units
$\hbar=c=1$, we use, the charge $e$ is a dimensionless quantity and
the Fermi constant $G_{F}$ has dimension $M^{-2}$ ($M$ is a mass).
Later we will discuss the origin of the dimension of the constant
$G_{F}$. We will see that the dimension of the constant $G_{F}$ means
that the Hamiltonian (\ref{bdecay}) is not a fundamental Hamiltonian
of interaction but is an {\em effective Hamiltonian}.

 Applying the methods of the Quantum Field Theory and
using the Hamiltonian (\ref{bdecay}), Fermi calculated the spectrum
of electrons emitted in the $\beta$-decay and suggested a method of
the measurement of the neutrino mass. For that he proposed to
investigate the shape of the electron spectrum in the region near the
maximal electron energy (which corresponds to the emission of non
relativistic neutrinos).\footnote{The same method of the measurement
of the neutrino mass was proposed  by Perrin\cite{Perrin}.}

It occurred that the investigation of the $\beta$-decay of tritium
\begin{equation}\label{tritium}
^{3}\mathrm{H}\to ^{3}\mathrm{He} +e^{-}+\bar\nu
\end{equation}
is one the most sensitive ways of the measurement of the neutrino
mass by the Fermi-Perrin method.\footnote{This is connected with the
fact that tritium has a convenient half-life $T_{1/2}=12.3$ years,
the energy release in the process (\ref{tritium}) is small ($Q=18.57$
keV), the nuclear matrix element of the process is a constant
($^{3}\mathrm{H}\to ^{3}\mathrm{He}$ is an allowed transition), etc.}

The electron spectrum for the allowed transitions  is determined
by the phase-space factor
\begin{equation}\label{phase}
    p_{e}E_{e}~p E,
\end{equation}
where $E_{e}$ and $E$ ($p_{e}$ and $p$) are the energies
(momenta) of the electron and the neutrino.

If we neglect the recoil  of the final nucleus, from
the conservation of the energy we have
\begin{equation}\label{phase1}
Q=T_{e} +E,
\end{equation}
where $T_{e}=E_{e}-m_{e}$ is the kinetic energy of the electron.

From (\ref{phase}) we obtain the following expression for the spectrum of the electrons in the decay
(\ref{tritium})
\begin{equation}\label{bspectrum}
\frac{d\Gamma}{dE}=Cp_{e}(T_{e}+m_{e})(Q-T_{e})
\sqrt{(Q-T_{e})^{2}-m^{2}_{\nu}}
~F(T_{e},Z) ,
\end{equation}
where $m_{\nu}$ is the neutrino mass,  $F(T_{e},Z)$ is the Fermi
function, which describes the Coulomb interaction of the final
electron and nucleus, and $C$ is a constant (which includes the
modulus-squared of the nuclear matrix element).

The neutrino mass enters into the expression for the $\beta$-spectrum
through the neutrino momentum $p= \sqrt{(Q-T_{e})^{2}-m^{2}_{\nu}}$.
From this expression it is obvious that the part of the spectrum in
which $Q-T_{e}\lesssim m_{\nu}$ is sensitive to the neutrino
mass.\footnote{In practice, for a neutrino mass $m_{\nu}\lesssim  1$
eV a much larger part of the spectrum is used for the analysis of
experimental data (in order to increase the luminosity of the
experiment).}

The largest contributions to the $\beta$-decay come from transitions
in which electron and (anti)neutrino are produced in states with
orbital momenta equal to zero ($S$-states).\footnote{The neutrino and
the electron are produced in the $\beta$-decay  in states with
definite momenta. Their wave function has the form
$e^{i\vec{p}_{e}\vec{x}+i\vec{p}\vec{x}}$. We have
$\vec{p}~\vec{x}\leq p~R$. where $R\simeq 1.2 \cdot 10^{-13}~A^{1/3}$
cm is the radius of a nucleus. Taking into account that the energies
of the neutrino and the electron produced in the $\beta$-decay are
not larger than a few MeV, we have $|\vec{p}~\vec{x}|\ll 1$ and
$|\vec{p}_{e}~\vec{x}|\ll 1$. Thus, in the first approximation
$e^{i\vec{p}_{e}\vec{x}+i\vec{p}\vec{x}}\simeq 1$. This approximation
corresponds to the emission of the electron and the neutrino into
$S$-states (allowed transition).} Such transitions are called
allowed. For allowed transitions it follows from the Fermi
Hamiltonian (\ref{bdecay}) that spins and parities of the initial and
final nuclei must be equal (Fermi selection rules):
\begin{equation}\label{Fermiselection}
    \Delta J=0,\quad \pi_{i}=\pi_{f}.
\end{equation}
Here $\Delta J=J_{f}-J_{i}$ ($J_{i}(J_{f})$ is the spin of the
initial (final) nucleus) and $\pi_{i}$ ($\pi_{f}$) is the parity of
the initial (final) nucleus.

From the conservation of the total momentum it follows that in the
case of an allowed transition which satisfies the Fermi selection
rule electron and (anti)neutrino are produced in a state with the
total spin $S$ equal to zero (singlet state). If electron and
(anti)neutrino are produced in the triplet state ($S=1$) in this case
for the allowed transition the total angular momentum of the final
state is equal to $J_{f}=J_{i}\pm 1$ or $J_{f}=J_{i}$ (for $J_{i}=0$
the total final angular momentum is equal to 1). We have in this case
\begin{equation}\label{GTselection}
    \Delta J=\pm1,0\quad \pi_{i}=\pi_{f}~~~( 0\to 0~\mathrm{is~forbidden}).
\end{equation}
The selection rules (\ref{GTselection}) are called  the Gamov-Teller
selection rules. They were introduced by Gamov and Teller in 1936
\cite{GamovTeller}.

In the  $\beta$-decay experiments, decays of nuclei which satisfy the
Fermi and Gamov-Teller selection rules were observed. Thus, the total
Hamiltonian of the $\beta$-decay must include not only the Fermi
Hamiltonian (\ref{bdecay}) but also an additional term (or terms).

 The Fermi Hamiltonian is the product of   vector$\times$vector.
The most general Hamiltonian of the Fermi type, in which only fields but not their derivatives enter, has the form of the sum of the products of
scalar$\times$scalar, vector$\times$vector, tensor$\times$tensor, axial$\times$axial  and
pseudoscalar$\times$pseudoscalar:
\begin{equation}
{\cal{H}}_{I}^{\beta}(x)= \sum_{i=S,V,T,A,P}G_{i}\, \bar{p}(x) O^{i} n(x)
\,~\bar{e}(x) O_{i}  \nu(x)  + \rm{ h.c.} \label{bdecay1}
\end{equation}
Here
\begin{equation}
O^{i}  \to  1~ (S),\, \gamma^{\alpha}~(V),\,\sigma^{\alpha\,\beta}~(T),\,
\gamma^{\alpha}\gamma_{5}~(A),\, \gamma_{5}~(P). \label{bdecay2}
\end{equation}
and $G_{i}$ are coupling constants, which have  dimensions
$[M]^{-2}$. \footnote{Dirac matrices $\gamma^{\alpha}$
($\alpha$=0,1,2,3) satisfy the relations
$\gamma^{\alpha}\gamma^{\beta}+\gamma^{\beta}\gamma^{\alpha}= 2
g^{\alpha\beta}$, where $g^{00}=1,g^{ii}=-1$ and  non diagonal
elements of $g^{\alpha\beta}$ are equal to zero. The matrix
$\gamma_{5}$ is determined as follows
$\gamma_{5}=-i\gamma_{0}\gamma_{1}\gamma_{2}\gamma_{3}$. It satisfies
the relations
$\gamma^{\alpha}\gamma_{5}+\gamma_{5}\gamma^{\alpha}=0$,
$\gamma_{5}\gamma_{5}=1 $. Sixteen matrices $1,~
\gamma^{\alpha},~\sigma_{\alpha\,\beta}=
\frac{1}{2}(\gamma^{\alpha}\gamma^{\beta}-\gamma^{\beta}\gamma^{\alpha}),~
\gamma^{\alpha}\gamma_{5},~ \gamma_{5}$ form a complete system of
4$\times$4 matrices. } The Hamiltonian (\ref{bdecay1}) describes all
$\beta$-decay data. Transitions, which satisfy the Fermi selection
rules, are due to $V$ and $S$ terms and transitions which satisfy the
Gamov-Teller selection rules, are due to $A$ and $T$ terms.

In the Fermi Hamiltonian (\ref{bdecay}) only one fundamental constant
$G_{F}$ enters. The Hamiltonian (\ref{bdecay1}) is characterized by
five (!) interaction constants. Analogy and economy which were the
basis of the Fermi theory were lost.

There was a general belief  that there are "dominant" terms in the
interaction (\ref{bdecay1}). Such terms were searched for many years
via analysis of the data of different $\beta$-decay experiments. This
search did not lead, however, to a definite result: some experiments
were in favor of $V$ and $A$ terms, other were in favor of $S$ and
$T$terms. Up to 1957 when the violation of parity in the
$\beta$-decay (and other weak processes) was discovered, the
situation with the Hamiltonian of the $\beta$-decay remained
uncertain.

\section{The first estimate of the neutrino-nucleus cross section}

\begin{center}
{\bf Introduction}
\end{center}

Very early physicists started to think about a possibility to detect
the neutrino and thus to prove directly its existence. However, in
the thirties and the forties there were no technical possibilities to
do this. We will discuss here the first paper in which the
neutrino-nucleous cross section was estimated.

\vskip0.5cm

In the Fermi Hamiltonian (\ref{bdecay}) $e(x)$, $\nu(x)$, $n(x)$ and
$p(x)$  are {\em quantum fields}. This means that the Hamiltonian
(\ref{bdecay}) allows one to calculate not only the probability of
the $\beta^{-}$-decay
\begin{equation}\label{betadecay}
    (A,Z)\to (A,Z+1) +e^{-}+\bar \nu
\end{equation}
but also the probabilities of the $\beta^{+}$ -decay and electron
capture
\begin{equation}\label{betadecay1}
    (A,Z)\to (A,Z-1) +e^{+}+ \nu,\quad e^{-}+(A,Z)\to \nu+ (A,Z-1),
\end{equation}
the cross sections of the neutrino reactions
\begin{equation}\label{betadecay2}
\bar \nu + (A,Z)\to e^{+}+(A,Z-1) ,
\end{equation}
\begin{equation}\label{betadecay3}
\nu + (A,Z)\to e^{-}+(A,Z+1)
\end{equation}
and other processes.

The first estimation of the cross section of the process
(\ref{betadecay2}) was obtained by Bethe and Peierls
\cite{BethePeierls} soon after the Fermi theory was proposed.

We will present here Bethe's and Peierls's arguments. At relatively
small MeV energies the nuclear matrix elements of the processes
(\ref{betadecay}) and (\ref{betadecay2}) are practically the same.
Since the $\beta$-decay width $\Gamma=\frac{1}{T_{1/2}}$ ($T_{1/2}$
is the half-life of the decay) and the cross section $\sigma$ of the
process (\ref{betadecay1}) are proportional to the modulus-squared of
the nuclear matrix elements, we have
\begin{equation}\label{betadecay3}
 \sigma=\frac{A}{T_{1/2}},
\end{equation}
where $A$ has dimension $(\mathrm{length})^{2}\times \mathrm{time}$. The authors suggested that "the longest length and time which can possibly be involved are $\frac{\hbar}{m_{e}c}$ and $\frac{\hbar}{m_{e}c^{2}}$" and found the following bound
\begin{equation}\label{BPbound}
    \sigma<\frac{\hbar^{3}}{m^{3}_{e}c^{4}T_{1/2}}
\end{equation}
From this inequality for  $T_{1/2}\simeq 3$ min Bethe and Peierls found
\begin{equation}\label{BPbound}
 \sigma  <10^{-44} ~\mathrm{cm}^{2}.
\end{equation}
 This bound corresponds to a neutrino absorption length
in solid matter larger than $10^{14}$ km. \footnote{ It is
interesting to compare this number with Pauli's original expectation:
"... I trustfully turn first to you, dear radioactive people, with
the question of how likely it is to find experimental evidence for
such a neutron if it would have the same or perhaps a 10 times larger
ability to get through [material] than a gamma-ray" (Pauli letter).}
On the basis of this estimate  Bethe and Peierls in their paper( with
the title The "Neutrino") concluded {\em "...there is no practically
possible way of observing the neutrino"}.

For comparison we  will present the current calculations
of the neutrino cross section. Let us consider the  process
\begin{equation}\label{antinu}
\bar\nu +p\to e^{+}+n.
\end{equation}
Using the modern Hamiltonian of the weak interaction for the cross
section of the process (\ref{antinu}) we have
\begin{equation}\label{antinu1}
    \sigma =4\frac{G^{2}_{F}}{\pi}p_{e}E_{e}\simeq 9.5\cdot  10^{-44}\frac{p_{e}E_{e}}{\mathrm{MeV^{2}}}\mathrm{cm}^{2},
\end{equation}
where $E_{e}$ and $p_{e}$ are the positron energy and the momentum.
Neglecting the recoil of the final neutron we have for the neutrino
energy $E$
\begin{equation}\label{antinu2}
E=E_{e}+\Delta,
\end{equation}
where $\Delta=m_{n}-m_{p}\simeq 1.3~ MeV$ is the neutron-proton mass difference. For (anti)neutrinos with the energy $E\simeq 3 $ MeV we find the value $\sigma\simeq 2.6\cdot 10^{-43}~\mathrm{cm}^{2}$ from (\ref{antinu1}). Correspondingly, the absorption length of (anti)neutrinos in water is given by \begin{equation}\label{antinu3}
L_{a}=\frac{1}{n\sigma}\simeq 6\cdot 10^{14}~\mathrm{km},
\end{equation}
where $n$ is the number density of protons (in the case of water $n\simeq 6.7\cdot 10^{22}\frac{1}{\mathrm{cm}^{3}}$). Thus, present-day calculations confirm  the Bethe and Peierls estimate.

After the Bethe and Peierls paper there was a general opinion that
the neutrino is an undetectable particle. The first physicist  who
challenged this opinion  was B. Pontecorvo \cite{BPonte46}. In 1946
he proposed a radiochemical method of neutrino detection and in
particular the $\mathrm{Cl}-\mathrm{Ar}$ method which is based on the
reaction
\begin{equation}\label{ClAr}
    \nu+^{37}\mathrm{Cl}\to e^{-}+^{37}\mathrm{Ar}.
\end{equation}
Many years later the $\mathrm{Cl}-\mathrm{Ar}$ method of neutrino
detection allowed R. Davis to observe solar neutrinos in the first
solar neutrino experiment \cite{Davis}. We will discuss solar
neutrinos and the Pontecorvo radiochemical method later.

\section{First ideas of $\mu-e$ universal weak interaction}

\begin{center}
{\bf Introduction}
\end{center}
With the idea of $\mu-e$ universality there appeared a notion of
universal weak interaction. The idea of universality was proposed,
however, at the time when the form of the weak interaction was not
known. It was, nevertheless, an extremely important general idea. We
will see later how it was implemented in the $V-A$ theory of  weak
interaction.

\vskip0.5cm

In 1947 B. Pontecorvo\cite{BPontmue} came to an idea of existence of
a universal weak interaction which governs not only processes in
which the electron-neutrino pair  takes part (like the nuclear
$\beta$-decay) but also processes in which the muon-neutrino pair
participates. The process of such a type is  $\mu$-capture
\begin{equation}\label{mucapture1}
\mu^{-}+(A,Z)\to \nu +(A,Z-1).
\end{equation}
B. Pontecorvo compared the probability of this process
and the probability  of the $K$-capture
\begin{equation}\label{Kcapt}
e^{-}+ (A,Z) \to \nu + (A,Z-1)
\end{equation}
and came to the qualitative conclusion that the constant of the
interaction of the muon-neutrino
pair with nucleons is of the same order as the Fermi constant.

 The idea of  $\mu-e$ universality of the weak interaction was
also proposed  by G. Puppi \cite{Puppi}. Puppi presented it in the
form of a triangle ("Puppi triangle"). He assumed that a universal
weak interaction includes not only Hamiltonians of the $\beta$-decay
and $\mu$-capture but also the Hamiltonian of the $\mu$-decay
\begin{equation}\label{mudecay}
\mu^{+}\to e^{+}+\nu +\bar\nu.
\end{equation}
Puppi suggested that different parts of the weak interaction
form a triangle with vertices
\begin{equation}\label{Puppi}
(\bar pn)-(\bar\nu e)-(\bar\nu \mu)
\end{equation}
 and the Hamiltonian of the weak interaction is given by  a sum of products
of different vertices. The idea of a universal weak interaction was
proposed also by O.Klein \cite{Klein} and Yang and Tiomno
\cite{YangTiomno}.

\section{Parity violation in the $\beta$-decay and other weak processes}
\begin{center}
{\bf Introduction}
\end{center}
Conservation of parity (invariance under space inversion, i.e. under
transition  from a right-handed to a left-handed system) was firmly
established for strong (hadronic) and electromagnetic processes. For
many years physicists thought that the invariance under space
inversion is a general law of nature valid for all interactions. The
discovery of violation of parity in the $\beta$-decay and other weak
processes was a great surprise. In the beginning it looked that this
discovery made the theory of the $\beta$-decay and other weak
processes more complicated. In reality, as we will see later, this
discovery allowed  building a simple, correct theory of the neutrino
and weak interaction.

The violation of parity in the weak interaction was one of the most
important discoveries in the physics of the XX century. In 1957 Lee
and Yang were awarded the Nobel Prize "for their penetrating
investigation of the so-called parity laws which has led to important
discoveries regarding the elementary particles".

\vskip0.5cm

Our understanding of the neutrino and the weak interaction has
drastically changed after it was discovered in 1957 that in the
$\beta$-decay, the decay $\mu^{+}\to e^{+}+\nu +\bar\nu$ and other
weak processes parity is not conserved.

The investigation of the decays of strange particles at the beginning
of the fifties created the so called $\theta-\tau$ puzzle.\footnote{
A strange particle which decays into $\pi^{+}$ and $\pi^{0}$ was
called $\theta^{+}$ ($\theta^{+}\to\pi^{+}+ \pi^{0}$) and a strange
particle  which decays into $\pi^{+}$ and $\pi^{-}$ and $\pi^{+}$ was
called $\tau^{+}$ ($\tau^{+}\to \pi^{+} +\pi^{-}+\pi^{+}$). From
experimental data it followed that the {\em masses and lifetimes of
$\theta^{+}$ and $\tau^{+}$ are the same.} The study of the Dalitz plot of the
decay of  $\tau^{+}$ showed that the total momentum  of the state
of ($\pi^{+},\pi^{-},\pi^{+}$) was equal to zero and the parity
(eigenvalue of the operator of the parity) was equal to -1. If
$\tau^{+}$ and  $\theta^{+}$ are the same particle in this case its
spin must be equal to zero. However, the parity of the two pions
produced in the $\theta^{+}$-decay is equal to +1 (the parity of two
pions is equal to $I^{2}_{\pi}(-1)^{l}=(-1)^{2}(-1)^{0}=1$, where
$I_{\pi}=-1$ is the internal parity of the pion and $l$ is the
orbital momentum of two pions). So if $\tau^{+}$ and  $\theta^{+}$
are the same particle (which is natural to assume because of the
equality of masses and lifetimes) we are confronted with the
following problem: the same particle decays into states with
different parities.} As one of the possible solutions of the
$\theta-\tau$ problem Lee and Yang \cite{LeeYang56}  put forward the
hypothesis of the non-conservation of parity (1956). They analyzed
existing experimental data and came to the conclusion that there was
an evidence that parity is conserved in the strong and
electromagnetic interactions, but there were no data which proved
that parity was conserved in the $\beta$-decay and other weak decays.
("...as for weak interactions parity conservation is so far only
extrapolated hypothesis unsupported by experimental evidence"
\cite{LeeYang56}). Lee and Yang proposed different experiments which
would allow to test the hypothesis of the parity conservation in weak
decays. The results of the first experiments in which large violation
of parity in weak processes was observed were published by Wu et al.
\cite{Wu} and Lederman et al. \cite{Lederman} at the beginning of
1957.

We will discuss first the experiment by Wu et al. in which the
$\beta$-decay of polarized $^{60}\mathrm{Co} $ was
investigated.\footnote{Polarization of a nucleus  is the average value
of its spin.} Let us consider the emission of the electron with
momentum $\vec{p}$ in the $\beta$-decay of a nucleus with
polarization $\vec{P}$. From the invariance under rotations
(conservation of the total momentum) it follows that the decay
probability can depend only on the scalar products
$\vec{p}\cdot\vec{p}$ and $\vec{P}\cdot\vec{p}$. Taking into account
that the decay probability  depends linearly on the polarization of a
nucleus we obtain the following general expression for the
probability of the emission of the electron with momentum $\vec{p}$
by a nucleus with polarization $\vec{P}$
\begin{equation}\label{decayprob}
w_{\vec{P}}(\vec{p})=w_{0}~(1+\alpha \vec{P}\cdot\vec{k})=
w_{0}~(1+\alpha P\cos\theta).
\end{equation}
Here $\vec{k}=\frac{\vec{p}}{p}$ is a unit vector in the direction of
the electron momentum,  $\theta$ is the angle between the vectors
$\vec{P}$ and $\vec{p}$, and $w_{0}$ and $\alpha$ are functions of
$p^{2}$.

Under the inversion of a coordinate system (change of directions of
all axes of the coordinate system) momentum $\vec{p}$ and
polarization $\vec{P}$ are transformed {\em differently}. Namely,
momentum is transformed as a vector
\begin{equation}\label{vector}
 p'_{i}=-p_{i}
\end{equation}
while polarization is transformed as a pseudovector\footnote{Notice
that momentum, coordinates, electric field etc. are vectors while
angular momentum, polarization, magnetic field etc. are
pseudovectors.}
\begin{equation}\label{pseudovector}
P'_{i}=+P_{i}.
\end{equation}
Here $p_{i}$ ($P_{i}$) are components of a vector of momentum
(pseudovector of polarization) in some right-handed system and
$p_{i}'$ ($P_{i}'$) are components of the same momentum (same
polarization) in the inverted (left-handed) system.

Relations (\ref{vector}) and (\ref{pseudovector}) mean that under the inversion the vector of momentum does not change its position in space while  polarization changes its direction to the opposite one.

From (\ref{vector}) and (\ref{pseudovector}) it follows that under
the inversion  the scalar product $\vec{P}\cdot\vec{p}$ is
transformed as a pseudoscalar (change sign)
\begin{equation}\label{pseudoscalar}
\vec{P'}\cdot\vec{p'}=-\vec{P}\cdot\vec{p}
\end{equation}
while $\vec{p}\cdot\vec{p}$ is transformed as a scalar
\begin{equation}\label{pseudoscalar}
\vec{p'}\cdot\vec{p'}=+\vec{p}\cdot\vec{p}.
\end{equation}
If the invariance under the inversion holds (parity is conserved), in
this case the decay probability in a right-handed system and in an
inverted left-handed system is the same
\begin{equation}\label{invariance}
 w_{\vec{P'}}(\vec{p'}) =w_{\vec{P}}(\vec{p}).
\end{equation}
From (\ref{decayprob}),  (\ref{vector}), (\ref{pseudovector}) and
(\ref{invariance}) we conclude that in the case of conservation of
parity $\alpha =0$ and the probability of the emission of the
electron by the polarized nucleus does not depend on the angle
$\theta$.\footnote{Let us consider the emission of  electrons by a
polarized nucleus at the angle $\theta$ in a right-handed system.
Conservation of parity  means that the probability of the emission of
electrons at the same angle  $\theta$ in the inverted left-handed
system must be the same. However, because polarization is a
pseudovector, the probability of the emission of electrons at the
angle $\theta$ in the left-handed system is equal to the probability
of the emission of  electrons in the  right-handed system at the
angle $\pi-\theta$. Thus, if parity is conserved the probabilities of
the emission of the electron at the angles $\theta$ and $\pi-\theta$
must be the same. This means that the pseudoscalar
$\vec{P}\cdot\vec{k}=Pk \cos\theta$  can not enter into the
probability.}

In the Wu et al. experiment \cite{Wu} it was found that the parameter
$\alpha$ was negative and $|\alpha| \geq 0.7$ (i.e. electrons are
emitted mainly in the direction opposite to the polarization of the
nucleus).\footnote{The sign of the asymmetry parameter $\alpha$
depends on the handedness of the system.  Conservation of parity
means that such parameters can not enter into measurable quantities.
After Wu et al. and other experiments we know that this is not the
case.} Thus, it was discovered that there was no invariance of the
$\beta$-decay interaction under inversion (parity in the
$\beta$-decay is not conserved).

The paper of Wu et al. \cite{Wu} was submitted  to Physical Review on
January 15, 1957. The same day another experimental paper
\cite{Lederman} on the observation of the violation of parity in weak
decays was submitted to the same journal. In the Lederman et al.
experiment \cite{Lederman} strong violation of parity in the chain of
the decays
\begin{equation}\label{1pidecay}
\pi^{+}\to \mu^{+}+\nu
\end{equation}
and
\begin{equation}\label{1mudecay}
\mu^{+}\to e^{+}+\nu+\bar\nu
\end{equation}
was observed.

If parity is violated, a muon produced in the decay (\ref{1pidecay})
will be polarized in the direction of the muon
momentum.\footnote{Muon possesses longitudinal polarization if the
probabilities  of the emission of the muon with positive and negative
helicities are different.  This could happen only in the case if
parity in the decay (\ref{1pidecay}) is violated.} Like in the case
of the $\beta$-decay, the dependence of the probability of the  decay
of polarized muons on the angle $\theta$ between muon polarization
and electron momentum has the general form  $(1+a\cos\theta)$, where
the second pseudoscalar term ($\alpha_{\mu}\vec{P}\cdot
k=a\cos\theta$) is due to non conservation of parity. In the Lederman
et al. experiment \cite{Lederman} a large asymmetry of $e^{+}$ was
found ($a\simeq -\frac{1}{2}$).

Let us discuss the Hamiltonian of the $\beta$-decay. The Hamiltonian
(\ref{bdecay1}) is a scalar. It conserves the parity. In order to
take into account the results of the Wu et al. and other experiments
we must assume that the  {\em Hamiltonian of the $\beta$-decay is the
sum of a scalar and a pseudoscalar}. In order to build such a
Hamiltonian we have to add to  five scalars which enter into the
Hamiltonian (\ref{bdecay1})  additional five pseudoscalars which are
formed from  products of the scalar $\bar{p}(x)  n(x)$ and
pseudoscalar $\bar{e}(x)\gamma_{5} \nu(x)$, vector
$\bar{p}(x)\gamma^{\alpha}  n(x)$ and pseudovector
$\bar{e}(x)\gamma^{\alpha}\gamma_{5}  \nu(x)$, etc. The most general
Hamiltonian of the $\beta$-decay takes the form
\begin{equation}
{\cal{H}}_{I}^{\beta}(x)= \sum_{i=S,V,T,A,P} \bar{p}(x) O_{i} n(x)
\,~\bar{e}(x) O^{i}(G_{i}+G'_{i}\gamma_{5})  \nu(x)  + \rm{ h.c.}, \label{bdecay3}
\end{equation}
 where the constants $ G_{i}$ characterize the scalar part of the Hamiltonian,
the constants $ G'_{i}$ characterize the pseudoscalar part and the
matrices $O^{i}$ are given by (\ref{bdecay2}).

The Hamiltonian (\ref{bdecay3}) is characterized by 10 (!)
fundamental interaction constants. From the Wu et al. experiment it
followed that scalar and pseudoscalar terms of the  Hamiltonian must
be of the same order. This means that the constants $ |G_{i}| $ and $
|G'_{i}|$ (at least for some $i$) must be of the same order.

In 1957-58 enormous progress in the development of the theory of the
$\beta$-decay and other weak processes was reached. Soon after the
discovery of the violation of parity the Hamiltonian of the weak
interaction took a simple form, compatible with the experimental
data.

The new development  of the theory of the weak interaction started
with {\em the two-component theory of the neutrino.}

\section{Massless two-component neutrino}
\begin{center}
{\bf Introduction}
\end{center}
The discovery of the violation of the parity in the $\beta$-decay and
other weak processes triggered enormous progress in the understanding
of the weak interaction. {\em This progress started with the theory
of the two-component neutrino.} This theory of the neutrino became
part of the universal $V-A$ theory of the weak interaction and the
unified theory of the electromagnetic and weak interaction (Standard
Model). The main idea of the two-component theory (left-handed
component of the neutrino field in the interaction Hamiltonian) was
generalized in the subsequent development of the theory of the weak
interaction.

The two-component theory was based on the assumption that the
neutrino is a massless particle. We know today that neutrinos have
small but different from zero masses and that the two-component
theory must be generalized.

\vskip0.5cm

Soon after the discovery of the parity violation Landau
\cite{Landau}, Lee and Yang \cite{LeeYang} and Salam \cite{Salam}
came to an idea of {\em a possible connection of the violation of
parity observed in the $\beta$-decay and other weak processes with
neutrinos.}

The neutrino field $\nu(x)$ satisfies the Dirac equation
\begin{equation}\label{Dequation}
    (i\gamma^{\alpha}~\partial_{\alpha}-m_{\nu})\nu(x)=0
\end{equation}
where $m_{\nu}$ is the neutrino mass.

Let us present the field $\nu(x)$ in the form of the sum of the
left-handed $\nu_{L}(x) = (\frac{1- \gamma_{5}}{2})\nu(x)$ and the
right-handed $\nu_{L}(x) = (\frac{1+ \gamma_{5}}{2})\nu(x)$ components:
\begin{equation}\label{Dequation1}
\nu(x) = \nu_{L}(x) + \nu_{R}(x).
\end{equation}

From (\ref{Dequation}) and (\ref{Dequation1}) we obtain two coupled
equations for $\nu_{L}(x)$ and $\nu_{R}(x)$
\begin{equation}\label{Dequation3}
    i\gamma^{\alpha}~\partial_{\alpha}\nu_{L}(x)-m_{\nu}\nu_{R}(x)=0\quad
i\gamma^{\alpha}~\partial_{\alpha}\nu_{R}(x)-m_{\nu}\nu_{L}(x)=0.
\end{equation}
Let us assume that $m_{\nu}=0$. In this case the equations
for $\nu_{L}(x)$ and $\nu_{R}(x)$  are decoupled and we have
\begin{equation}\label{Dequation4}
    i\gamma^{\alpha}~\partial_{\alpha}~\nu_{L,R}(x)=0.
\end{equation}
Thus, for $m_{\nu}=0$, the neutrino field can be $\nu_{L}(x)$ (or
$\nu_{R}(x)$). Such a theory can be valid only if parity is violated.
In fact, under the inversion of coordinates the field $\nu(x)$ is
transformed as follows:
\begin{equation}\label{Ptransform}
\nu'(x')=\eta\gamma^{0}\nu(x).
\end{equation}
Here $x'=(x^{0}-\vec{x})$ and $\eta$ is a phase factor. From (\ref{Ptransform}) we have
\begin{equation}\label{Ptransform1}
\nu_{L(R)}'(x')=\eta\gamma^{0}\nu_{R(L)}(x).
\end{equation}
Thus, under the inversion a left-handed  (right-handed ) component of
the field is transformed into a (right-handed)  (left-handed)
component. This means that  equations (\ref{Dequation4}) are not
invariant under the inversion.\footnote{Equations (\ref{Dequation4})
for massless spin 1/2 particle were considered by H. Weil in 1929
\cite{Weil}. However, as they did not conserve parity they were
rejected. In \cite{Pauli33}, Pauli wrote "...because the equation for
$\nu_{L}(x)$ ($\nu_{R}(x))$ is not invariant under space reflection
it is not applicable to the physical reality".}

From the investigation of  the high-energy part of
the tritium $\beta$-spectrum it was found only an upper bound for the neutrino
mass which was much smaller than the mass of the electron, a particle  emitted in the
$\beta$-decay together with the neutrino. In 1957 {\em Landau \cite{Landau},
Lee and Yang \cite{LeeYang} and Salam \cite{Salam} assumed
 that the neutrino mass was equal to zero} and that the neutrino
field is $\nu_{L}(x)$ (or $\nu_{R}(x)$). For reasons, which will be
clear later, this theory is called the two-component neutrino theory.

There were two major consequences of the two-component neutrino theory.
\begin{enumerate}
 \item Parity is strongly violated in the $\beta$-decay and
in other processes in which neutrino(s) participate.

The most general Hamiltonian of the $\beta$-decay in the case of
parity violation is given by  expression (\ref{bdecay3}).  Five
interaction constants  $G_{i}$ characterize  the scalar part of the
Hamiltonian and five interaction constants $G'_{i}$  characterize the
pseudoscalar part ($i=S,V,T,A,P$).

In the case of the two-component theory these constants are connected
by the relations
\begin{equation}\label{2compnu}
G'_{i}=-G_{i}\quad(\mathrm{if~neutrino~ field~ is~ \nu_{L}(x)})
\end{equation}
and
\begin{equation}\label{2compnu1}
G'_{i}=G_{i}\quad(\mathrm{if~neutrino~ field~ is~ \nu_{R}(x)}).
\end{equation}
The most general Hamiltonian of the $\beta$-decay takes the form
\begin{equation}\label{twocompH}
 {\cal{H}}_{I}^{\beta}(x)= \sum_{i=S,V,T,A,P}G_{i}\, \bar{p}(x) O_{i} n(x)
\,~\bar{e}(x) O^{i}(1\mp\gamma_{5}) \nu(x)  + \rm{ h.c.}.
\end{equation}
From this expression it follows that effects of violation of parity
in the $\beta$-decay will be large (maximal).
\item The neutrino helicity (projection of the spin onto the direction of momentum)
is equal to -1 (+1) in the case if the neutrino field is $\nu_{L}(x)$
($\nu_{R}(x)$).\footnote{The spinor $u^{r}(p)$ which describes a
massless neutrino with momentum $p$ and helicity $r$ satisfies the
equations $\gamma\cdot
p~u^{r}(p)=(\gamma^{0}p^{0}-\vec{\gamma}\vec{p})~u^{r}(p)=0,\quad
\vec \Sigma\cdot \vec k ~u^{r}(p)= r\,~u^{r}(p)$. Here $\vec \Sigma
=\gamma_{5} \gamma^{0}\vec \gamma $ is the spin operator  and $\vec
k$ is the unit vector in the direction of the momentum $\vec p$. From
these equations we find $\gamma_{5}\, u^{r}(p)= r u^{r}(p)$. In
$\nu_{L}(x)$ ($\nu_{R}(x)$) the spinor $u^{r}(p)$ is multiplied by
the projection operator $\frac{1- \gamma_{5}}{2}$ ($\frac{1-
\gamma_{5}}{2}$ ). We have $\frac{1- \gamma_{5}}{2}\,
u^{r}(p)=\frac{1- r}{2}~u^{r}(p)$ ($\frac{1+ \gamma_{5}}{2}\,
u^{r}(p)=\frac{1+ r}{2}~u^{r}(p)$). Thus, $r=-1$ ($r=1$) in the case
that the neutrino field is $\nu_{L}(x)$ ($\nu_{R}(x)$).}
\end{enumerate}
The theory we are discussing is called the two-component neutrino
theory by the following reason. In the general case  of a Dirac
particle with spin 1/2 there are four states with momentum $\vec{p}$
and energy $E_{p}=\sqrt{p^{2}+m^{2}}$: two particle states with
helicities $\pm 1$ and two antiparticle states with helicities $\pm
1$.
 In the two-component theory with the neutrino field
$\nu_{L}(x)$ ($\nu_{R}(x)$) only the state of the neutrino with
helicity -1 (+1) and the state of the antineutrino with helicity +1
(-1) are allowed.

It is easy to see that in the processes in which a two-component
neutrino is emitted large (maximal) violation of parity will be
observed. In fact, let $w^{R}_{r}$ be the probability to emit a
neutrino with helicity $r$ in a right-handed system. This probability
is equal to the probability of the emission of a neutrino with
helicity $-r$ in a left-handed system
\begin{equation}\label{leftright}
 w^{R}_{r}=w^{L}_{-r}.
\end{equation}
If the parity is conserved
\begin{equation}\label{leftright1}
 w^{R}_{r}=w^{L}_{r}.
\end{equation}
From (\ref{leftright}) and (\ref{leftright1}) it follows that in the case of the conservation of parity the probabilities of the emission of neutrinos with helicities $r$ and $-r$ must be equal
\begin{equation}\label{leftright2}
 w^{L,R}_{r}=w^{L,R}_{-r},\quad \mathrm{i.~e.}\quad w^{L,R}_{1}=w^{L,R}_{-1}
\end{equation}
In the case of the two-component neutrino theory $w_{1}=0$ (or
$w_{-1}=0$). Thus, in the two-component theory  relation
(\ref{leftright2}) is maximally violated.

We will discuss now the difference between a neutrino and an
antineutrino. In general particles and antiparticles have the same
mass but different (in sign) charges. There exist in nature different
conserved charges. The most familiar is the electric charge $Q$. The
electric charges of the proton, neutron, electron, neutrino, for
example, are equal to 1, 0, -1, 0, respectively.\footnote{ Usually,
charges of particles are expressed in the unit of the proton charge.}

Another conserved charge is the baryon number $B$. The baryon numbers of the
proton, neutron, electron, neutrino are equal to 1,1,0,0, respectively.

We will be interested here mainly in the lepton number $L$. The
lepton numbers of the proton, neutron, electron, neutrino are defined
as 0,0,1,1, respectively.

Particles like a proton and a neutron which possess a baryon number
are called baryons. An electron, a neutrino and other particles which
possess the lepton number are called leptons.

According to the Quantum Field Theory, the existence of a particle
with a mass $m$ and some charges implies the existence of an
antiparticle which has the same mass $m$ and opposite charges. This
general consequence of the Quantum Field Theory was confirmed by
numerous experiments. For example, the antiparticles of the proton
$p$ and the neutron $n$  are the antiproton $\bar p$  ( $Q=-1,
~B=-1,~L=0$) and the antineutron $\bar n$ ($Q=0, ~B=-1,~L=0$). The
positron $e^{+}$ is the antiparticle of the electron. Its mass is
equal to $m_{e}$,~ $Q=1,~ B=0,~L=-1$. The antiparticle of the
neutrino is the antineutrino $\bar \nu$. It has $Q=0,~ B=0,~ L=-1$.

Due to the conservation of the lepton number, for example, in the
$\beta$-decay of the neutron an electron and an antineutrino are
emitted
$$n\to p +e^{-}+\bar\nu.$$

There exist in nature also neutral particles with all charges equal
to zero. Examples are the $\gamma$-quantum, the $\pi^{0}$-meson, etc.
In the case of such particles there is no notion of antiparticles (or
particles and antiparticles are identical).

In 1937 \cite{Majorana} great Italian physicist E. Majorana proposed
a theory of truly neutral particles with spin equal to 1/2 (which
today are called Majorana particles). E. Majorana was not satisfied
with the existing at that time theory of electrons and positrons in
which positrons were  considered as holes in the Dirac sea of the
states of electrons with negative energies. He wanted to formulate
the symmetrical theory in which there is no notion of states with
negative energies. In the paper "Symmetrical theory of electron and
positron" \cite{Majorana} he came also to a theory of spin 1/2
particles in which particles and antiparticles are identical.
Majorana wrote in the paper \cite{
Majorana}: "A generalization of
Jordan-Wigner quantization method allows not only to give symmetrical
form to the electron-positron theory but also to construct an
essentially new theory for particles without electric charge
(neutrons and hypothetical neutrinos)".

It is an open problem if the neutrino is a truly neutral Majorana
particle or a Dirac particle which possesses a lepton number. This is
one of the most important problems of modern neutrino physics and we
will discuss it later. Now we notice that if the lepton number is
conserved, the neutrino is a Dirac particle and $L(\nu)=-L(\bar
\nu)=1$. If there is no conserved lepton number, the neutrino is a
truly neutral Majorana particle. In the case of the Majorana neutrino with
$m\neq 0$ there
are only two states with momentum $\vec{p}$ and energy
$E_{p}=\sqrt{p^{2}+m^{2}}$: states with helicities $\pm 1$. Let us
notice that for the massless neutrino the theory of the two-component
Dirac neutrino and the Majorana neutrino are equivalent.

Before finishing this section we would like to notice that Landau
\cite{Landau}, Lee and Yang \cite{LeeYang} and Salam \cite{Salam} had
different arguments in favor of the two-component neutrino theory.

Landau assumed that the neutrino mass was equal to zero and for the
neutrino field he chose  $\nu_{L}(x)$ (or $\nu_{R}(x)$) assuming $CP$
invariance of the weak interaction ($C$ is charge conjugation, i.e.
the operation of transition from particles to antiparticles). Lee and
Yang assumed that the neutrino is a particle with helicity equal to
-1 (or +1). This is possible only if the neutrino mass is equal to
zero, parity is violated and the neutrino field is $\nu_{L}(x)$ (or
$\nu_{R}(x)$). Salam assumed invariance of the equation for the
neutrino field under $\gamma_{5}$-transformation
($\nu\to\gamma_{5}\nu$). From this requirement it follows that the
neutrino mass is equal to zero and the neutrino field is $\nu_{L}(x)$
(or $\nu_{R}(x)$).

\section{Measurement of neutrino helicity. Goldhaber et al. experiment}

Soon after  the two-component neutrino theory  had been proposed, the
neutrino helicity was determined from the results of  the spectacular
Goldhaber, Grodzins and Sunyar experiment \cite{Goldhaber}.

In this experiment the neutrino helicity was obtained from the measurement of the circular polarization of $\gamma$'s produced in the chain of reactions
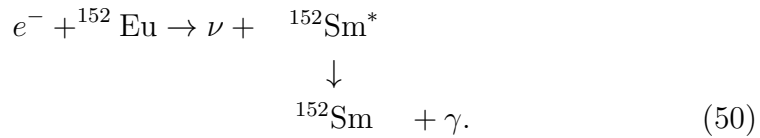
\begin{eqnarray}
e^- + ^{152}\rm {Eu} \to \nu + \null & ^{152}\rm {Sm}^* & \nonumber
\\
& \downarrow & \nonumber
\\
& ^{152}\rm{Sm} & \null + \gamma.\label{GGS}
\end{eqnarray}
The spins of $^{152}\rm {Eu} $ and $^{152}\rm{Sm}$ are equal to zero
and the spin of $^{152}\rm {Sm}^*$ is equal to one. Since the orbital
momentum of the initial electron is equal to zero (K-capture), from
the conservation of the projection of the total angular momentum on
the neutrino momentum we have
$$\frac{1}{2}h +m=\pm \frac{1}{2},$$
where $h$ is the neutrino helicity and $m$ is the projection of the spin of $^{152}\rm {Sm}^*$. From this relation we have
\begin{equation}\label{GGS1}
m=0,-1~~\mathrm{for}~~ h=1, \quad m=0,+1~~\mathrm{for}~~ h=-1.
\end{equation}
 Thus, the circular polarization of $\gamma$'s emitted in the direction of the $^{152}\rm {Sm}^*$ momentum is equal to the helicity of the neutrino. In the Goldhaber et al. experiment, the circular polarization of resonantly  scattered $\gamma$'s was measured (only $\gamma$'s emitted in the direction of motion of $^{152}\rm {Sm}^*$ satisfy the resonance condition). The authors concluded "... our result is compatible with 100\% negative helicity  of neutrino emitted in orbital electron capture".

Thus, the Goldhaber et al. experiment confirmed the two-component
neutrino theory.  It was established that from the two possibilities
for the neutrino field ($\nu_{L}(x) ~\mathrm{or}~\nu_{R}(x)$) the
first possibility was realized.

\section{Universal current $\times$ current V-A theory}

\begin{center}
{\bf Introduction}
\end{center}

The $V-A$ current$\times$current theory of the weak interaction
signified a great progress in the understanding of the weak
interaction and neutrino.  The Feynman and Gell-Mann,  Marshak and
Sudarshan idea of the left-handed components of all fields in the CC
Hamiltonian was triggered mainly by some experimental data, success
of the two-component neutrino theory and great intuition. The idea of
the left-handed components complemented with the idea of the
universality of the weak interaction made it possible to build the
simplest possible CC Hamiltonian of the weak interaction which is
characterized by only one (Fermi) constant. The authors of this
theory were courageous enough to state that some experimental data
which existed at that time but contradicted this theory were wrong.
Further experiments showed that the authors were correct:
current$\times$current $V-A$ theory is in perfect agreement with all
existing  CC data.

\vskip0.5cm

The most general Hamiltonian of the $\beta$-decay in the case of the
two-component neutrino is given by expression (\ref{twocompH}). It
includes five terms (S,V,T,A,P). There were many attempts to
determine the dominant terms of the Hamiltonian from the data of
different $\beta$-decay experiments. However, during many years  the
situation was contradictory. From the measurement of the angular
electron-neutrino correlation in the decay $^{6}\mathrm{He}\to
^{6}\mathrm{Li}+e^{-}+\bar\nu$ and from other data it followed that
S,T terms must be the dominant ones. On the other hand,
 the data on  the measurement of electron-neutrino correlation in the decay $^{35}\mathrm{Ar}\to ^{35}\mathrm{Cl}+e^{+}+\nu$ and other data were in favor of V,A terms.

In this uncertain experimental situation in 1958 two fundamental
theoretical papers by Feynman and Gell-Mann \cite{FeyGel} and Marshak
and Sudarshan \cite{MarSud} appeared. These authors proposed a
principle which allowed them to build the simplest possible universal
theory of the $\beta$-decay and other weak processes.

This theory was a generalization of the two-component neutrino
theory. Feynman and Gell-Mann,  Marshak and Sudarshan assumed that
{\em in the Hamiltonian of the weak interaction there enter not only
the left-handed component of the massless neutrino field but
left-handed components of all  fermion fields.} Thus, Feynman and Gell-Mann, Marshak
and Sudarshan  assumed that there has no connection between the mass
of a particle and the fact that the left-handed component of its
field enters into the Hamiltonian of the weak interaction. Feynman
and Gell-Mann assumed that $(\frac{1-\gamma_{5}}{2})\psi_{a}(x)$
enters into the Hamiltonian of the weak interaction because this
field satisfies a second order equation and could be considered as a
fundamental field.  Marshak and Sudarshan came to  left-handed
components from the requirement of $\gamma_{5}$ invariance of the
interaction (invariance under the change $\psi_{a}(x)\to
-\gamma_{5}\psi_{a}(x)$).

The Hamiltonian of the $\beta$-decay has in this case the form
\begin{equation}\label{VAHam}
 {\cal{H}}_{I}^{\beta}(x)= \sum_{i=S,V,T,A,P}G_{i}\, \bar{p}_{L}(x) O_{i} n_{L}(x)
\,~\bar{e}_{L}(x) O^{i} \nu_{L}(x)  + \rm{ h.c.}.
\end{equation}
We have
\begin{equation}\label{VAHam1}
\bar{p}_{L}(x) O_{i} n_{L}(x)=\bar{p}(x)\frac{1+\gamma_{5}}{2}
 O_{i}\frac{1-\gamma_{5}}{2} n(x).
\end{equation}
Using the algebra of the Dirac matrices $\gamma$'s it is easy to show that
\begin{equation}
\frac{1+\gamma_{5}}{2}
\left(1;\,\sigma_{\alpha\,\beta};\,\gamma_{5}\right)
\frac{1-\gamma_{5}}{2} =0\,. \label{VAHam2}
\end{equation}
Thus, $S$, $T$ and $P$ terms do not enter into the Hamiltonian
(\ref{VAHam1}). Moreover, $A$ and $V$ terms are connected by the
relation:
\begin{equation}
\frac{1+\gamma_{5}}{2}\, \gamma_{\alpha}\gamma_{5}\,
\frac{1-\gamma_{5}}{2} =-\frac{1+\gamma_{5}}{2}\gamma_{\alpha}\,
\frac{1-\gamma_{5}}{2}. \label{VAHam3}
\end{equation}
The Hamiltonian of the $\beta$-decay takes the simplest possible form
\begin{eqnarray}
{\cal{H}}_{I}^{\beta}(x)&=& \frac {G_{F}}{\sqrt{2}}4\,
\bar{p}_{L}(x)\gamma_{\alpha}  n_{L}(x) \,~
\bar{e}_{L}(x)\gamma^{\alpha}  \nu_{L}(x)  + \,h.c. \nonumber \\
&=& \frac {G_{F}}{\sqrt{2}}\, \bar{p}(x)\gamma_{\alpha}(1-\gamma_{5})n(x)
\, \bar{e}(x)\gamma^{\alpha}(1-\gamma_{5})  \nu (x) + \rm{h.c.}.
\label{VAHam4}
\end{eqnarray}
  The Hamiltonian (\ref{VAHam4}), like the Fermi Hamiltonian(\ref{bdecay}),
is characterized by only one  interaction constant $G_{F}$.
\footnote{Interesting that the title of  the Feynman and Gell-Mann
paper is  "Theory of the Fermi interaction".}
 There is, however, a crucial difference between the Hamiltonian
(\ref{VAHam4}) and the Fermi Hamiltonian. In the Hamiltonian
(\ref{VAHam4}) left-handed components of all fields enter. This means
that the Hamiltonian (\ref{VAHam4}) unlike the Fermi Hamiltonian does
not conserve parity.

What about numerous experiments from which it  followed that S and T
terms are the dominant terms of the Hamiltonian of the $\beta$-decay?
In the Feynman and Gell-Mann paper it is written "These theoretical
arguments seem to the authors to be strong enough to suggest that the
disagreement with $^{6}\mathrm{He}$ recoil experiment and with some
other less accurate experiments indicates that these experiments are
wrong". In fact,
 subsequent  experiments did
not confirm the results of the experiments which indicated in  favor
of  the dominance of S and T terms. The Hamiltonian (\ref{VAHam4})
described data of all experiments concerning the study of the
$\beta$-decay.

With the Feynman-Gell-Mann, Marshak-Sudarshan prescription
(left-handed components of all fields enter into the Hamiltonian of
the weak interaction), which lead to the unique expression for the
interaction Hamiltonian, it was natural to implement the idea of the
universal weak interaction which we discussed before.

For the Hamiltonian of the decay
\begin{equation}
\mu^{+} \to e^{+}+ \nu + \bar \nu.
\label{mudec}
\end{equation}
we have in this case
\begin{equation}\label{mudecH}
{\cal{H}}_{I}^{\mu\to e\nu\bar\nu}(x)=\frac {G_{F}}{\sqrt{2}}4\,
\bar{e}_{L}(x)\gamma_{\alpha}  \nu_{L}(x) ~
\bar\nu_{L}(x)\gamma^{\alpha}  \mu_{L}(x)  + h.c.
\end{equation}
From (\ref{mudecH}) it follows that the lifetime of the muon is given
by the expression $\tau_{\mu}=\frac{192
\pi^{3}}{G^{2}_{F}m^{5}_{\mu}}$, where $m_{\mu}$ is the mass of the
muon. Feynman and Gell-Mann demonstrated that if we take for $G_{F}$
the  value obtained from the superallowed $0^{+}\to 0^{+}$
$\beta$-decay of $^{14}\mathrm{O}$, we will find perfect agreement
with the experimental lifetime. This was an important confirmation of
the hypothesis of the universality of the weak interaction.

This agreement was also an evidence in favor of the conserved vector
current (CVC) hypothesis \cite{GerZeld}. According to this hypothesis
the weak vector current is the  "charged" component of the isovector
current which is conserved due to isotopic invariance. The
conservation of the vector current ensures the fact that the Fermi
constant is not renormalized by the strong interaction.

 From the
$\mu-e$ universality it followed that the Hamiltonian of the
$\mu$-capture and other connected processes can be obtained from
(\ref{VAHam4}) by the change $e(x)\to \mu(x)$. We have
\begin{equation}
{\cal{H}}_{I}^{\mu}(x)= \frac {G_{F}}{\sqrt{2}}4\,
\bar{p}_{L}(x)\gamma_{\alpha}  n_{L}(x) \,~
\bar{\mu}_{L}(x)\gamma^{\alpha}  \nu_{L}(x)  + \rm{h.c.}.
\label{VAHam5}
\end{equation}
At the time when Feynman and Gell-Mann, Marshak and Sudarshan wrote
their papers there was a contradiction to
 the idea of $\mu-e$
universality of the weak interaction with the  data on the
measurement of the width of the decay $\pi^{+}\to e^{+}+\nu$. From
(\ref{VAHam4}) and (\ref{VAHam5})it follows that the ratio  of the
decay widths $R=\frac{\Gamma(\pi^{+}\to e^{+}\nu)}{\Gamma(\pi^{+}\to
\mu^{+}\nu)}$ is given by the expression
\begin{equation}\label{ratio}
R=\frac{m^{2}_{e}}{m^{2}_{\mu}}
\frac{(1-\frac{m^{2}_{e}}{m^{2}_{\pi}})^{2}}
{(1-\frac{m^{2}_{\mu}}{m^{2}_{\pi}})^{2}}\simeq 1.2\cdot 10^{-4}.
\end{equation}
On the other hand in the experiment \cite{Anderson} the decay
$\pi^{+}\to e^{+}+\nu$ was not observed and  it was found
$R<10^{-5}$. Feynman and Gell-Mann wrote "This is a very serious
discrepancy. The authors have no idea on how it can be resolved".

In 1958 a new experiment on measurement of the $\pi^{+}\to
e^{+}+\nu$-decay was performed at CERN \cite{Fidecaro}. In this
experiment, perfect agreement with prediction (\ref{ratio}) of the
universal Feynman and Gell-Mann, Marshak and Sudarshan theory was
obtained.\footnote{When this result  was obtained Feynman was
visiting CERN. The news reached him when he was queuing  in the CERN
cafeteria. It is said that when Feynman learnt about the $\pi\to
e\nu$-news he started to dance.}

In order to unify interactions (\ref{VAHam4}), (\ref{mudecH}) and
(\ref{VAHam5}) Feynman and Gell-Mann introduced  the $\mu-e$
symmetric weak current
\begin{equation}
j^{\alpha}=2 ~(\bar{p}_L \gamma^{\alpha} n_L \ + \ \bar{\nu}_{ L}
\gamma ^{\alpha} e_L \ + \ \bar{\nu}_{  L} \gamma^{\alpha} \mu_L)
\label{CC}
\end{equation}
and assumed that the total Hamiltonian of the weak interaction had
the current$\times$current form
\begin{equation}
{\cal{H}}_I = \frac{G_{F}}{\sqrt{2}} \ j^{\alpha} \ j^{+}_{\alpha},
\label{CCH}
\end{equation}
where  $G_{F}$ was the Fermi constant. Two remarks are in order.
\begin{enumerate}
  \item The hadron part of the current has the form
 $$j^{\alpha} =v^{\alpha} - a^{\alpha},$$  where $v^{\alpha}=\bar{p}
\gamma^{\alpha} n$ and $a^{\alpha}=\bar{p} \gamma^{\alpha}\gamma_{5}
n$ are the vector and axial currents.\footnote{This is the reason why
the Feynman and Gell-Mann, Marshak and Sudarshan theory is called the
$V-A$ theory.} Notice that Fermi $\beta$-transitions of nuclei are
due to the vector current and Gamov-Teller transitions are due to the
axial current.

\item The current $j^{\alpha}$ provides transitions $n\to p$,
$e^{-}\to \nu$, etc. in which $\Delta Q=Q_{f}-Q_{i}=1$
($Q_{i}(Q_{f})$ is the initial (final) charge). By this reason the
current $j^{\alpha}$ is called the charged current (CC).
\end{enumerate}
There are two types of terms in the Hamiltonian (\ref{CCH}):
nondiagonal and diagonal. The nondiagonal terms have the form
\begin{eqnarray}
{\cal{H}}_I^{nd}= \frac{G_{F}}{\sqrt{2}} 4\,\{[(\bar{p}_L
\gamma^{\alpha} n_L)( \bar{e}_L \gamma_{\alpha} \nu_{L}) \ + \rm{
h.c.}] \ + \ \nonumber\\ \ [(\bar{p}_L \gamma^{\alpha} n_L)
(\bar{\mu}_L \gamma_{\alpha} \nu_{ L}) \ + \rm{ h.c.}] \ + \
\nonumber\\
 \ [(\bar{e}_L \gamma^{\alpha} \nu_{L}) (\bar{\nu}_{ L}
\gamma_{\alpha} \mu_L) \ + \rm{ h.c.}]\} \label{ndiagH}
\end{eqnarray}
The first term of this expression is the Hamiltonian of $\beta$-decay
of the neutron $n\to p +e^{-}+\bar\nu$, of the process $\bar{\nu} + p
\to e^+ + n$ and other processes. The second term of (\ref{ndiagH})
is the Hamiltonian of the process $\mu^{-}+ p \to \nu + n$,  of the
neutrino process $\nu + n \to \mu^- + p$ and other processes.
Finally, the third term of (\ref{ndiagH}) is the Hamiltonian of the
$\mu$-decay (\ref{mudec}), of the process $\nu +e^{-}\to \mu^{-}+\nu$
and other processes.

The diagonal terms of the Hamiltonian (\ref{CCH}) are given by
\begin{equation}
{\cal{H}}^{d}= \frac{G_{\rm{F}}}{\sqrt{2}} ~
4[(\bar{\nu}_{L}\gamma^{\alpha} e_L)
  (\bar{e}_L \gamma_{\alpha} \nu_{L})
+(\bar{\nu}_{ L} \gamma^{\alpha} \mu_L)
  (\bar{\mu}_L \gamma_{\alpha} \nu_{ L})
+(\bar{p}_L \gamma^{\alpha} n_L)
  (\bar{n}_L \gamma_{\alpha} p_L)]
\label{ndiagH1}
\end{equation}
The first term of (\ref{ndiagH1}) is the Hamiltonian of the processes
of elastic scattering of a neutrino and an antineutrino on an
electron
\begin{equation}
\nu + e \to \nu + e  \label{nuescat}
\end{equation}
and
\begin{equation}
\bar \nu + e \to \bar\nu + e, \label{barnuescat}
\end{equation}
of the process $e^{+}+ e^{-}\to\bar \nu +\nu$  and other processes.
Such processes were not known in the fifties. Their existence and the
cross sections of these  processes were {\em predicted by the
current$\times$current  theory.}

The cross sections of the processes (\ref{nuescat}) and
(\ref{barnuescat}) are very small (at MeV's energies of the order
$10^{-45}\mathrm{cm}^{2} $).  The observation of such processes was a
challenge. After many years of efforts, the cross section of the
process (\ref{barnuescat}) was measured by F. Reines et al.
\cite{Reinesnue} in an experiment with antineutrinos from a reactor.
At that time, the Glashow-Weinberg-Salam
Standard Model already existed. According to the
Standard model, to the matrix elements of the processes
(\ref{nuescat}) and (\ref{barnuescat})there contribute the
Hamiltonian (\ref{ndiagH1}) and an additional (so called neutral
current) Hamiltonian. The result  of the experiment by F.Reines et
al. was in agreement with the Standard Model.

In the  Feynman and Gell-Mann  and Marshak and Sudarshan papers
decays of the $\Lambda$-hyperon and other strange particles were also
briefly discussed. However, weak interaction of the strange particles
was fully included into the current $\times$ current Hamiltonian in
1963 by N. Cabibbo \cite{Cabibbo}. We will discuss Cabibbo's
contribution to the theory of weak interaction later.

\section{Intermediate vector $W$ boson}

\begin{center}
{\bf Introduction }
\end{center}
The great Yukawa idea that the interaction between nucleons is due to
the exchange of a meson (which allowed him to predict the
$\pi$-meson) was applied by Klein to the short range weak
interaction. Klein assumed that the weak decay of the neutron was due
to the exchange of a heavy charged vector boson between $(np)$ and
$(e\nu)$ pairs. It is impressive that this general quantum idea very
early in the thirties allowed one to anticipate the existence of a
very heavy particle which could be observed only many years later
after modern high-energy accelerators were built.

\vskip0.5cm

In the  Feynman and Gell-Mann paper, which we discussed in the
previous section, it was mentioned that the current$\times $current
Hamiltonian of the weak interaction (\ref{CCH}) could originate from
the exchange of a heavy intermediate charged vector meson.( "We have
adopted the point of view  that the weak interactions all arise from
the interaction of a current $J_{\alpha}$ with itself, possibly via
an intermediate charged vector meson  of  high mass" \cite{FeyGel}.)
We will discuss now the hypothesis of a charged intermediate vector
boson. Let us assume that there exists a charged vector  $W^{\pm}$
boson  and that the Lagrangian of the weak interaction  has the form
of a scalar product of the current $j^{\alpha}$ given by
Eq.(\ref{CC}) and the vector field $W_{\alpha}$
\begin{equation}\label{WH}
{\cal{L}_{I}}=-\frac{g}{2\sqrt{2}}~j_{\alpha}~W^{\alpha}+\rm{h.c.},
\end{equation}
where $g$ is a dimensionless interaction constant.

Let us notice that the Lagrangian (\ref{WH}) has the form analogous
to the Lagrangian of the electromagnetic interaction
\begin{equation}\label{1EH}
\mathcal{L}^{EM}_{I}=-ej^{EM}_{\alpha}A^{\alpha}.
\end{equation}
where $j^{EM}_{\alpha}$ is the electromagnetic current, $A^{\alpha}$
is the electromagnetic field and $e$ is the electric charge.

If the Lagrangian of the weak interaction has the form (\ref{WH}), in
this case the $\beta$-decay of the neutron proceeds in the following
three steps: 1. neutron produces the virtual $W^{-}$-boson and is
transferred into proton; 2. the virtual $W^{-}$-boson propagates; 3.
the virtual $W^{-}$-boson decays into a electron and an antineutrino.
In the Feynman diagram, the propagator of the $W$-boson contains a
factor $\frac{-1}{Q^{2}- m^{2}_{W}}$, where $Q=p_{n}-p_{p}$ is the
momentum transfer and $m_{W}$ is the mass of the $W$-boson. If the
$W$-boson is a heavy particle (say, with a mass which is much larger
than the mass  of the proton), in this case $Q^{2}$ in the
$W$-propagator can be safely neglected and the matrix element of the
$\beta$-decay of the neutron can be obtained from the Hamiltonian
(\ref{CCH}) in which the Fermi constant is given by the relation
\begin{equation}\label{Fconstant}
    \frac{G_{\rm{F}}}{\sqrt{2}}=\frac{g^{2}}{8~m^{2}_{W}}.
\end{equation}
 In a similar way it can be shown that in the region of relatively small energies,
the  matrix elements of all weak processes with the virtual
(intermediate) charged $W$-boson can be obtained from the
current$\times$current Hamiltonian (\ref{CCH}) in which the Fermi
constant is given by relation (\ref{Fconstant}).\footnote{From the
point of view of the theory with the $W$-boson, the
current$\times$current Hamiltonian with the Fermi constant
(\ref{Fconstant}) is the effective Hamiltonian of the  weak
interaction.}

Thus, the theory with the vector $W^{\pm}$-boson could explain the
current$\times$current structure of the weak interaction Hamiltonian
and the fact that  the Fermi constant has the dimension $[M]^{-2}$.

We know today that the intermediate charged $W^{\pm}$-boson exists. The
$W^{\pm}$-boson is one of the heaviest particles: its mass is equal
to $m_{W}\simeq 80.4~\mathrm{GeV}$. For the discovery of the
$W^{\pm}$-boson and the $Z^{0}$-boson (see later) in 1984 C. Rubbia
and S. van der Meer were awarded the Nobel Prize. As we will see
later, the Lagrangian (\ref{WH}) is  part of the total Lagrangian of
the Standard Model.

The first idea of the charged vector boson, mediator of the weak
interaction, was discussed by O. Klein \cite{Klein} in 1938, soon
after the Fermi $\beta$-decay paper had appeared. Fermi built the
first Hamiltonian of the $\beta$-decay by  analogy with
electrodynamics. O.Klein noticed that the analogy would be more
complete if a charged vector boson (analog of the $\gamma$-quantum)
existed and the  weak interaction originated from an interaction
which (like the electromagnetic interaction) had the form of a
product of a current and a vector field. In order to build such a
theory O.Klein assumed gauge invariance.

\section{The first method of neutrino detection}

As we discussed before, because of the extreme smallness of the cross
section for the absorption of neutrinos by nuclei for many years most
physicists considered the neutrino as an undetectable particle.

The first method of neutrino detection was proposed by B. Pontecorvo in 1946 \cite{BPonte46}.  He  wrote "The object of this note is to show that the experimental observation of an inverse $\beta$ process
produced by neutrino is not out of the question with the modern experimental facilities, and to suggest a method which might make an experimental observation feasible."

Pontecorvo proposed radiochemical methods of neutrino detection.
Let us consider, as an example, the reaction
\begin{equation}\label{nurec}
\nu+^{37}\mathrm{Cl}\to e^{-}+ ^{37}\mathrm{Ar}.
\end{equation}
The $^{37}\mathrm{Ar}$ atoms decay (via K-capture) with a life-time
of about 34 days.

After irradiation of a  target (containing $^{37}\mathrm{Cl}$) by
neutrinos for a relatively long time (say, one month) a few
radioactive atoms of $^{37}\mathrm{Ar}$ could be produced. As argon
is a nobel gas, atoms of $^{37}\mathrm{Ar}$ can be extracted from the
target and can be placed into a proportional counter in which their
decay will be detected. This is the main idea of Pontecorvo's
radiochemical method. He discussed different reactions. He considered
the Cl-Ar reaction (\ref{nurec}) as very appropriate for the neutrino
detection (a large volume of liquid Carbon Tetra-Chloride can be used
as a target, $^{37}\mathrm{Ar}$ atoms have a convenient life-time,
etc.).

In the report \cite{BPonte46} B.Pontecorvo also pointed out the
following intensive sources of neutrinos which existed at that time:
\begin{enumerate}
  \item The Sun. The flux of the solar neutrinos is equal to
$ 6\cdot 10^{10} \frac{\nu}{\mathrm{cm}^{2}\mathrm{sec}}$
  \item Nuclear reactors during operation.
\footnote{Pontecorvo wrote "Probably this is the most convenient
neutrino source".} The total flux of (anti) neutrinos from a reactor
is approximately equal to $ 2\times
10^{20}\frac{\bar\nu}{\mathrm{sec}}$ per $GW _{thermal}$.
  \item Radioactive sources which can be prepared in  reactors.
\end{enumerate}
Pontecorvo's radiochemical method of neutrino detection was used in
solar neutrino experiments. The first experiment in which solar
neutrinos were detected was performed by R. Davis and collaborators
\cite{Davis}. In this experiment solar neutrinos were detected via
the observation of the $\mathrm{Cl}-\mathrm{Ar}$ reaction
(\ref{nurec}). In 2002, R. Davis was awarded the Nobel Prize for this
experiment.

\section{Detection of neutrino. Reines and Cowan experiment}

\begin{center}
{\bf Introduction }
\end{center}
The pioneering reactor neutrino experiment by Reines and Cowan proved
the Pauli-Fermi hypothesis of the neutrino. The value of the cross
section measured in this experiment confirmed the correctness of the
V-A theory of weak interaction. This experiment opened a new era in
neutrino physics: era of experiments with reactor neutrinos.
\vskip0.5cm

The first proof of the existence of neutrinos was obtained in 1953-59
in the F.Reines and C.L. Cowan experiments \cite{Reines}. In these
experiments antineutrinos from the Savannah River reactor\footnote{In
the beginning Reines and Cowan planned to do an experiment with
neutrinos from an atomic bomb explosion. Later they understood that
an experiment with reactor antineutrinos was simpler and  feasible.
Reines remembered in his Nobel lecture "I have wondered since why it
took so long for us to come to this now obvious conclusion and how it
escaped others around us with whom  we talked..."} were detected
through the observation of the process
\begin{equation}\label{RC}
\bar \nu+p \to e^{+} +n.
\end{equation}
Antineutrinos are produced in a reactor in a chain of $\beta$-decays
of neutron-rich nuclei, products of the fission of uranium and
plutonium. The energies of antineutrinos from a reactor are $\lesssim 10$
MeV. About $2.3\cdot 10^{20}$ antineutrinos  per second were emitted by the
Savannah River reactor. The flux of $\bar\nu_{e}$'s in
the Reines and Cowan experiment was about $10^{13}
\rm{cm}^{-2}s^{-1}$.

A liquid scintillator \footnote{Reines and Cowan were the first who
understood that the phenomenon of scintillation of organic  liquids,
discovered at that time,  could be employed in order  to build a big
(1 $m^{3}$) detector which was necessary to detect neutrinos.}
($1.4\cdot 10^{3}$ liters) loaded with $\rm{CdCl}_{2}$ was used as a
target in the experiment. A positron produced in the process
(\ref{RC}), slowed down in the scintillator and annihilated with an
electron, producing two $\gamma$- quanta with energies of $\simeq
0.51$ MeV and opposite momenta.

A neutron, produced in the process (\ref{RC}), slowed down in the
target and was captured by $\rm{Cd}$ within about 5 $\mu$s, producing
a $\gamma$-quantum in the process $n+^{108}\rm{Cd}\to
^{109}\rm{Cd}+\gamma$ (at small energies the cross section of this
process is very large). The $\gamma$-quanta were detected by 110
photomultipliers. Thus, the signature of the $\bar\nu$-event in the
Reines and Cowan experiment was two $\gamma$-quanta from the
$e^{+}-e^{-}$-annihilation in coincidence with a delayed
$\gamma$-quantum from the neutron capture by cadmium. For the cross
section of the process (\ref{RC}) the value
\begin{equation}\label{RC3}
    \sigma_{\nu}= (11\pm 2.6)~10^{-44}~\rm{cm}^{2}
\end{equation}
was obtained in the latest measurements \cite{Reines}. This value was  in agreement with the predicted value.

In the V-A current$\times$current theory  the cross section of the
process (\ref{RC}) is connected with the life-time $\tau_{n}$ of the
neutron by the relation
\begin{equation}\label{RC1}
\sigma(\bar \nu_{e}p \to
e^{+}n)=\frac{2\pi^{2}}{m^{5}_{e}f\tau_{n}}p_{e}E_{e},
\end{equation}
where $E_{e}\simeq E_{\bar\nu}-(m_{n}-m_{p})$ is the energy of the
positron, $p_{e}$ is the positron momentum, f=1.686
is the
phase-space factor, $m_{n}, m_{p}, m_{e}$ are the masses of the
neutron, proton and electron, respectively.  From (\ref{RC1}) for the
cross section of the process (\ref{RC}), averaged over the
antineutrino spectrum, the value
\begin{equation}\label{RC2}
  \bar  \sigma(\bar \nu_{e}p \to e^{+}n)\simeq 9.5 \cdot
    10^{-44}~\rm{cm}^{2}
\end{equation}
was found. In 1995 the Nobel Prize in Physics was awarded to F. Reines
"for the detection of the neutrino".\footnote{Clyde  Cowan died in 1974.}

\section{Lepton number conservation. Davis experiment}

The particle which is produced in the $\beta$-decay together with the
electron is called the antineutrino. It is a direct consequence of
the quantum field theory that an antineutrino can produce a positron
in the process  (\ref{RC}) and other similar processes. Can
antineutrinos produce electrons in weak-interaction processes with
nucleons? This question was studied in an experiment which was
performed in 1956 by Davis  \cite{Davis1} at the Savannah River
reactor. This was the first application of Pontecorvo's radiochemical
method. Radioactive $^{37}\rm{Ar}$ atoms which could be produced in
the process
\begin{equation}\label{Davis}
\bar\nu +^{37}\rm{Cl}\to e^{-}+   ^{37}\rm{Ar}
\end{equation}
were searched for in this experiment. No $^{37}\rm{Ar}$ atoms were
found. For the cross section of the process (\ref{Davis}) the
following upper bound was obtained:
$$\sigma(\bar\nu +^{37}\rm{Cl}\to e^{-}+   ^{37}\rm{Ar})<0.9\cdot 10^{-45}~\mathrm{cm}^{2}.$$
This bound is about five times smaller than the
 cross section of the corresponding reaction with the neutrino.

 Thus, it was established that antineutrinos from a reactor can produce
positrons (the Reines-Cowan experiment) but can not produce electrons (the Davis experiment).

This result can be explained if we assume that there exists
{\em conserving lepton number $L$} and $\nu$
and $e^{-}$ have the same values of $L$. Let us put $L(\nu)=L(e^{-})=1$.
By definition, the lepton numbers of antiparticles are opposite to the lepton numbers of particles. We have
$L(\bar\nu)=L(e^{+})=-1$.
 We also assume that
the lepton numbers of proton, neutron and other hadrons are equal to
zero. The conservation of the lepton number explains the negative result of
the Davis experiment.

\section{Discovery of muon neutrino. The Brookhaven neutrino experiment}
\begin{center}
{\bf Introduction}
\end{center}

The discovery of the second (muon) neutrino was a great event in
physics. It was proved that two different neutrinos $\nu_{e}$ and
$\nu_{\mu}$ corresponded to  two different (in mass) leptons $e$ and
$\mu$. Now we know that with the discovery of $\nu_{\mu}$ it was
established that in addition to the first family of leptons
($\nu_{e},e$) there existed
the second family ($\nu_{\mu},\mu$).

The Brookhaven neutrino experiment was the first experiment with high
energy neutrinos originating from decays of pions, kaons and muons
produced at accelerators. As we will see later,  important
discoveries were made in high energy accelerator neutrino
experiments.

\vskip0.5cm

When the universal V-A  theory of  weak interaction was formulated by
Feynman and Gell-Mann, Marshak and Sudarshan they considered only one
type of neutrinos.

 There was, however, an idea of many physicists that neutrinos which
take part in the weak interaction together with an electron and a
muon could be different.\footnote{Pontecorvo \cite{Ponte62}
remembered "....for people working with muons in the old times, the
question about different types of neutrinos has always been present.
True, later on many theoreticians forgot all about it and some of
them "invented" again the two neutrinos....} Let us call neutrinos
which participate in weak processes together with electrons and
neutrinos which participate in weak processes together with muons,
correspondingly, the electron and muon neutrinos ($\nu_{e}$ and
$\nu_{\mu}$). The charged current of the current$\times$current
theory takes in this case the form
\begin{equation}
j^{\alpha}=2 ~(\bar{p}_L \gamma^{\alpha} n_L \ + \ \bar{\nu}_{\mu L}
\gamma ^{\alpha} e_L \ + \ \bar{\nu}_{ e L} \gamma^{\alpha} \mu_L)
\label{CC1}
\end{equation}
 {\em Are
$\nu_{e}$ and $\nu_{\mu}$ the same or different particles?} The
answer to this fundamental question was obtained in the famous
Brookhaven neutrino experiment in 1962 \cite{Brookhaven}.

The first indication that  $\nu_{e}$ and $\nu_{\mu}$ are different
particles was obtained from the data on the search for the decay
$\mu\to e \gamma$. If $\nu_{e}$ and $\nu_{\mu}$ are identical
particles, the $\mu\to e \gamma$ decay is allowed. The probability of
the decay $\mu\to e \gamma$  in the theory with the $W$-boson was
calculated in \cite{Feinberg}\footnote{Such a theory is a non
renormalizable one. In \cite{Feinberg} the cut-off $\Lambda\simeq
m_{W}$ was applied.} soon after the V-A theory has been proposed. It
was found that the ratio $R$ of the probability of the decay
$\mu^{+}\to e^{+} \gamma$ to the probability of the decay $\mu^{+}\to
e^{+}+\nu+\bar\nu$ was given by
\begin{equation}\label{muegamma}
    R\simeq\frac{\alpha}{24~ \pi}\simeq 10^{-4}
\end{equation}
The decay $\mu\to e \gamma$ was not observed in experiment. At the
time of the Brookhaven experiment, for the upper bound of the ratio
$R$ much smaller than (\ref{muegamma}) value
\begin{equation}\label{muegamma1}
    R< 10^{-8},
\end{equation}
has been found.

A direct proof of the existence of the second (muon) type of the
neutrino was obtained by L.M. Lederman, M. Schwartz, J. Steinberger
et al. in the first experiment with accelerator neutrinos in 1962.
The idea of the experiment was proposed by B.Pontecorvo in 1959
\cite{Ponte59}.\footnote{B. Pontecorvo came to the idea of such an
experiment thinking about a possible neutrino program at Meson
Factories which were under construction at different places. "At the
Laboratory of Nuclear Problems of JINR (Dubna) in 1958 a proton
relativistic cyclotron was being designed with a beam energy 800 MeV
and beam current 500 A. I started to think about experimental
research program for such an accelerator"\cite{Ponte62}. The Dubna
Meson Factory eventually was not built.}

A beam of $\pi^{+}$'s  in the Brookhaven experiment was obtained by
the bombardment of a Be target by protons with an energy of 15 GeV.
In the decay channel (about 21 m long) practically all $\pi^{+}$'s
decay. After the channel there was  shielding material (13.5 m of
iron), in which charged particles were absorbed. After the shielding
there was neutrino detector (aluminium spark chamber, 10 tons) in
which the production of charged leptons was observed.

The dominant decay channel of the $\pi^{+}$-meson is
\begin{equation}\label{BNL1}
\pi^{+}\to \mu^{+} +\nu_{\mu}.
\end{equation}
According to the universal $V-A$ theory, the ratio $R$ of the width
of the decay
\begin{equation}\label{BNL2}
\pi^{+}\to e^{+} +\nu_{e}
\end{equation}
to the width of the decay (\ref{BNL1}) is given by the relation (\ref{ratio}).
Thus, the decay  $ \pi^{+}\to e^{+} \nu_{e}$
 is strongly suppressed with respect to the decay
$\pi^{+}\to \mu^{+} +\nu_{\mu}$.

The reason for this
suppression can be easily understood. Indeed, let us consider the
decay (\ref{BNL2}) in the rest frame of the pion. The helicity of the
neutrino is equal to -1. If we neglect the mass of the $e^{+}$, the
helicity of the positron will be equal to +1 (the helicity of the
positron will be  in this case the same as the helicity of the
antineutrino). Thus, the projection of the total angular momentum on
the neutrino momentum will be equal to -1. The spin of the pion is
equal to zero and consequently the process (\ref{BNL2}) in the limit
$m_e \to 0$ is forbidden. These arguments explain the appearance of
the small factor $(\frac{m_e}{m_\mu})^2 $ in (\ref{ratio}). From
(\ref{ratio}) it follows that  the neutrino beam in the Brookhaven
experiment was practically a pure $\nu_{\mu}$ beam (with a small
admixture of $\nu_{e}$ from decays of muons and kaons).

Neutrinos, emitted in the decay (\ref{BNL1}), produce $\mu^{-}$ in
the process
\begin{equation}\label{BNL4}
\nu_{\mu}+N \to  \mu^{-}+X.
\end{equation}
If $\nu_{\mu}$ and $\nu_{e}$ were  the same particles, neutrinos from
the decay (\ref{BNL1}) would produce also $e^{-}$ in the reaction
\begin{equation}\label{BNL4}
\nu_{\mu}+N \to  e^{-}+X.
\end{equation}
Due to the $\mu-e$ universality of the weak interaction one could
expect  to observe in the detector  practically equal
numbers of muons and electrons.

In the Brookhaven experiment 29 muon events were detected. The
observed six electron candidates could be explained by the
background. The measured cross section was in agreement with the
$V-A$ theory. Thus, it was proved that {\em $\nu_{\mu}$ and $\nu_{e}$
were different particles.}

In 1963, with the invention of the magnetic horn at the CERN
laboratory the intensity and purity of neutrino beams were greatly
improved. In a more precise 45 ton spark-chamber experiment and in
the large bubble chamber experiment the Brookhaven result  was fully
confirmed.

The results of the Brookhaven and other experiments suggested that
the total electron and muon  lepton numbers $L_{e}$ and
$L_{\mu}$, which are called flavor lepton numbers, were conserved:
\begin{equation}\label{Lemucons}
    \sum_{i} L^{(i)}_{e}=\rm{const};~~\sum_{i}L^{(i)}_{\mu}=\rm{const}
\end{equation}
The flavor lepton numbers of particles are given in  Table I. The lepton
numbers of antiparticles are opposite to the lepton numbers of the
corresponding particles.
\begin{center}
 Table I
\end{center}
\begin{center}
 Flavor lepton numbers of particles
\end{center}

\begin{center}
\begin{tabular}{|cccc|}
\hline Lepton number & $\nu_{e}\,~ e^{-}$ & $\nu_{\mu}\,~  \mu^{-}$
& \rm{hadrons}, $\gamma$
\\
\hline $L_{e}$ & 1 & 0 & 0
\\
$L_{\mu}$ & 0 & 1 & 0
\\
\hline
\end{tabular}
\end{center}
We know now that the notion of the flavor lepton number is an
approximate one. It is valid only if we neglect small neutrino
masses. The conservation laws (\ref{Lemucons}) are violated in
neutrino transitions (oscillations) which are  due to small neutrino
masses and neutrino mixing. Later we will discuss neutrino masses,
mixing and oscillations in detail.

In 1988 the Nobel Prize was awarded to L. Lederman, M. Schwartz and
J. Steinberger "for the neutrino beam method and the demonstration of
the doublet structure of the leptons through the discovery of the
muon neutrino".

\subsection{Strange particles. Quarks. Cabibbo current}
\begin{center}
{\bf Introduction}
\end{center}

With the idea of quarks, physics of elementary particles and, in
particular, physics of the weak interaction and of the neutrino was
changed. If the fundamental weak interaction is {\em the interaction
of quarks and leptons}, the phenomenological rules $|\Delta S|=1$ and
$\Delta Q=\Delta S$, which were established for semi-leptonic decays
of strange particles, have a natural explanation. The prediction of
the charmed quark was motivated by the  Cabibbo mixture of quarks,
and the Cabibbo-GIM mixture of quarks implied a symmetry between
lepton and quark terms in the charged weak current. This symmetry was
based on the fact that the number of  lepton pairs ( ($\nu_{e},
e^{-}$) and ($\nu_{\mu}, \mu^{-}$)) was equal to the number of quark
pairs (($u, d$) and ($c,s$)). Taking into account that fields of $d$
and $s$ quarks are mixed it was natural to extend the lepton-quark
symmetry of the charged current and to assume that neutrinos are also
mixed. This implies the assumption that neutrinos have small, nonzero
masses.

\vskip0.5cm

The current$\times$current Hamiltonian (\ref{CCH}) with CC current
(\ref{CC1}) is the Hamiltonian of such processes in which $e$,
$\nu_{e}$, $\mu$, $\nu_{\mu}$, $p$, $n$, $\pi^{\pm}$ and other non
strange particles take part. The strange particles were discovered in
cosmic rays in the fifties. Their decays were studied in detail in
accelerator experiments. From the investigation of the semi-leptonic
decays
\begin{eqnarray}
K^+ \to \mu^+ + \nu_\mu, ~~~~ \Lambda \to n + e^- + \bar{\nu}_e ,
\nonumber \\
\Sigma^- \to n + e^- + \bar{\nu}_e,~~~~\Xi^- \to \Lambda + \mu^- +
\bar{\nu}_\mu  \nonumber
\end{eqnarray}
and others  the following {\em  phenomenological rules} were
formulated.

I. The strangeness $S$ in the decays of strange particles is changed
by one $$|\Delta S|=1.$$
Here $\Delta S =S_f - S_i$, where $ S_i$
 ($ S_f $)  is the initial (final) total strangeness of the hadrons.
As an example, according to this rule, the decay $\Xi^{-}\to
p+\pi^{-}+e^{-}+\bar\nu_{e}$, in which $\Delta S=2$, is forbidden.
From the data of experiments for the ratio $R$ of the width of the
decay  $\Xi^{-}\to n+\pi^{-}+e^{-}+\bar\nu_{e}$ to the total decay
width of $\Xi^{-}$ the following upper bound was obtained: $R<4\cdot
10^{-4}$. \footnote{Here and below we present data given in "The
Review of Particle Physics"\cite{PDG}.}

II. The semi-leptonic  decays of strange particles obey the rule
$$\Delta Q = \Delta S.$$
Here $\Delta Q = Q_f - Q_i$ where $ Q_i$ ($ Q_f $) is the initial
(final) total electric charge of hadrons (in the unit of the proton
charge). According to this rule, the decay $\Sigma^{+}\to
n+e^{+}+\nu_{e}$ is forbidden. From experimental data it follows that
$R<5\cdot 10^{-6}$.

III. The decays of  strange particles are suppressed with respect to
the decays of non-strange particles.

In 1964, Gell-Mann and Zweig made the assumption that  strange and
nonstrange hadrons are bound states of  $u$, $d$ and $s$ quarks. The
quantum numbers of the quarks are presented in  Table II.
\begin{center}
 Table II.
\end{center}
\begin{center}
 Quantum numbers of quarks ( $Q$ is the charge, $S$ is the strangeness, $B$
 is the baryon number)
\end{center}
\begin{center}
\begin{tabular}{|cccc|}
\hline Quark & $ Q$ & $ S$ & $B$
\\
\hline $u$ & 2/3 & 0 & 1/3
\\
\hline $d$ & -1/3 & 0 & 1/3
\\
\hline $s$ & -1/3 & -1 & 1/3
\\
\hline
\end{tabular}
\end{center}
From the point of view of the theory of quarks $p$, $n$, $\Lambda$,
$\Sigma^{+}$, $\Xi^{-}$ and other baryons are bound states of three
quarks ($p=(uud),~n=(udd),~\Lambda=(uds),~ \Sigma^{+}=(uus),~
\Xi^{-}=(dss)$, etc.) and $\pi^{+}$, $K^{-}$, $\bar K^{0}$ and other
mesons are bound states of a quark and an antiquark ($\pi^{+}=(u\bar
d),~K^{-}=(s \bar u),~ \bar K^{0}=(d\bar s)$, etc.).

One of the first argument
s in favor of the quark structure of the
hadrons was obtained from the study of the weak decays of strange
particles. In expression (\ref{CC1}) for the charged current enter
the fields of protons and neutrons. If a proton and a neutron are
bound states of quarks we can assume  that the fundamental
weak interaction is the interaction of charged leptons, neutrinos and
quarks.

Let us build hadronic charged currents from the quark fields. The
current (\ref{CC1}) changes the charge by one. If we  accept the
Feynman-Gell-Mann, Marshak-Sudarshan  prescription (the left-handed
components of the fermion fields enter into the weak current) there
are only two possibilities to build such  currents from the fields of
$u$, $d$ and $s$ quarks:
\begin{equation}\label{qCC1}
 j^{\Delta S =0}_{\alpha}=2\bar{u}_L \gamma_{\alpha} d_L~~~\rm{and}~~~j^{\Delta S =0}_{\alpha}= 2\bar{u}_L \gamma_{\alpha}
    s_L.
\end{equation}
The first current changes the charge by one  and does not change
the strangeness ($\Delta Q =1 $, $\Delta S =0) $. The second current
changes the charge by one and the strangeness by one ($\Delta Q =1, $
$\Delta S =1 $). The matrix elements of these currents automatically
satisfy  $|\Delta S|=1$ and $\Delta Q = \Delta S$ rules.

In order to take into account the rule III (suppression of the decays
with the change of the strangeness with respect to the decays in
which the strangeness is not changed) and describe decays of
different strange particles  N. Cabibbo \cite{Cabibbo} introduced a
parameter $\theta_{C}$, which later was  called the Cabibbo angle,   and assumed
that the hadronic charged current was a combination of currents
$j^{\Delta S =0}_{\alpha}$ and $j^{\Delta S =1}_{\alpha}$ with the
coefficients $ \cos\theta_{C}$ and $ \sin\theta_{C}$,
respectively. Assuming a  weak universality Cabibbo
suggested that in the total hadronic current
\begin{equation}\label{Cabcur}
j^{h}_{\alpha}=aj^{\Delta S =0}_{\alpha}+bj^{\Delta S =1}_{\alpha}
\end{equation}
the real coefficients $a$ and $b$ would satisfy the condition
$a^{2}+b^{2}=1$. From this condition it follows that
$a=\cos\theta_{C}$ and $b=\sin\theta_{C}$. The Cabibbo paper was
written before the quark hypothesis appeared. He assumed that the
current which did not change strangeness and the current which
changed the strangeness by one are the $1+i2$ and $4+i5$ components
of the $SU(3)$ octet current. Cabibbo found that with the parameter
he introduced it was possible to describe all data on the
semi-leptonic decays of mesons and baryons. From  analysis of the
data he found that $\sin\theta_{C}\simeq 0.2$.

The Cabibbo current had the form
\begin{equation}
j^{\rm{Cabbibo}}_{\alpha}= 2~(\cos{\theta_C}~ \bar{u}_L
\gamma_{\alpha} d_L + \sin{\theta_C}~ \bar{u}_L \gamma_{\alpha} s_L)=
\bar{u}_L \gamma_{\alpha} d_L^{\mathrm{mix}},
\label{CabC}
\end{equation}
where
\begin{equation}
d^{\mathrm{mix}}_L(x) = \cos{\theta_C} d_L(x) + \sin{\theta_C} s_L(x).
\label{mix1}
\end{equation}
The total charged current took the form
\begin{equation}
j_{\alpha}=2~(\bar{\nu}_{eL} \gamma_{\alpha} e_L + \bar{\nu}_{\mu L}
\gamma_{\alpha} \mu_L + \bar{u}_L \gamma_{\alpha} d_L^{\mathrm{mix}}).
\label{qCC2}
\end{equation}
As it is seen from this expression, the lepton and quark terms have
the same form and enter into the current with the same coefficients
(universality). However, there was  asymmetry in the current
(\ref{qCC2}): there are two lepton terms and one quark term. This
asymmetry was connected with the fact that four leptons ($e,
\nu_{e},\mu, \nu_{\mu}$) and only three quarks ($u,d,s$) were known
at that time.

\subsection{Charmed quark. Quark mixing}
Some years later it was discovered that the charged current (\ref{qCC2}) creates some problems. Namely, the current  (\ref{qCC2}) generates  neutral currents which change the strangeness by one. Such a neutral current induces the decay
\begin{equation}\label{Kpi}
K^+ \to \pi^+ + \nu + \bar{\nu}.
\end{equation}
with a decay rate which is many orders of magnitude larger than
the upper bound obtained in experiments.

The solution of the problem was proposed in 1970 by Glashow,
Illiopulos and Maiani (GIM) \cite{GIM}. They  assumed that there
existed a fourth "charmed" quark $c$ with the charge 2/3 and there
was an additional term in the weak charged current
\begin{equation}\label{GIM}
j^{\rm{GIM}}_{\alpha}= 2~(-\sin{\theta_C}~ \bar{c}_L
\gamma_{\alpha} d_L + \cos{\theta_C}~ \bar{c}_L \gamma_{\alpha} s_L)
\end{equation}
into which the field of the new quark $c$ enters. In the theory with this additional current a "dangerous" neutral current which changes the strangeness does not appear.

 The total weak charged currents take the form
\begin{equation}
j_{\alpha}(x) = 2~(\bar{\nu}_{eL}(x) \gamma_{\alpha} e_L(x) + \bar{\nu}_{\mu
L}(x) \gamma_{\alpha} \mu_L(x) + \bar{u}_L(x) \gamma_{\alpha} d^{\mathrm{mix}}_L(x) +
 \bar{c}_L(x) \gamma_{\alpha} s^{\mathrm{mix}}_L(x)),
\label{qCC}
\end{equation}
where
\begin{eqnarray}
d^{\mathrm{mix}}_L(x) & = &\cos{\theta_C} d_L(x) + \sin{\theta_C} s_L(x)
\nonumber \\
s^{\mathrm{mix}}_L(x) & =&  -\sin{\theta_C}d_L(x) + \cos{\theta_C} s_L(x)\,.
\label{mix2}
\end{eqnarray}
Thus, the field of $d$ and $s$ quarks, which have the same charge
(-1/3) and differ in their masses, enter into the charged current
(\ref{qCC}) in the form of the orthogonal combinations
$d^{\mathrm{mix}}_L(x)$ and $s^{\mathrm{mix}}_L(x)$ ({\em "mixed
form"}). The Cabibbo angle $\theta_C$ is the mixing angle.

 With the additional $c$-quark the numbers of leptons and quarks are
equal and there is symmetry between lepton and quark terms in the
current (\ref{qCC}). It would be, however, a full lepton-quark
symmetry of the charged current if the neutrino masses were different
from zero and the fields of neutrinos with definite masses, like the
fields of quarks, enter into the CC in a mixed form
\begin{eqnarray}
\nu_{\mu L }(x)& = &\cos{\theta}\nu_{1 L }(x)  + \sin{\theta}\nu_{2
L }(x)
\nonumber \\
\nu_{e L }(x) & =&  -\sin{\theta}\nu_{1 L }(x) + \cos{\theta} \nu_{2
L }(x), \label{numix}
\end{eqnarray}
where $\nu_{1}(x)$ and $\nu_{2}(x)$ are the fields of  neutrinos with
masses $m_{1}$ and $m_{2}$ and $\theta$ is the neutrino mixing angle
(generally different from $\theta_C$).

We know now that mixing of quarks exists, neutrino masses are
different from zero and neutrino mixing (in a more general form; see
later) is confirmed by experiment. The lepton-quark symmetry
arguments we presented above were early arguments in favor of the
neutrino masses and mixing put forward in the seventies (see
\cite{BilPonte78}).

If the $c$-quark, a constituent of hadrons, exists, in this case a
new family of "charmed" particles must exist. This prediction  was
perfectly confirmed by experiment. In 1974 the  $J/\Psi$ particles
($m_{J/\Psi} \simeq$ 3096.9 MeV), bound states of $(c-\bar c)$, were
discovered. In 1976, $D^{+}= (c\bar d)$, $D^{-}= (\bar cd)$
($m_{D^{\pm}}\simeq 1868.6$ MeV),
 $D^{0}=(c \bar u)$,  $\bar D^{0}=(\bar c u)$ ($m_{D^{\pm}}\simeq 1864.8 $ MeV) were discovered. Later many charmed mesons and baryons were found in experiment. All data obtained from the investigation of weak decays and neutrino reactions were in
agreement with the current$\times$current theory with the current
given by (\ref{qCC}).

\section{Discovery of the third charged lepton $\tau$.
The third family of leptons and quarks}

 We do not know why the muon, the particle which has the same interaction as the electron but with
a mass 206.8 times larger than the electron mass, exists (the masses
of the electron and the muon are $m_{e}=0.51$ MeV and $m_{\mu}=105.6$
MeV)\footnote{The question which was put many years ago by the Nobel
Prize winner I.Rabi  "Who ordered the $\mu$-meson?" still has no
answer. Now we can also ask who ordered  $s$, $c$
and other quarks}. In such a situation it was natural to ask whether more
heavier than $\mu$ (sequential) lepton(s) exist.

 The answer to this question was obtained in  experiments
which were performed in 1975-77 by M. Perl et al. at the
$e^{+}-e^{-}$ collider at Stanford \cite{Perl}. In these experiments,
the third lepton $\tau^{\pm}$ was discovered.\footnote{In 1995 M.
Perl was awarded the Nobel Prize "for the discovery of the tau
lepton".} The $\tau$-lepton decays into an electron (muon) and two
neutrinos, pion(s) and neutrino etc. Its mass $m_{\tau}=1776.8$ MeV.

Let us combine a charged lepton, a neutrino and quark fields in the
following way
\begin{enumerate}
\item $(\nu_{e}, e^{-})\quad (u, d)$.
 \item $(\nu_{\mu}, \mu^{-})\quad (c,s)$.
 \end{enumerate}
In the first group (family, generation) enter the fields of the
lightest leptons and quarks, and in the second family enter fields of
heavier leptons and quarks\footnote{For quark masses we have:
$m_{u}=(1.5-3.3)$ MeV, $m_{d}=(3.5-6.0)$ MeV, $m_{s}=104^{+26}_{-34}$
MeV,  $m_{c}=1.27^{+0.07}_{-0.11}$ GeV.}.

The discovery of the $ \tau $ could mean that there exists a third
family of leptons and quarks. In this case a third type of the
neutrino $\nu_{\tau}$, which takes part in weak processes together
with  $ \tau $, and an additional pair of quarks (the top quark $t$
with electric charge 2/3 and bottom quark $b$
 with electric charge -1/3) must exist.

Let us notice that at the time when the $\tau$-lepton was discovered,
the Standard Model of the electroweak interaction which we will
discuss later existed. Due to the $SU(2)\times U(1)$ symmetry of this
theory the existence of the $ \tau $ {\em requires} the existence of
$\nu_{\tau}, t, b$.

All these expectations were  confirmed by experiment. In
1977 $\Upsilon$-particles, a bound state of $(b-\bar b)$, were
discovered at the Fermilab ( $m_{\Upsilon}=9460.3 $ MeV). Later
$B^{+}=(b\bar u)$ ($m_{B^{+}}=5279.2$ MeV) $B^{0}=(d\bar b)$
($m_{B^{0}}=5279.5$ MeV) and other $B$-mesons,
$\Lambda^{0}_{b}=(ubd)$ ($m_{\Lambda_{b}}=5629.2$ MeV) and other
bottom baryons were detected and studied in many experiments. The
mass of the $b$ quark is equal to $m_{b}=4.20^{+0.17}_{-0.07}$ GeV.
In 1995, at the Fermilab the $t$-quark was discovered. The $t$-quark
is the heaviest known elementary particle ($m_{t}=(171.2\pm 2.1)$
GeV). The third type of neutrino $\nu_{\tau}$, the partner of the
$\tau$-lepton, was observed in 2000 in an experiment performed by the
DONUT Collaboration at Fermilab\cite{Donut}. In this experiment the
production of $\tau$  in the process $\nu_{\tau}+(A,Z)\to \tau
+.....$ was observed. At the energy of the experiment the
$\tau$-lepton  decays, producing predominantly a single charged
particle at an average distance of 2 mm from the production point.
Nuclear emulsion was used to detect the $\tau$ production. A
signature of the event in the emulsion was a track with a kink.

In the case of three generations the charged current takes the form
\begin{eqnarray}
j^{CC}_{\alpha}(x) &=& 2~(\bar{\nu}_{eL}(x) \gamma_{\alpha} e_L(x) +
\bar{\nu}_{\mu L}(x) \gamma_{\alpha} \mu_L(x) +
\bar{\nu}_{\tau L}(x)
\gamma_{\alpha} \tau_L(x)\nonumber\\&+& \bar{u}_L (x)\gamma_{\alpha} d^{\mathrm{mix}}_L (x)+
 \bar{c}_L (x)\gamma_{\alpha} s^{\mathrm{mix}}_L(x) +
 \bar{t}_L (x)\gamma_{\alpha} b^{\mathrm{mix}}_L(x) ).
\label{qCC3}
\end{eqnarray}
The Cabibbo-GIM mixing of quarks (\ref{mix2}) was generalized for the
case of three families of quarks  by Kobaya
shi and Maskawa in 1973
\cite{KobMas}. They assumed that "mixed" fields
$d^{\mathrm{mix}}_L(x), s^{\mathrm{mix}}_L(x), b^{\mathrm{mix}}_L(x)$
were connected with the left-handed components of the fields of $d$,
$s$ and $b$ quarks by the unitary transformation:
\begin{equation}\label{mix3}
d^{\mathrm{mix}}_L(x) =\sum_{q=u,s,b}V_{uq}~q_{L}(x),~~s^{\mathrm{mix}}_L(x)
=\sum_{q=u,s,b}~V_{cq}q_{L}(x),~~b^{\mathrm{mix}}_L(x)
=\sum_{q=u,s,b}V_{tq}~q_{L}(x).
\end{equation}
The unitary $3\times3$ matrix $V$ is called the
Cabibbo-Kobayashi-Maskawa (CKM) mixing matrix. The matrix $V$ is
characterized by three mixing angles and one phase which is
responsible for $CP$ violation.

 Kobayashi and Maskawa showed
that in the case of two generations of quarks it is impossible to
explain $CP$ violation which was observed in decays of neutral $K$-
mesons. This was a main motivation for the assumption of the
existence of the third generation of quarks made in 1973 before the
$\tau$-lepton, $b$ and $t$ quarks were discovered. In 2008 Kobayashi and Maskawa were awarded the Nobel
Prize for "for the discovery of the origin of the broken symmetry
which predicts the existence of at least three families of quarks in
nature".

 On the basis of the lepton-quark symmetry it was natural
to assume that the neutrino fields $\nu_{e}, \nu_{\mu}, \nu_{\tau}$
were also mixed (see \cite{BilPonte78}):
\begin{equation}\label{numix1}
\nu_{lL}(x)=\sum^{3}_{i=1}U_{li}~\nu_{iL}(x)\quad l=e,\mu,\tau
\end{equation}
Here $U$ is the unitary $3\times3$ neutrino mixing matrix.

In the theory with the intermediate vector boson $W^{\pm}$
the Lagrangian
of the CC weak interaction has the form
\begin{equation}\label{CCH1}
{\cal{L}^{CC}_{I}}(x)=-\frac{g}{2\sqrt{2}}~j^{CC}_{\alpha}(x)~W^{\alpha}(x)
+\rm{h.c.}
\end{equation}
where the charged current $j^{CC}_{\alpha}(x)$ is given by
expression (\ref{qCC3}).

\section{Number of families of quarks and leptons}
How many families of quarks and leptons exist in nature? The answer
to this fundamental question was obtained in experiments made at SLC
(Stanford) and LEP (CERN). In these experiments the width of the
decay
\begin{equation}\label{Zwidth}
Z^{0}\to \nu_{l}+\bar\nu_{l},\quad l=e,\mu,\tau,...
\end{equation}
was determined
. The $Z^{0}$-boson has a mass $m_{Z}=91.1876\pm 0.0021$
GeV. Different decay modes of the $Z^{0}$-boson ($Z^{0}\to
l^{+}+l^{-}~~(l=e,\mu,\tau),\quad Z^{0}\to \mathrm{hadrons}$ ) were
investigated in detail at ($e^{+}-e^{-}$) colliders.

Neglecting small neutrino masses we have
\begin{equation}\label{Zwidth1}
\sum_{l}\Gamma(Z^{0}\to \nu_{l}\bar\nu_{l})=n_{\nu_{f}}\Gamma(Z^{0}\to \nu\bar\nu),
\end{equation}
where $n_{\nu_{f}}$ is the number of neutrino-antineutrino pairs and
$ \Gamma(Z^{0}\to \nu\bar\nu)$ is the width of the decay of the
$Z^{0}$ into a neutrino-antineutrino  pair (this width is known from
the Standard Model calculations).

From (\ref{Zwidth1}) we find the following relation:
\begin{equation}\label{Zwidth2}
n_{\nu_{f}}=\frac{\sum_{l}\Gamma(Z^{0}\to \nu_{l}\bar\nu_{l})}{\Gamma(Z^{0}\to l\bar l)}\left (\frac{\Gamma(Z^{0}\to l\bar l)}{\Gamma(Z^{0}\to \nu\bar\nu)}\right)_{SM}
\end{equation}
The first ratio is measured in experiments. The second ratio is known
from the SM calculations ($\left(\frac{\Gamma (Z^{0}\to
\nu\bar\nu)}{\Gamma(Z^{0}\to l\bar l)}\right)_{SM}=1.991\pm 0.001$).

From the data of four LEP experiments it was found \cite{PDG}
\begin{equation}\label{Zwidth3}
n_{\nu_{f}}=2.984\pm 0.008.
\end{equation}
Thus, it was established that the number of different types of
neutrinos was equal to three (only $\nu_{e}, \nu_{\mu}, \nu_{\tau}$
exist in nature). Each family of leptons and quarks has his own
neutrino. We conclude that {\em only three families of leptons and
quarks exist in nature}.\footnote{From these data we can not exclude,
however, that there exist neutral leptons with masses larger than
$\frac{m_{Z}}{2}$ which can not be produced in decays of the $Z^{0}
$-bosons. Thus, we can not exclude the existence of new families in
which instead of neutrinos such heavy neutral leptons are present.}

\section{Unified theory of weak and electromagnetic interactions. The Standard Model}
\begin{center}
{\bf Introduction}
\end{center}

The unified theory of  weak and electromagnetic interactions
(Standard Model) is a theory of interaction of neutrinos, charged
leptons and quarks with the $W^{\pm}$, $Z^{0}$ bosons and
$\gamma$-quanta in a wide range of energies.  This theory was
 confirmed by numerous experiments including very precise
LEP (CERN) experiments.

The SM is based on the spontaneously broken local gauge $SU(2)\times
U(1)$ symmetry and it is built in such a way to include the charged
current of the phenomenological $V-A$ theory  and the electromagnetic
interaction of charged leptons and quarks.

The SM predicts the existence of the $W^{\pm}$ and  $Z^{0}$ bosons.
This prediction was confirmed by
experiment. In 1984, C. Rubbia and S. Van der Meer were
awarded the Nobel Prize "for their decisive contributions to the
large project, which led to the discovery of the field particles W
and Z, communicators of weak interaction".

Taking into account radiative corrections, for masses and decay
widths of the $W^{\pm}$ and  $Z^{0}$ bosons from the Standard Model
it was obtained
 \cite{PDG}
 \begin{equation}\label{masswidth1}
(m_{W})_{SM}=(80.420 \pm 0.031)~\mathrm{GeV},~~(\Gamma_{W})_{SM}=
(2.0910 \pm 0.0007 )~\mathrm{GeV}
\end{equation}
\begin{equation}\label{masswidth2}
(m_{Z})_{SM}=(91.1874 \pm 0.0021)~\mathrm{GeV},~~(\Gamma_{Z})_{SM}=
(2.4954 \pm 0.0009)~\mathrm{GeV}.
\end{equation}
These values are in agrement with the measured masses and decay
widths:
\begin{equation}\label{masswidth3}
m_{W}=(80.384 \pm 0.014)~\mathrm{GeV},~~\Gamma_{W}=(2.085 \pm
0.042)~\mathrm{GeV}
\end{equation}
\begin{equation}\label{masswidth4}
m_{Z}=(91.1876 \pm 0.0021)~\mathrm{GeV},~~\Gamma_{W}= (2.4952 ï¿-
0.0023)~\mathrm{GeV}
\end{equation}
The Standard Model predicts a new class of  weak interactions:
neutral currents. Numerous experimental data confirm this
prediction. The standard neutral current is diagonal in quark,
charged lepton and neutrino fields and is characterized by
$\sin^{2}\theta_{W}$. The values of this parameter determined from
different data ($e^{+}-e^{-}$, deep inelastic neutrino-nucleon
scattering, P-odd asymmetry in deep-inelastic electron-nucleon
scattering, etc.) are compatible with each other. From the fit of all
data it was found
\begin{equation}\label{Wangle}
\sin^{2}\theta_{W}=0.23108 \pm 0.00005.
\end{equation}
The Standard Model provides a natural framework for quark mixing.
However, the SM can not predict the masses of quarks and charged
leptons and the CKM mixing angles.

Neutrino masses are not of the Standard Model Higgs origin. For the
generation of small neutrino masses and neutrino mixing a new (or
additional) mechanism is needed.

In 1979 S. Glashow, S. Weinberg, and A. Salam were awarded the Nobel
prize "for their contributions to the theory of the unified weak and
electromagnetic interaction between elementary particles, including,
inter alia, the prediction of the weak neutral current".\footnote{ In
his Nobel lecture S. Weinberg remembered how he came to  the idea of
the electroweak theory. He tried to apply ideas of local gauge
invariance and spontaneous symmetry breaking to strong interaction of
hadrons. These attempts failed. "At some point at the fall of 1967,
while driving to my office at MIT, it occurred to me that I had been
applying the right ideas to the wrong problem. It is not the
$\rho$-meson that is massless: it is the photon. And its partner is
not the $A1$, but the massive intermediate vector boson...The weak
and electromagnetic interactions could then be described in a unified
way..."}. In 1999 G. t'Hooft and M Veltman were awarded the Nobel
Prize "for elucidating the quantum structure of electroweak
interactions in physics".

\vskip0.5cm

In 1967-68, S. Weinberg \cite{Weinberg67} and A. Salam \cite{Salam68}
proposed a new theory which unified the weak and electromagnetic
interactions into one electroweak interaction. They built such a
theory for the electron neutrino and the electron. Later all three
families of leptons and quarks were included in the theory. It is
called the Standard Model (SM).

The Standard Model  predicted a new class of the weak interaction
(Neutral currents), the $W^{\pm}$ and $Z^{0}$ vector bosons and the
masses of these particles, the existence of the scalar Higgs boson,
etc. All predictions of the Standard Model are in perfect agreement
with existing experimental data.

Recently it was anounced by the CMS \cite{CMS} and ATLAS \cite{Atlas}
collaborations, working at the LHC accelerator at CERN, that they
discovered a Higgs-like boson with mass equal to  125 GeV. In order
to prove that the discovered new particle is the Standard Model Higgs
boson it is necessary to determine its spin and parity.(The SM Higgs
boson must be scalar particle with spin equal to zero and positive
parity). This will be done in futuure experiments.

{\em Neutrinos played an extremely  important role in the
establishment of the SM.} In neutrino experiments  fundamental
parameters of the theory were determined. Neutrinos  played  also an
important role in the establishment of the quark structure of
nucleons and  its investigation.

The V-A current$\times$current theory of the weak interaction, which
we considered in the previous sections, has been a very successful
theory. It allowed  one to describe all experimental data, which
existed in the sixties. However, the current$\times$current theory
and also the theory with the intermediate $W^{\pm}$ vector boson were
unrenormalizable theories. The infinities at  higher orders of
perturbation theory could not be excluded in these theories by the
renormalization of the masses and other physical parameters.

This was the main reason why, in spite of great phenomenological
successes, the current$\times$current theory of the weak interaction
and the theory with the intermediate vector boson were not considered
as satisfactory. The Standard Model was born in the sixties in an
attempt to build a renormalizable theory of weak interaction. The
only renormalizable physical theory that was known at that time was
quantum electrodynamics. The renormalizable theory of the weak
interaction  was built in the framework of {\em the unification of
the weak and electromagnetic (electroweak) interactions}. This theory
was proposed by  Weinberg \cite{Weinberg67}  and Salam
\cite{Salam68}. The same theory with the unification of the weak and
electromagnetic  interactions but without the mechanism of
spontaneous symmetry breaking (see later) was proposed by Glashow in
1961 \cite{Glashow61}. Weinberg and Salam suggested that the SM would
be a renormalizable theory but they did not prove that. The
renormalizability of the SM was proved in 1971 by  t'Hooft
\cite{Hooft}.

We will briefly discuss now the Standard Model of the electroweak
interactions.\footnote{This section requires some knowledge of the
Quantum Field Theory (QFT) and a group theory.  Readers who are not
familiar with QFT and (or) the theory of groups can simply follow the
main ideas of the SM, skipping technical details.} The Standard Model
is based on
\begin{enumerate}
\item Phenomenological V-A theory of the weak interaction.
\item Local gauge $SU(2)\times U(1)$ invariance
of the Lagrangian  of  fields of massless quarks, leptons and vector
bosons \footnote{The $SU(2)$ group is a group of unitary $2\times 2$
matrices with determinant equal to 1; the phase factors
$e^{i\Lambda}(x)$ ($\Lambda (x)$ is a function of $x$) form a $U(1)$
group.}.

\item Minimal interaction of fermions and vector bosons.

\item Spontaneous breaking of  symmetry and the
Higgs mechanism of the generation of masses of quarks and leptons.

  \item Unification of the weak and electromagnetic interactions
into one electroweak interaction.

\end{enumerate}

The minimal group which ensures the CC interaction  of
leptons and quarks with $W^{\pm}$-bosons  is the local $SU(2)$ group
if we assume that left-handed components of the fields of quarks and
leptons form doublets (the meaning of primes will be clear later)
\begin{eqnarray}
\psi_{1L}=\left(
\begin{array}{c}
u'_L \\
d'_L \\
\end{array}
\right),~ \psi_{2L}=\left(
\begin{array}{c}
c'_L \\
s'_L \\
\end{array}
\right),~ \psi_{3L}(x)=\left(
\begin{array}{c}
t'_L \\
b'_L \\
\end{array}
\right)\label{1SU2dub}
\end{eqnarray}
and
\begin{eqnarray}
\psi_{eL}=\left(
\begin{array}{c}
\nu'_{eL} \\
e'_L\\
\end{array}
\right),~ \psi_{\mu L}=\left(
\begin{array}{c}
\nu'_{\mu L}\\
\mu'_L \\
\end{array}
\right),~ \psi_{\tau L}=\left(
\begin{array}{c}
\nu'_{\tau L}\\
\tau'_L\\
\end{array}
\right)\label{lepSU2dub}
\end{eqnarray}
We assume also that the right-handed components of the fields of
quark and leptons are singlets.

From the local $SU(2)$ invariance it follows that the minimal
interaction includes only the left-handed components of quark and
lepton fields and has the form \begin{equation}\label{interL1}
\mathcal{L}_{I}(x) = \left( -\frac{g}{2\,\sqrt{2}}\,
j^{CC}_{\alpha}(x)\,W^{\alpha}(x) + \rm{h.c}\right)
-g\,j^{3}_{\alpha}(x)\,A^{\alpha 3}(x)~.
\end{equation}
Here
\begin{equation}\label{1CC}
j^{CC}_{\alpha} =2( \bar{u'}_L \gamma_{\alpha} d'_L +
 \bar{c'}_L \gamma_{\alpha} s'_L +
 \bar{t'}_L \gamma_{\alpha} b'_L )+2\sum_{l=e,\mu,\tau}\bar\nu'_{lL}\gamma_{\alpha} l'_L
\end{equation}
is the charged current of the quarks and leptons,
\begin{equation}\label{2CC}
j^{3}_{\alpha}=\sum_{a=1,2,3} \bar \psi_{aL} \frac{1}{3}\tau_{3}\gamma_{\alpha}  \psi_{aL}+\sum_{l=e,\mu,\tau} \bar \psi_{lL} \frac{1}{3}\tau_{3}\gamma_{\alpha}  \psi_{lL},
\end{equation}
($\tau_{3}$ is the third Pauli matrix) and $g$ is a constant
connected with the $SU(2)$ group. The field $A^{\alpha 3}(x)$ is the
field of neutral vector particles.

We would like to unify the weak and electromagnetic interactions on
the basis of the local gauge invariance. The first term of
(\ref{interL1}) is the Lagrangian of the CC weak interaction.
However, the second term violates parity and cannot be identified
with the Lagrangian of the electromagnetic interaction.

In order to unify the weak interaction (which maximally violates
parity) and the electromagnetic interactions (which conserve parity)
in one electroweak interaction  we must enlarge the symmetry group.
The Standard Model is based on the local gauge $SU(2)\times U(1)$
invariance. This is a minimal enlargement of the $SU(2)$ group which
generates the charge current weak interaction.

The $U(1)$ group is the group of the hypercharge $Y$ which is
determined by the Gell-Mann-Nishijima relation
\begin{equation}\label{GelNish}
Q=I_{3}+ \frac{1}{2}\,Y,
\end{equation}
where $Q$ is the electric charge (in the unit of the proton charge)
and $I_{3}$ is the third component of the isotopic spin.

 The invariance under the additional
$U(1)$ group can be realized if in addition to the vector
$W^{\alpha}$ field (field of vector $W^{\pm}$ bosons) and the field
of  neutral vector particles $A^{\alpha 3}$ the {\em  field of
neutral vector particles $B^{\alpha }$ exists.}

 The Lagrangian of the minimal interaction takes the form
\begin{equation}\label{interL4}
\mathcal{L}_{I}(x) = \left( -\frac{g}{2\,\sqrt{2}}\,
j^{CC}_{\alpha}(x)\,W^{\alpha}(x) + \rm{h.c}\right)
+\mathcal{L}^{0}_{I}(x).
\end{equation}
Here
\begin{equation}\label{interL5}
\mathcal{L}^{0}_{I}(x)=-g\,j^{3}_{\alpha}(x)\,A^{\alpha 3}(x)-g'~(j^{\rm{EM} }_{\alpha}(x) - j^{3}_{\alpha}(x))~B^{\alpha}(x)
\end{equation}
is the Lagrangian of interaction of quarks and
neutral vector particles,
\begin{equation}\label{EMcurr}
j^{\rm{EM} }_{\alpha} = (\frac{2}{3})\sum_{q=u,c,t}\bar q'
\,\gamma _{\alpha}\,q'+(-\frac{1}{3})\sum_{q=d,s,b}\bar
q' \,\gamma _{\alpha}\,q'+(-1)\sum_{l=e,\mu,\tau}\bar l'\,\gamma _{\alpha}\,l
\end{equation}
is the electromagnetic current of the quarks and leptons and $g'$ is
a constant connected with the $U(1)$ group.

Up to now we considered fields of massless particles. The Standard
Model is based on {\em the Higgs mechanism of the generation of
masses} of the $W^{\pm}$- and $Z^{0}$ bosons, quarks, lepton and
other particles. We will assume that in our system enter scalar
complex fields of charged and neutral particles ($\phi_{+}$ and
$\phi_{0}$) and that these fields form the $SU(2)$ doublet
\begin{equation}\label{Higgs}
\phi ={\phi _{+}\choose\phi _{0}}.
\end{equation}
 The Lagrangian of the Higgs field is chosen in such a way that the
energy of the field reaches a minimum when the value of the field is
different from zero.  This means that the Higgs vacuum is not an
empty state. Moreover due to the symmetry there are many (infinite)
degenerate vacuum states. If we choose a definite vacuum field, say,
\begin{equation}\label{Higgs1}
\phi_{0} ={0\choose\frac{v}{\sqrt{2}}}
\end{equation}
we will violate the symmetry ($v$ is a constant).
Such a violation is called spontaneous.

Before spontaneous violation of the symmetry we had a massless
complex (charged) $W_{\alpha}$  vector field and two massless real
(neutral) vector fields $A^{3}_{\alpha}$ and  $B_{\alpha}$. After
spontaneous violation of the symmetry, the masses of the $W^{\pm}$
and $Z^{0}$ bosons are generated. The field of $Z^{0}$ bosons  is the
following combination of $A^{3}_{\alpha}$ and  $B_{\alpha}$ fields:
\begin{equation}\label{Z}
Z_ {\alpha}=\frac{g}{\sqrt{g^{2}+g'^{2}}}A^{3}_{ \alpha}-
\frac{g'}{\sqrt{g^{2}+ g'^{2}}}B_{ \alpha}.
\end{equation}
For the masses of the $W^{\pm}$ and $Z^{0}$ bosons we have the
following relations:
\begin{equation}\label{mass}
m^{2}_{W} =\frac{1}{4}\, g^{2}\, v ^{2},\quad
m^{2}_{Z}=\frac{1}{4}\,(g^{2}+g'^{2})~ v ^{2}.
\end{equation}
After spontaneous violation of the symmetry the mass of particles,
quanta of the field
\begin{equation}\label{A}
A_ {\alpha}=\frac{g'}{\sqrt{g^{2}+g'^{2}}}A^{3}_{ \alpha}+
\frac{g}{\sqrt{g^{2}+ g'^{2}}}B_{ \alpha},
\end{equation}
which is an orthogonal to  $Z_{\alpha}$, remain
equal to zero.

Let us introduce the weak (Weinberg) angle $\theta_{W}$ by the relation
\begin{equation}\label{Wangle}
\frac{g'}{g}=\tan \theta_{W}.
\end{equation}
We have
\begin{equation}\label{AZ}
A_ {\alpha}=\cos\theta _{W} B_{ \alpha}+\sin\theta _{W} A^{3}_{ \alpha},\quad Z_ {\alpha}=-\sin\theta _{W} B_{ \alpha}+\cos\theta _{W} A^{3}_{ \alpha}.
\end{equation}
From (\ref{interL5}) and (\ref{AZ}) we find the following expression
for the Lagrangian of interaction of quarks and leptons with neutral
vector particles
\begin{equation}\label{1.4interL4}
\mathcal{L}^{0}_{I}=-\frac{g}{2\cos\theta_{W}}\,j^{\rm{NC}}_{\alpha}\,Z^ {\alpha} -g\sin \theta_{W}\,j^{\rm{EM}}_{\alpha}\, A^{ \alpha}~,
\end{equation}
where
\begin{equation}\label{NC}
j^{\rm{NC}}_{\alpha}=2~ j^{3}_{\alpha} -2~\sin^{2}\theta_{W}\,j^{\rm{EM}}_{\alpha}.
\end{equation}
From (\ref{1.4interL4}) we can draw the following important
conclusions:
\begin{enumerate}
  \item The second term of (\ref{1.4interL4}) is  the Lagrangian
of the electromagnetic interaction of quarks and charged leptons if
{\em the constants $g$ and $\sin \theta_{W}$  satisfy the following
(unification) condition}:
 \begin{equation}\label{unif}
    g~\sin \theta_{W}=e,
 \end{equation}
where e is the proton charge.
\item The unification of the weak and electromagnetic interaction
is possible if in addition to the charged  vector $W^{\pm}$-boson
there exists a neutral  vector  $Z^{0}$-boson with a mass larger than
the mass of the $W^{\pm}$-boson (see  relation (\ref{mass})). As a
consequence of the unification {\em a new (neutral current)
interaction of quarks, charged leptons and neutrinos with the
$Z^{0}$-boson appears.}
\end{enumerate}
The Fermi constant is given by the relation
\begin{equation}\label{1Fermi}
\frac{G_{F}}{\sqrt{2}}=\frac{g^{2}}{8m^{2}_{W}}.
\end{equation}
From this relation and (\ref{mass}) it follows that the parameter $v$
(vacuum expectation value), which characterizes the scale of the
$SU(2)\times U(1)$ symmetry breaking, is given by
\begin{equation}\label{vev}
v=(\sqrt{2}~G_{F})^{-1/2}\simeq 246 ~\mathrm{GeV}.
\end{equation}
From the unification condition (\ref{unif}) and  relations
(\ref{mass}) it follows that the masses of the $W$ and $Z$ bosons are
given by the following relations
\begin{equation}\label{WZmasses}
m_{W} =\left(\frac{\pi\,\alpha}{\sqrt{2}\,G_{F}}\right
)^{1/2}\,\frac{1}{\sin\theta_{W}},~~~ m_{Z}
=\left(\frac{\pi\,\alpha}{\sqrt{2}\,G_{F}}\right)^{1/2}\,\frac{1}{\sin\theta_{W}\cos\theta_{W}},
\end{equation}
where $\alpha=\frac{e^{2}}{4\pi}$ is the fine structure constant.

The value of the parameter $\sin\theta_{W}$ can be determined from the
study of neutral current (NC) processes. Thus, {\em the Standard Model
 predicts the masses of the $W^{\pm}$ and $Z^{0}$ bosons}. This prediction
is in an agreement with experimental data.

We will now briefly discuss a  much less predictive part of the SM,
the Higgs mechanism of the generation of masses of quarks and
leptons. In order to generate the masses of fermions we need to
assume that  the total Lagrangian of the system contains an
$SU(2)\times U(1)$ invariant Lagrangian of a Yukawa interaction of
fermions and Higgs boson. For example, the Lagrangian
\begin{equation}\label{Yint}
\mathcal{L}_{Y}^{\rm{down}}(x)=-\frac{\sqrt{2}}{v}\,\sum_{a=1,..q'_{R}=d'_{R},..}
\bar\psi_{aL}(x)~M_{a;q}^{\rm{down}}~ q'_{R}(x)~\phi(x) +\rm{h.c.},
\end{equation}
after spontaneous violation of the symmetry generates masses of the
$d$, $s$ and $b$ quarks. The matrix $M^{\rm{down}}$ in (\ref{Yint})
is a complex $3\times 3$ matrix. The Standard Model does not put any
constraints on this  matrix. After the diagonalization of the matrix
$M^{\rm{down}}$ and another similar matrix $M^{\rm{up}}$  we find
\begin{equation}\label{udmix}
q'_{L}=\sum _{q=d,...}V^{\rm{down}}_{q'_{L}q_{L}}~q_{L}~~(q'_{L}=d'_{L},...)\quad
q'_{L}=\sum _{q=u,...}V^{\rm{up}}_{q'_{L}q_{L}}~q_{L}~~(q'_{L}=u'_{L},...)
\end{equation}
Here $V^{\rm{down}}$ and $V^{\rm{up}}$ are the unitary $3\times3$
matrices and $q_{L}$ is the left-handed component of the field of
$q$-quark with mass $m_{q}$ ($q=u,c,t,d,s,b$).

For the lepton fields we have
\begin{equation}\label{lepmix}
l'_{L}=\sum _{l=e,\mu,\tau}V^{\rm{lep}}_{l'_{L}l_{L}}~l_{L}~~
(l'_{L}=e'_{L},\mu'_{L},\tau'_{L}),
\end{equation}
where $l_{L}$ is the left-handed component of the field of the lepton $l$ with mass $m_{l}$ ($l=e,\mu,\tau$).

Similar relations connect the primed right-handed components of the
fields of quarks and leptons and the right-handed components of
fields of quarks and leptons with definite masses.

It is important that the matrices  $V^{\rm{down}}$ and $V^{\rm{up}}$
are {\em different and unitary}. From (\ref{1CC}), (\ref{udmix}) and
(\ref{lepmix}) we obtain the following expression for the charged
current:
\begin{equation}\label{finCC}
j^{CC}_{\alpha} =2( \bar{u}_L \gamma_{\alpha} d^{\mathrm{mix}}_L +
 \bar{c}_L \gamma_{\alpha} s^{\mathrm{mix}}_L+
 \bar{t}_L \gamma_{\alpha} b^{\mathrm{mix}}_L )+2\sum_{l=e.\mu,\tau}\bar\nu_{lL}\gamma_{\alpha}l_{L}.
\end{equation}
Here
\begin{equation}\label{finqCC1}
q^{\mathrm{mix}}_L =\sum_{q=d,s,b}V_{q^{\mathrm{mix}}_{L}q_{L}}q_{L},\quad    q^{\mathrm{mix}}_L=d^{\mathrm{mix}}_L, s^{\mathrm{mix}}_L, b^{\mathrm{mix}}_L,
\end{equation}
where
\begin{equation}\label{5CKM}
 V= V^{\rm{L,up\dag}}V^{\rm{L,down}}
\end{equation}
is the {\em $3\times3$ unitary mixing Cabibbo-Kobayashi-Maskawa mixing matrix} and
\begin{equation}\label{nu}
\nu_{lL} =\sum_{l_{1}=e,\mu,\tau} (V^{\rm{lep}})^{\dag}_{l_{L}l_{1L}}\nu'_{l_{1}L}.
\end{equation}
Taking into account the unitarity of the matrices which connect
$L$($R$)-components of primed fields with $L$($R$)-components of
fields of particles with definite masses it is easy to show that in
the neutral current and in the electromagnetic current we must change
primed fields of quarks, leptons and neutrinos by the corresponding
physical non primed fields. This means that the NC of the SM does not
change strangeness, charm etc.

From discussion of the Higgs mechanism for quarks and charged leptons
we could make the following conclusions
\begin{enumerate}
\item The Higgs mechanism provides a {\em natural framework for the unitary
CKM mixing of quarks} in the charged current. It leaves electromagnetic
and neutral currents  diagonal over  fields.
\item However, the Standard Model  cannot predict
masses of quarks and charged leptons and  CKM mixing angles.
In the SM these quantities are parameters which have
to be determined from experimental data.
\end{enumerate}
What about neutrino masses and mixing in the Standard Model? Many
people claim that in the Standard Model neutrinos are massless
two-component particles. {\em If we assume that there are no
right-handed fields $\nu'_{lR}$}, in this case the corresponding
Yukawa interaction can not be built and flavor neutrinos $\nu_{lL}$
will be massless two-component particles. But this is equivalent to
assume from the very beginning that neutrinos are the Landau, Lee and
Yang and Salam two-component massless particles.\footnote{Originally,
the Standard Model was built with massless two-component neutrinos.
It was natural in 1967 for the authors of the Standard Model to make
this assumption. }

We can generate neutrino masses by the standard Higgs mechanism in
the same way as masses of quarks and charged leptons were generated.
In this case neutrino masses would be proportional to the parameter
$v$ and we could expect that they are of the same order of magnitude
as the masses of other fermions, partners of neutrinos.

Let us consider for illustration the masses of the quarks
and leptons of the third family. We have
\begin{eqnarray}\label{3family}
 m_{t}\simeq 1.7 \cdot 10^{2}~\mathrm{GeV} \quad m_{b}\simeq 4.7 ~\mathrm{GeV} \nonumber \\
  m_{3}\leq 2.2~10^{-9}~\mathrm{GeV} \quad m_{\tau}\simeq 1.8~\mathrm{GeV}
\end{eqnarray}
The masses of $t$, $b$ and $\tau$ differ by not more than 2 orders of
magnitude. The neutrino masses differ from the masses of quarks and
charged leptons by (at least) (9-11) orders of magnitude. {\em It is
very unlikely that the masses of quarks, leptons and neutrinos are of
the same Higgs origin. For neutrino masses a new (or additional)
mechanism is needed.} A possible mechanism of the generation of small
neutrino masses will be discussed briefly later.

\section{Neutrino and discovery of Neutral Currents}

\begin{center}
{\bf Introduction }
\end{center}
The discovery of the neutral currents in the Gargamelle neutrino
experiment at CERN in 1973 opened a new era in the physics of the
weak and electromagnetic interactions. The Gargamelle result was the
first confirmation of the  approach based on the idea of the
unification of these interactions.

At the beginning of the seventies the Glashow-Weinberg-Salam model
was considered as a correct strategy and one of the possible models.
However, after the Gargamelle discovery of NC neutrino processes,
detailed investigations of the effects of NC in deep inelastic
electron(muon)-nucleon scattering and in atomic transitions, the
discovery of $W^{\pm}$ and $Z^{0}$ bosons and precise measurement of
their masses, high precision studies  of different electroweak
processes at the $e^{+}-e^{-}$ colliders SLC(Stanford) and LEP (CERN)
fully confirmed the minimal Glashow-Weinberg-Salam model. This model
became the Standard Model of the weak and electromagnetic
interactions. It perfectly describes the existing electroweak data.

Up to now, however, there is no proof of the correctness of the
standard Higgs mechanism. The search for the scalar Higgs boson and
the investigation of the mechanism of the symmetry breaking are first
priority problems for experiments at the LHC collider at CERN.

\vskip0.5cm

Neutral currents were discovered in 1973 at CERN. This was the first
confirmation of the unified theory of  weak and electromagnetic
interactions.

Due to the exchange of the $W$-boson between lepton and quark
vertices $\nu_{\mu}$ ($\bar \nu_{\mu}$) produce $\mu^{-}$ ($\mu^{+}$)
in the inclusive processes
\begin{equation}\label{CCnuinclusiv}
\nu_{\mu}+N\to \mu^{-} +X,\quad \bar\nu_{\mu}+N\to \mu^{+} +X.
\end{equation}
Here $X$ means any possible final state of hadrons. If $Q^{2}\ll m^{2}_{W}$ ($Q^{2}$ is the square of the momentum transfer) the effective Hamiltonian of the processes (\ref{CCnuinclusiv}) has the form
\begin{equation}\label{CCnuinclusiv2}
\mathcal{H}^{CC}=\frac{G_{F}}{\sqrt{2}}2 \bar \mu_{L} \gamma^{\alpha}
\nu_{\mu L}j^{CC}_{\alpha}+\mathrm{h.c.},
\end{equation}
where $j^{CC}_{\alpha}$ is the quark charged current and $G_{F}$ is
the Fermi constant (he Fermi constant has the following numerical
value $G_F = 1.166364(5)\cdot 10^{-5}\mathrm{GeV}^{-2}$). In the
seventies the CC processes (\ref{CCnuinclusiv}) were intensively
studied in neutrino experiments at the  Fermilab and CERN. These
experiments were very important for the establishment of the quark
structure of the nucleon.

If in addition to the CC interaction there exists also the
NC interaction, in this case processes
\begin{equation}\label{NCnuinclusiv}
\nu_{\mu }+N\to \nu_{\mu } +X,\quad \bar\nu_{\mu}+N\to \bar\nu_{\mu}
+X
\end{equation}
induced by the exchange of the $Z$-boson between neutrino and quark vertices
become possible. The signature of such processes is  hadrons in the
final state (no muons).  The  effective SM Hamiltonian
of the processes (\ref{NCnuinclusiv})  has the form
\begin{equation}\label{NCnuinclusiv1}
\mathcal{H}^{NC}=\frac{G_{F}}{\sqrt{2}}2 \bar \nu_{\mu L}
\gamma^{\alpha} \nu_{\mu L}j^{NC}_{\alpha}+\mathrm{h.c.},
\end{equation}
where $j^{NC}_{\alpha}$ is the neutral current of quarks. Thus, in the framework of the Standard Model CC and NC interactions are characterized by the same Fermi constant. We can  expect that the cross section of the processes
(\ref{CCnuinclusiv}) and (\ref{NCnuinclusiv}) are comparable.

The processes (\ref{NCnuinclusiv})  were observed in the large bubble
chamber Gargamelle at CERN in 1973. The  bubble chamber Gargamelle
(4.8 m long, 2 m in diameter, filled with 18 tonnes of liquid Freon)
was built specially for the study of neutrino processes. At the first
meeting of the collaboration in Milan (1968), where the neutrino
program was discussed, the search for NC induced processes had the
eighth priority. The main aim of the experiment was an investigation
of the structure of a nucleon through the  observation of CC
processes (\ref{CCnuinclusiv}). The Glashow-Weinberg-Salam model was
considered at that time  only as one of the possibilities.

At the beginning of 1973 one event of the NC process
\begin{equation}\label{NCelastic}
\nu_{\mu}+e\to \nu_{\mu} +e.
\end{equation}
was found in the Gargamelle chamber. Taking into account that the
background for (\ref{NCelastic}) is very small (less than 1\%) this
one event motivated the intensive search for hadronic NC-induced
processes (\ref{CCnuinclusiv}) which have cross sections about two
orders of magnitude larger than the cross section of the NC leptonic
process (\ref{NCelastic}).

The main problem in the search for hadronic NC processes was a
background from neutrons produced in CC neutrino interactions in the
surrounding materials. The proof of the neutrino origin of NC
hadronic events followed from the fact that the ratio of selected NC
events and CC events did not depend on the longitudinal and radial
distances, whereas hadronic events of the neutron origin would have
shown strong dependence on the distance. Obviously,  the large size
of the bubble chamber was very important for the detection of NC
events. In the first Gargamelle publication \cite{Gargamelle} for the
ratio $R$ of the number of NC and CC events the following values were
given
\begin{equation}\label{Gargamelle}
 R_{\nu}=0.21\pm 0.03,\quad   R_{\bar\nu}=0.45\pm 0.09.
\end{equation}
In the beginning these data were confirmed by the HPWF  collaboration
working at the Fermilab. However, later the HPWF collaboration
modified their detector and for the ratio $R_{\nu}$ they announced a
result compatible with zero ($R_{\nu}=0.05\pm 0.05$). For about one
year many people at CERN and other places did not believe in the
correctness of the Gargamelle result.

By the middle of 1974 the Gargamelle collaboration doubled their
statistics and confirmed their original result. The HPWF
collaboration made a new measurement and also confirmed the
Gargamelle finding. This result was confirmed by other Fermilab
neutrino experiments. {\em The discovery of the neutral currents was
firmly established.}

The eighties and nineties were the years of intensive study of
different NC-induced processes. The effects of neutral currents were
observed in the experiments on the measurement of the asymmetry in
the deep inelastic scattering of polarized electrons (and muons) on
an unpolarized nucleon target and on the study of  atomic processes
\footnote{In such experiments, the effect of interference of diagrams
with the exchange of $\gamma$ and $Z$ was revealed.}, in experiments
on the study of  $\nu_{\mu}(\bar\nu_{\mu})+e\to
\nu_{\mu}(\bar\nu_{\mu}) +e$ processes, etc. All these data were in
perfect agreement with the SM. The values of the parameter
$\sin^{2}\theta_{W}$ obtained from the data of different experiments
are in a good agreement with each other. From the measurement of the
cross sections of NC neutrino reactions (\ref{NCnuinclusiv})  and CC
neutrino reactions (\ref{CCnuinclusiv}) it was obtained \cite{NuTeV}
\begin{equation}\label{Weinbangle1}
\sin^{2}\theta_{W} = 0.2277\pm 0.0016.
\end{equation}

\section{Neutrino masses, mixing and oscillations}

\subsection{Earliest ideas of neutrino masses and oscillations}

\begin{center}
{\bf Introduction }
\end{center}
The earliest ideas of neutrino masses, mixing and oscillations were
based on arguments like an analogy between weak interactions of
leptons and hadrons (quarks), the Nagoya model with the neutrino as a
constituent of the proton and other baryons, etc.

After the great success of the theory of the two-component massless
neutrinos, for many years these ideas were not shared by the majority
of physicists.

Of course, it was absolutely unknown in the seventies whether
neutrinos had small masses and, if they had masses, whether they were
mixed. However, understanding of neutrino oscillations as an
interference phenomenon made it clear  (S. Bilenky and B. Pontecorvo
\cite{BilPonte77}) that
\begin{enumerate}
  \item Experiments on the search for neutrino oscillations
constitute the most sensitive way to look for small  neutrino
mass-squared differences.
  \item Experiments with neutrinos from different facilities
are sensitive to different values of neutrino mass-squared
differences. Neutrino oscillations must be searched for in all
possible neutrino experiments (solar, atmospheric, reactor,
accelerator).
\end{enumerate}
This strategy was summarized in \cite{BilPonte77}. After many years
of efforts it brought success.
\vskip0.5cm

The first idea of neutrino masses and oscillations was suggested in
1957-58 by B. Pontecorvo \cite{BPonte57,BPonte58}. At that time the
Gell-Mann and Pais \cite{GMPais} theory of $K^{0}\rightleftarrows
\bar K^{0}$ mixing and oscillations was confirmed by experiment.
Pontecorvo was fascinated by the idea of particle-mixing and
oscillations and thought about a possibility of oscillations in the
lepton world. In such a way he came to the idea of neutrino
oscillations which was a very courageous idea at the time when there
was a common opinion that the neutrino is a two-component massless
particle.

Before discussing neutrino oscillations let us briefly consider
($K^{0}-\bar K^{0}$) mixing and oscillations which were studied in
detail in many experiments. $K^{0}$ and $\bar K^{0}$ are particles
with strangeness equal to +1 and -1, respectively. They are produced
in hadronic processes ($\pi_{-}+p\to  K^{0} +\Lambda$ etc) in which
the strangeness is conserved. For  the states of $K^{0}$ and $\bar
K^{0}$ we have
\begin{equation}\label{Keigenstates}
H_{0}~|K^{0}\rangle =m~ |K^{0}\rangle,\quad H_{0}~|\bar K^{0}\rangle =m ~|\bar K^{0}\rangle.
\end{equation}
Here $H_{0}$ is the sum of the free Hamiltonian and Hamiltonians of
the strong and electromagnetic interactions, $ |K^{0}\rangle $ and $
|\bar K^{0}\rangle $ are states of $K^{0}$ and $\bar K^{0}$ (in the
rest frame) and $m$ is their mass. The arbitrary phase of $
|K^{0}\rangle $ and $ |\bar K^{0}\rangle $ can be chosen in such a
way that
\begin{equation}\label{Keigenstates1}
|\bar K^{0}\rangle =CP ~| K^{0}\rangle,
\end{equation}
where $C$ is the operator of the charge conjugation and
$P$ is the operator of the space inversion.

The weak interaction does not conserve strangeness. Eigenstates of
the total Hamiltonian, which includes the Hamiltonian of the weak
interaction, are superpositions
\begin{equation}\label{Keigenstates2}
| K_{S}^{0}\rangle=p~|K^{0}\rangle+q~|\bar K^{0}\rangle,~~~
| K_{L}^{0}\rangle=p~|K^{0}\rangle-q~|\bar K^{0}\rangle.
\end{equation}
From the normalization condition of the states
$| K_{S,L}^{0}\rangle$
it follows that the coefficients $p$ and $q$ satisfy the condition $|p|^{2}+
|q|^{2}=1$. From (\ref{Keigenstates2}) we find the following  relations
\begin{equation}\label{Kmixture}
| K^{0}\rangle=\frac{1}{2p}~(|K^{0}_{S}\rangle+|\bar K^{0}_{L}\rangle),\quad
| \bar K^{0}\rangle=\frac{1}{2q}~(|K^{0}_{S}\rangle-|\bar K^{0}_{L}\rangle).
\end{equation}
Thus, the states of particles with definite strangeness $K^{0}$ and
$\bar K^{0}$ are superpositions ("mixtures") of the states of
particles with definite masses and widths $K_{S}^{0}$ and $\bar
K_{L}^{0}$, eigenstates of the total effective nonhermitian
Hamiltonian $H$ :
\begin{equation}\label{Keigenstates4}
H~| K_{S,L}^{0}\rangle=\lambda_{S,L}~| K_{S,L}^{0}\rangle.
\end{equation}
Here
\begin{equation}\label{Keigenstates5}
\lambda_{S,L}=m_{S,L}-\frac{i}{2}\Gamma_{S,L},
\end{equation}
where $m_{S,L}$ and $\Gamma_{S,L}$ are the mass and the total width
of $ K_{S}^{0}$ ($ K_{L}^{0}$). From experimental data it follows
that the lifetimes of $K^{0}_{S}$ (Short-lived) and $K^{0}_{L}$
(Long-lived) are given by \cite{PDG}
\begin{equation}\label{Keigenstates6}
\tau_{S}=\frac{1}{\Gamma_{S}}=(0.8953\pm 0.0005)\cdot 10^{-10}s,\quad
\tau_{L}=\frac{1}{\Gamma_{L}}=(5.116\pm 0.021)\cdot 10^{-8}s.
\end{equation}
States with definite masses and widths are evolved in proper time $t$
as follows:
\begin{equation}\label{evolution}
|K_{S}^{0}\rangle_{t}=e^{-i\lambda_{S}t}~
|K_{S}^{0}\rangle,\quad |K_{L}^{0}\rangle_{t}=e^{-i\lambda_{L}t}~
|K_{L}^{0}\rangle.
\end{equation}
Let us consider the evolution in time of a state $|K^{0}\rangle$
which describes $K^{0}$-particles produced in a strong process. We
will neglect small effects of $CP$ violation. In this case we have
$p=q=\frac{1}{\sqrt{2}}$ and
\begin{equation}\label{Keigenstates3}
| K_{S}^{0}\rangle\simeq | K_{1}^{0}\rangle=\frac{1}{\sqrt{2}}(|K^{0}\rangle+|\bar K^{0}\rangle),~~| K_{L}^{0}\rangle\simeq | K_{2}^{0}\rangle=\frac{1}{\sqrt{2}}(|K^{0}\rangle-|\bar K^{0}\rangle).
\end{equation}
Thus, in the case of $CP$-conservation we have the following mixing relations
\begin{equation}\label{Kmixtures}
|K^{0}\rangle=\frac{1}{\sqrt{2}}(|K_{1}^{0}\rangle+| K_{2}^{0}\rangle),\quad|\bar K^{0}\rangle=\frac{1}{\sqrt{2}}
(|K_{1}^{0}\rangle-| K_{2}^{0}\rangle).
\end{equation}
From (\ref{evolution}) and (\ref{Kmixtures}) we find
\begin{equation}\label{evolution1}
|K^{0}\rangle_{t}= \frac{1}{\sqrt{2}}(e^{-i\lambda_{S}t}~
|K_{1}^{0}\rangle+e^{-i\lambda_{L}t}~
|K_{2}^{0}\rangle)=g_{+}(t)|K^{0}\rangle+g_{-}(t)|\bar K^{0}\rangle,
\end{equation}
where
\begin{equation}\label{evolution2}
g_{+}(t)=\frac{1}{2}(e^{-i\lambda_{S}t}+e^{-i\lambda_{L}t}),
\quad g_{-}(t)=\frac{1}{2}
(e^{-i\lambda_{S}t}-e^{-i\lambda_{L}t}).
\end{equation}
The state $|\bar K^{0}\rangle_{t}$ depends on time in a similar way
\begin{equation}\label{evolution3}
|\bar K^{0}\rangle_{t}\simeq \frac{1}{\sqrt{2}}(e^{-i\lambda_{S}t}~
|K_{1}^{0}\rangle-e^{-i\lambda_{L}t}~
|K_{2}^{0}\rangle)=g_{+}(t)|\bar K^{0}\rangle+g_{-}(t)| K^{0}\rangle.
\end{equation}
Thus, because of the mixing (\ref{Kmixture}) at $t>0$ {\em the states
$|K^{0}\rangle_{t}$ and $|\bar K^{0}\rangle_{t}$ are superpositions
of $| K^{0}\rangle$ and $|\bar K^{0}\rangle$.} The probability of the
transition $ K^{0} \to \bar K^{0}$ during the time $t$ is given by
the expression
\begin{equation}\label{evolution4}
P(K^{0} \to \bar K^{0}; t)=|g_(t)|^2=\frac{1}{4}(e^{-\Gamma_S t}+e^{-\Gamma_L t}-
2e^{-\frac{1}{2}(\Gamma_S+\Gamma_L) t}\cos\Delta m t),
\end{equation}
where $\Delta m=m_L-m_S$. Thus, the oscillating term  of the
probability is determined by  the mass difference of the $ K_{L}^{0}$
and $ K_{S}^{0}$ mesons. Let us stress that this term originates from
the interference of the exponents in (\ref{evolution3}).

The study of the $t$-dependence of the probability $P(K^{0} \to \bar
K^{0}; t)$ in the region  $\Delta m ~t\geq 1$ allows one to determine
the mass difference $\Delta m$. From the analysis of the experimental
data it was found\footnote{This value is many orders of magnitude
smaller than the masses of the neutral kaons ($ m_{K^0}=497.614\pm
0.022\mathrm{MeV}$)}
\begin{equation}\label{evolution5}
\Delta m=(3.483 \pm 0.006) \cdot 10^{-6}\mathrm{eV}.
\end{equation}
The measurement of  such a small quantity became possible because of
the interference nature of the $ K^{0} \to \bar K^{0}$ oscillations.

We will  discuss now Pontecorvo's idea of neutrino oscillations.
Pontecorvo believed in the existence of  symmetry between weak
interaction of leptons and hadrons and  he came first to the idea of
muonium -antimuonum  oscillations \cite{BPonte57} which in the
framework of the lepton-hadron symmetry are analogous to
$K^{0}\rightleftarrows \bar K^{0}$ oscillations. (muonium is the
bound state $(\mu^{+}-e^{-})$ and antimuonum is the bound state of
$(\mu^{-}-e^{+})$).
 In  \cite{BPonte57}, Pontecorvo also mentioned
neutrino oscillations. This was soon after the two-component theory
of a massless neutrino was proposed and the neutrino helicity was
measured in the Goldhaber et al. experiment. Only one type of
neutrino was known at that time. Everybody believed that there were
only two neutrino states: $\nu_{L}$ and $\bar\nu_{R}$. Pontecorvo
assumed that
\begin{enumerate}
  \item Neutrinos had small masses.
  \item Lepton number was not conserved.
  \item Additional neutrino states $\bar\nu_{L}$ and
 $\nu_{R}$ existed so that $\nu_{L}$ could be transferred into
$\bar\nu_{L}$ and $\bar\nu_{R}$ could be transferred into $\nu_{R}$.
\end{enumerate}
Pontecorvo wrote in
\cite{BPonte57}:
``If the theory of two-component neutrino theory was not valid
(which is hardly probable at present)
and if the conservation law for neutrino charge took not place,
neutrino $\to$ antineutrino transitions in vacuum would be in principle possible.''

A special paper on neutrino oscillations \cite{BPonte58} was
published by B. Pontecorvo in 1958. At that time R. Davis was doing
an experiment with reactor antineutrinos \cite{Davis1} with the aim
to test the conservation of the lepton number $L$. Davis searched for
the production of $^{37}\rm{Ar}$ in the process
\begin{equation}
\bar{\nu} +  ^{37}\rm{Cl} \to e^- +  ^{37}\rm{Ar},
\label{02}
\end{equation}
which is evidently forbidden if $L$ is conserved. A rumor reached B.
Pontecorvo that Davis had seen some events (\ref{02}). B.Pontecorvo
who had earlier been thinking about neutrino oscillations was very
excited with a possibility to explain Davis "events" by
$\bar\nu_{R}\to \nu_{R} $ oscillations.

He wrote: ``Recently the question was discussed~\cite{BPonte57}
whether there exist other {\em mixed} neutral particles beside the
$K^0$ mesons, i.e., particles that differ from the corresponding
antiparticles, with the transitions between particle and antiparticle
states not being strictly forbidden. It was noted that the neutrino
might be such a mixed particle, and consequently there existed a
possibility of real neutrino $\rightleftarrows$ antineutrino
transitions in vacuum, provided that lepton (neutrino) charge was not
conserved. This means that the neutrino and antineutrino are {\em
mixed} particles, i.e., a symmetric and antisymmetric combination of
two truly neutral Majorana particles $\nu_1$ and $\nu_2$ of different
combined parity''.

So basically by analogy with the $K^{0}- \bar K^{0}$ mixing
(\ref{Kmixtures}) Pontecorvo assumed that
\begin{equation}\label{numixtures}
|\bar\nu_{R}\rangle=\frac{1}{\sqrt{2}}(|\nu_{1}\rangle+| \nu_{2}\rangle),\quad|\nu_{R}\rangle=\frac{1}{\sqrt{2}}
(|\nu_{1}\rangle-|\nu_{2}\rangle),
\end{equation}
where $|\nu_{1,2}\rangle$ are states of Majorana neutrinos
$\nu_{1,2}$ with masses $m_{1,2}$.\footnote{If the lepton number $L$
is violated, there is no way to distinguish a neutrino and an
antineutrino: they are the same particles. A theory of such particles
was proposed by E. Majorana in 1937 \cite{Majorana}.}

In contrast to  $K_{S,L}^{0}$ the neutrinos $\nu_{1,2}$ are stable
particles \footnote{No indications in favor of neutrino decays were
found.}. From (\ref{numixtures}) we find (in the lab. system)
\begin{equation}\label{numixtures1}
|\bar\nu_{R}\rangle_{t}=
\frac{1}{\sqrt{2}}(e^{-iE_{1}t}|\nu_{1}\rangle+e^{-iE_{2}t}| \nu_{2}\rangle)=\frac{1}{2}(g_{+}(t)|\bar\nu_{R}\rangle
+g_{-}(t)|\nu_{R}\rangle).
\end{equation}
Here
\begin{equation}\label{numixtures2}
g_{\pm}(t)=(e^{-iE_{1}t}\pm e^{-iE_{2}t}),\quad E_{i}=\sqrt{p^{2}+m^{2}_{i}}\simeq p+\frac{m^{2}_{i}}{2E},
\end{equation}
where $p$ is the neutrino momentum. In neutrino experiments we have $p\gg m_{i}$ and  $p\simeq E$ ($E$ is the neutrino energy).

From (\ref{numixtures1}) and (\ref{numixtures2}) for the transition
probabilities we obtain the following expressions:
\begin{equation}\label{numixtures3}
P(\bar\nu_{R}\to \nu_{R})=\frac{1}{2}(1-\cos\frac{\Delta m^{2}L}{2E}),\quad P(\bar\nu_{R}\to \bar\nu_{R})=1-\frac{1}{2}(1-\cos\frac{\Delta m^{2}L}{2E}),
\end{equation}
where $\Delta m^{2}= m_{2}^{2}- m_{1}^{2}$  and $L\simeq t$ is the distance between neutrino source and neutrino detector.

Thus, in the case of the neutrino oscillations the probability for a
reactor antineutrino to survive $P(\bar\nu_{R}\to \bar\nu_{R})$
depends on the distance $L$. B. Pontecorvo wrote in
\cite{BPonte58}:``...the cross section of the production of neutrons
and positrons in the process of the absorption of antineutrinos from
a reactor by protons would be smaller than the expected cross
section. It would be extremely interesting to perform  the
Reines-Cowan experiment at different distances from reactor''.

Pontecorvo obviously did not know the value of the neutrino
mass-squared difference $\Delta m^{2}$. If it is relatively large,
the cosine terms in (\ref{numixtures3}) disappear due to averaging
over neutrino energies and distance. In this case $P(\bar\nu_{R}\to
\bar\nu_{R})=P(\bar\nu_{R}\to \nu_{R})=\frac{1}{2}$. Discussing this
case he wrote :"...a beam of neutral leptons consisting mainly of
antineutrinos when emitted from a nuclear reactor, will consist at
some distance $L$ from the reactor of half neutrinos and half
antineutrinos.''

If $\Delta m^{2}$ is very small, in this case the cosine terms are
practically equal to one and the effect of oscillations of reactor
antineutrinos  could not be observed. Pontecorvo noticed in
\cite{BPonte58}: "...effect of transformation of neutrino into
antineutrino and vice versa may be unobservable in laboratory, but
will certainly occur, at least, on an astronomic scale".

Let us stress again that the proposal of neutrino oscillations
immediately after the great success of the two-component neutrino
theory and in the situation when only one type of neutrino was known
was a very nontrivial one. The Pontecorvo paper was written at the
time when the Davis reactor experiment was not yet finished and
candidate-events (\ref{02}) existed. In order to explain them he had
to assume that $\nu_{R}$ interacts with matter. He wrote:"... it is
impossible to conclude a priori that the antineutrino beam which at
first is essentially incapable of inducing the reaction in question
transforms itself into a beam in which a definite fraction of
particles can induce such reaction".

In spite of the fact that the candidate-events (\ref{02}) disappeared
and only an upper bound for the cross section of the process
(\ref{02}) was found in the Davis experiment, Pontecorvo continued to
believe in neutrino oscillations. He liked the idea that neutrinos
(antineutrinos) produced in weak processes can oscillate into
antineutrinos (neutrinos)  which have no (standard) weak interaction.
He proposed to name such noninteracting neutrinos {\em sterile}. The
idea of sterile neutrinos is very popular nowadays.

The program of the study of oscillations of reactor antineutrinos,
which was outlined by B. Pontecorvo in the very first paper on
neutrino oscillations, was realized in the KamLAND experiment about
40 years later. We will discuss this experiment in the next
subsection.

After the first paper on the neutrino oscillations Pontecorvo
continued to think about this fascinating phenomenon. His belief in
neutrino masses was based on the fact that there was no principle
(like gauge invariance for photon) which requires the neutrino to be
a massless particle.

In the sixties, B. Pontecorvo discussed the problem of neutrino
masses with L. Landau, one of the authors of the two-component
neutrino theory. Landau agreed with Pontecorvo that after the V-A
theory, which was based on the assumption that the left-handed
components of all fields entered into weak interaction Hamiltonian,
neutrino was not longer special particles and it would be natural for
neutrinos to have small masses.

After the discovery of the second neutrino $\nu_{\mu}$ Pontecorvo
applied his idea of neutrino oscillations to the case of  two types
of neutrinos $\nu_{e}$ and $\nu_{\mu}$.  In the second paper on
neutrino oscillations published in 1967 \cite{BPonte67} Pontecorvo
considered $\nu_{e}\rightleftarrows \nu_{\mu}$,
$\nu_{e}\rightleftarrows \bar\nu_{eL}$ (sterile),
$\nu_{e}\rightleftarrows \bar\nu_{\mu L}$ (sterile), etc.
 oscillations and applied the idea of neutrino oscillations to solar neutrinos.

At that time R. Davis started his famous experiment on the detection
of the solar neutrinos in which the radiochemical method of neutrino
detection, proposed by B.Pontecorvo in 1946, was used. Solar
neutrinos were detected in this experiment via the the observation of
the  reaction
\begin{equation}
\nu_e +  ^{37}\rm{Cl} \to e^- +  ^{37}\rm{Ar}.
\label{03}
\end{equation}
In the paper \cite{BPonte67} B. Pontecorvo wrote: "From an
observational point of view the ideal object is the sun. If the
oscillation length is smaller than the radius of the sun region
effectively producing neutrinos, (let us say one tenth of the sun
radius $R_\odot$ or 0.1 million km for $^8B$ neutrinos, which will
give the main contribution in the experiments being planned now),
direct oscillations will be smeared out and unobservable. The only
effect on the earth's surface would be that the flux of observable
sun neutrinos must be two times smaller than the total (active and
sterile) neutrino flux."

The first Davis results was obtained at the end of the sixties (see \cite{1Davis}). It was found that the
upper bound of the observed flux of the solar $\nu_{e}$'s was (2-3)
times smaller than the predicted flux. This result created "the solar
neutrino problem". In  \cite{BPonte67} {\em  Pontecorvo envisaged the
solar neutrino problem.} He understood, however, that the prediction
of the flux of high-energy $^8B$ neutrinos, which gave the major
contribution to the event rate in the Davis experiment, was an
extremely difficult problem: "Unfortunately, the relative weight of
different thermonuclear reactions in the sun and its central
temperature are not known well enough to permit a comparison of the
expected and observed solar neutrino intensities." It took many years
of research to prove that the observed depletion of fluxes of solar
neutrinos are effects of neutrino transitions due to neutrino masses,
mixing and interaction of neutrinos with matter which we will briefly
discuss later.

The first phenomenological scheme of neutrino mixing was proposed by
V. Gribov and B. Pontecorvo in 1969 \cite{GPonte69}. They assumed
that only the left-handed flavor fields $\nu_{eL}(x)$ and $\nu_{\mu
L}(x)$ entered into the total Lagrangian. There was a widespread
opinion at that time that in this case neutrino masses must be equal
to zero. V. Gribov and B. Pontecorvo showed that this is not the case
{\em if the total lepton number $L$ is violated.} In this case
\begin{eqnarray}\label{Mjmixing}
\nu_{eL}(x)&=&\cos\theta\nu_{1L}(x) + \sin\theta \nu_{2L}(x)\nonumber\\
 \nu_{\mu L}(x)&=&-\sin\theta\nu_{1L}(x) + \cos\theta \nu_{2L}(x),
 \end{eqnarray}
where $\nu_{1}(x)$ and $\nu_{2}(x)$ are the fields of {\em Majorana
neutrinos with masses $m_{1}$ and $m_{2}$} and $\theta$ is the mixing
angle.

The scheme of two-neutrino mixing, proposed by V. Gribov and B.
Pontecorvo, was the minimal one. In this scheme:
\begin{itemize}
  \item The only possible oscillations are $\nu_{e}\rightleftarrows \nu_{\mu }$.
  \item There are no sterile neutrinos.
  \item To four states of flavor neutrinos and antineutrinos
(left-handed $\nu_{e}$, $ \nu_{\mu }$ and right-handed $\bar\nu_{e}$,
$ \bar\nu_{\mu }$) there correspond four states of two massive
Majorana neutrinos with helicities $\pm 1$.
\end{itemize}
In  \cite{GPonte69}  the following general expression for the
two-neutrino survival probability in vacuum was obtained
\footnote{Expression (\ref{numixtures3}) corresponds to the case of
maximal mixing $\theta=\frac{\pi}{4}$.}
\begin{equation}\label{Mjmixing1}
 P(\nu_{e}\to \nu_{e})=1-\frac{1}{2}\sin^{2}2\theta ~ (1-\cos\frac{\Delta m^{2}L}{2E}).
\end{equation}
In \cite{GPonte69} and later in \cite{BF69} the effect of vacuum
$\nu_{e}\rightleftarrows \nu_{\mu }$ oscillations on the flux of
solar $\nu_{e}$'s on the earth was discussed.

In the eighties the Cabibbo-GIM mixing (\ref{mix2}) of $d$ and $s$
quarks was fully established. In the papers
\cite{BilPonte76,FM76,ES76}  neutrino mixing
\begin{eqnarray}\label{Dmixing}
\nu_{eL}(x)&=&\cos\theta\nu_{1L}(x) + \sin\theta \nu_{2L}(x)\nonumber\\
 \nu_{\mu L}(x)&=&-\sin\theta\nu_{1L}(x) + \cos\theta \nu_{2L}(x)
 \end{eqnarray}
was introduced on the basis of the lepton-quark analogy. The main
ideas were the following:
\begin{enumerate}
  \item Neutrinos like all other fundamental
fermions (charged leptons and quarks) are massive particles.
  \item The mixing is a general feature of
gauge theories with a mass generation mechanism based on the
spontaneous violation of symmetry. Thus, it is quite natural to
assume that {\em fields of neutrinos like fields of quarks enter into
the charged current in a mixed form.}
      \end{enumerate}
In (\ref{Dmixing}), $\nu_{1}(x)$ and $\nu_{2}(x)$ are the fields of
neutrinos with masses $m_{1}$ and $m_{2}$. However, in contrast to
the Gribov-Pontecorvo scheme, in this scheme the total lepton number
is conserved and {\em $\nu_{1,2}$ are the Dirac particles} (like
quarks). In \cite{BilPonte76,FM76,ES76} possible neutrino
oscillations in reactor and accelerator neutrino experiments were
discussed.

As we have seen earlier, the initial ideas of neutrino masses, mixing
and oscillations  were based on  symmetry (analogy) of weak
interactions of leptons and hadrons (and later leptons and quarks).
In 1962 Maki, Nakagawa and Sakata \cite{MNS}  introduced the neutrino
mixing in the framework of the Nagoya model in which the proton and
other baryons were considered as bound states of neutrinos and a
vector boson $B^{+}$, "a new sort of matter". At that time the
Brookhaven experiment, in which it was proved that $\nu_{e}$ and
$\nu_{\mu}$ were different particles, was not yet finished. However,
there was an indication, based on the fact the decay $\mu^{+}\to
e^{+}+\gamma$ was not observed, that $\nu_{e}$ and $\nu_{\mu}$ were
different types of neutrinos determined by the weak charged current
\begin{equation}\label{weakCC}
j_{\alpha}^{CC}=2(\bar \nu_{eL}\gamma_{\alpha}e_{L}+
\bar \nu_{\mu L}\gamma_{\alpha}\mu_{L}).
\end{equation}
The authors wrote: "We assume that there exists a representation
which defines the true neutrinos $\nu_{1}$ and $\nu_{2}$ through
orthogonal transformation"
\begin{eqnarray}\label{MNS}
\nu_{1}&=&\cos\delta\nu_{e} - \sin\delta \nu_{\mu}\nonumber\\
\nu_{2}&=&\sin\delta\nu_{e} + \cos\delta \nu_{\mu},
\end{eqnarray}
where $\delta$ is the Cabibbo angle. In  \cite{MNS} it was assumed
that the  "true neutrino" $\nu_{1}$ was a constituent of baryons and
possessed some  mass $m_{1}$. Further, the authors assumed that there
existed an additional interaction of $\nu_{2}$ with a field of heavy
particles $X$ which ensured the difference of masses of $\nu_{2}$ and
$\nu_{1}$.

In contrast to \cite{BPonte58,BPonte67,GPonte69}, in \cite{MNS} the
quantum phenomenon of neutrino oscillations, based on the difference
of phases which were gained in propagation of neutrinos with definite
masses,
 was not considered. Nevertheless $\nu_{e}\to \nu_{\mu}$
transitions were discussed in \cite{MNS}. The authors wrote: "Weak neutrinos
\begin{eqnarray}\label{MNS1}
\nu_{e}&=&\cos\delta\nu_{1} + \sin\delta \nu_{2}\nonumber\\
\nu_{\mu}&=&-\sin\delta\nu_{1} + \cos\delta \nu_{2}
\end{eqnarray}
are not stable due to the occurrence of virtual transition
$\nu_{e}\rightleftarrows \nu_{\mu}$ caused by this additional
interaction with $\nu_{2}$". Moreover, in connection with the
Brookhaven neutrino experiment they noticed : ..." a chain of
reactions
\begin{eqnarray}\label{MNS2}
\pi^{+}&\to &\mu^{+}+\nu_{\mu}\nonumber\\
\nu_{\mu}+Z&\to & (\mu^{-}~\mathrm{and/or})~e^{-}
\end{eqnarray}
 is useful to check the two-neutrino hypothesis only when
\begin{equation}\label{MNS3}
   | m_{\nu_{2}}- m_{\nu_{1}}|\leq ~\mathrm{eV}
\end{equation}
under the conventional geometry of the experiments. Conversely, the
absence of $e^{-}$ will be able not only to verify the two-neutrino
hypothesis but also to provide an upper limit of the mass of the
second neutrino $\nu_{2}$ if the present scheme should be accepted."

The papers \cite{GPonte69,BilPonte76,FM76,ES76,MNS} were written at
the time when only two types of flavor neutrinos $\nu_{e}$  and
$\nu_{\mu}$ were known. In  \cite{GPonte69} it was assumed that there
was no conserved lepton number and neutrinos with definite masses
$\nu_{1}$ and $\nu_{2}$ were truly neutral Majorana particles. In
\cite{BilPonte76,FM76,ES76,MNS} it was assumed that the total lepton
number $L$ was conserved and $\nu_{1}$ and $\nu_{2}$ were Dirac
particles ($L(\nu_{i})=1,~~L(\bar\nu_{i})=-1$). After the discovery
of the $\tau$-lepton it was natural to assume that there existed (at
least) three different types of neutrinos. The mixing relations
(\ref{Mjmixing}) and (\ref{Dmixing}) were generalized for an
arbitrary number $n$ of flavor neutrinos in the following way (see
\cite{BilPonte77}):
\begin{equation}\label{nmixing}
\nu_{lL}=\sum^{n}_{i=1}U_{li}\nu_{iL}, \quad l=e,\mu,...
\end{equation}
Here $U$ is a unitary $n\times n$ matrix ($U^{\dag}U=1$).  The matrix
$U$ is called the mixing matrix. As we will see later, the existing
neutrino oscillation data can be described if we assume that $U$ is
the $3\times 3$ matrix. This matrix is usually called the
Pontecorvo-Maki-Nakagawa-Sakata (PMNS) mixing matrix in order to pay
tribute to the pioneering contribution of these authors to the
neutrino mixing and oscillations.

The mixing (\ref{nmixing}) is not the most general one. In the most general case we have (see \cite{BilPonteNC})
\begin{eqnarray}\label{nmixing1}
\nu_{lL}&=&\sum^{n+n_{st}}_{i=1}U_{li}\nu_{iL}\nonumber\\
\nu_{sL}&=&\sum^{n+n_{st}}_{i=1}U_{si}\nu_{iL},
\end{eqnarray}
where the index $s$ takes $n_{st}$ values and $U$ is a unitary
$(n+n_{st})\times (n+n_{st})$ mixing matrix.  The fields $\nu_{sL}$
are fields of the sterile neutrinos which have no standard weak
interaction. Due to the mixing (\ref{nmixing1}) transitions between
flavor neutrinos $\nu_{l}\rightleftarrows \nu_{l'}$ as well as
transitions between flavor and sterile neutrinos
$\nu_{l}\rightleftarrows \nu_{sL}$ are possible.

In spite of the fact that in the seventies some plausible arguments
for small nonzero masses were given and a general phenomenological
theory of neutrino mixing and oscillations was developed there was no
so much interest in neutrino masses and oscillations at that time:
the idea of massless two-component neutrinos was still the dominant
one. In the first review on neutrino oscillations \cite{BilPonte77}
only about  ten neutrino oscillation papers existing at that time
were referred to.

\subsection{Neutrino oscillations at the time when neutrino masses started
to be considered as a signature of the physics beyond the SM}

\begin{center}
{\bf Introduction }
\end{center}

In the eighties and  nineties with new solar neutrino experiments and
the increase in the number of detected atmospheric neutrino events
the evidence in favor of neutrino masses and oscillations, coming
from these experiments, became stronger and stronger. However, the
interpretation of the data of solar neutrino experiments depended on
the Standard Solar Model. In experiments with neutrinos of
terrestrial origin (reactor and accelerator neutrinos) no positive
indications in favor of neutrino oscillations were found. In 1998,
the situation with neutrino oscillations drastically changed.

\vskip0.5cm

The situation with neutrino masses and the mixing problem drastically
changed at the end of the seventies with appearance of the models of
grand unification (GUT). In these models leptons and quarks enter
into the same multiplets, and the generation of masses of quarks and
charged leptons in some models naturally lead to nonzero neutrino
masses. At that time the famous seesaw mechanism of the neutrino mass
generation \cite{seesaw}, which could explain the smallness of the
neutrino masses with respect to the masses of quarks and charged
leptons, was proposed.

After the appearance of the GUT models and the seesaw mechanism of
neutrino mass generation, {\em masses and mixing of neutrinos started
to be considered as a signature of the physics beyond the Standard
Model.} The problem of neutrino masses and oscillations started to
attract more and more attention of theoreticians and
experimentalists. Several short-baseline\footnote{Distances between
sources and detectors in these experiments were a few hundred meters
or less.} experiments on the search for neutrino oscillations with
reactor and accelerator neutrinos were performed in the eighties. No
positive indications in favor of oscillations in these experiments
with artificially produced neutrinos were found at that
time.\footnote{Recently, fluxes of $\bar\nu_{e}$'s from reactors were
recalculated. It occurred that the fluxes are (3-4) \% higher than
the fluxes used in the analysis of old reactor neutrino oscillation
data. Thus, these data nowadays are interpreted as an indication in
favor of short-baseline neutrino oscillations. New reactor and
accelerator neutrino experiments are under preparation in order to
check the hypothesis of short-baseline oscillations.}

On the other hand, indications in favor of oscillations of solar
neutrinos were strengthened in the eighties. The second solar
neutrino experiment Kamiokande was performed \cite{Kamiokande}. In
this experiment high energy solar neutrinos from the decay
$^{8}\mathrm{B}\to ^{8}\mathrm{Be}+e^{+}+\nu_{e}$ were detected via
the observation of the recoil electrons from the elastic
$\nu+e\to\nu+e$ scattering.  The ratio of the observed flux of the
solar neutrinos to the predicted flux  obtained in the Kamiokande
experiment was about 1/2.

In the Kamiokande and IMB water Cherenkov detectors high energy muons
and electrons, produced by atmospheric muon and electron neutrinos
were detected.\footnote{Atmospheric neutrinos are produced mainly in
decays of pions produced in processes of interactions of cosmic rays
in the atmosphere, and muons which  are produced in decays of pions
($\pi^{\pm}\to \mu^{\pm} +\nu_{\mu}(\bar\nu_{\mu})$, $\mu^{\pm}\to
e^{\pm}+\nu_{e}(\bar\nu_{e})+\bar\nu_{\mu}(\nu_{\mu})$).} It was
found in these experiments that the ratio of the numbers of the
$\nu_{\mu}$ and $\nu_{e}$ events was significantly smaller than the
(practically model independent) predicted ratio
\cite{Kamatmospheric}. This effect was  called the atmospheric
neutrino anomaly. The anomaly could be explained by the disappearance
of $\nu_{\mu}$  due to transitions of $\nu_{\mu}$ into other neutrino
states.

At the beginning of the nineties two new solar neutrino experiments
GALLEX \cite{Gallex} and SAGE \cite{Sage} were performed. In these
experiments, like in the first Davis experiment, Pontecorvo's
radiochemical method of neutrino detection was used. Solar
$\nu_{e}$'s were detected via the observation of radioactive
$^{71}\rm{Ge}$ atoms produced in the process
\begin{equation}\label{Ga}
\nu_{e}+  ^{71}\rm{Ga} \to e^{-}+  ^{71}\rm{Ge}.
\end{equation}
There are three main sources of $\nu_{e}$'s in the sun
\begin{enumerate}
  \item The $pp$ reaction $p+p\to d +e^{+}+ \nu_{e}$ ($E\leq 0.42 ~\mathrm{MeV}$).
  \item The $^{7}\mathrm{Be}$ capture
$e^{-}+^{7}\mathrm{Be}\to ^{7}\mathrm{Li}+\nu_{e}$ ($E= 0.86
~\mathrm{MeV}$).
  \item The $^{8}\mathrm{B}$ decay
$^{8}\mathrm{B}\to ^{8}\mathrm{Be}+e^{+}+\nu_{e}$ ($E\leq 15
~\mathrm{MeV}$).
\end{enumerate}
The threshold of the $\mathrm{Cl}-\mathrm{Ar}$ reaction (\ref{nurec})
is equal to 0.81 MeV. Thus, in the Davis experiment mainly
$^{8}\mathrm{B}$ neutrinos can be detected. The threshold of the
reaction (\ref{Ga}) is equal to 0.23 MeV. This means that in the
GALLEX and SAGE experiments neutrinos from all reactions of
thermonuclear cycles in the sun including low-energy neutrinos from
the  $pp\to de^{+}\nu_{e}$ reaction were detected. This reaction
gives the largest contribution to the flux of the solar neutrinos.
The flux of the $pp$ neutrinos can be connected with the luminosity
of the sun and can be predicted in a model independent way.

The event rates measured in the GALLEX and SAGE experiments were
approximately  two times smaller than  the predicted rates. Thus, in
these experiments additional important evidence was obtained in favor
of the  disappearance of solar $\nu_{e}$ on the way from the central
region of the sun, where solar neutrinos are produced, to the earth.

Solar $\nu_{e}$'s are produced in the central region of the sun and
on the way to the earth pass about $7\cdot 10^{5}$ km of the solar
matter. It was discovered in the nineties \cite{MSW} that for
neutrino propagation in matter  not only masses and mixing but also
coherent interaction are important. This interaction gives an
additional contribution to the Hamiltonian of a neutrino in matter
which is determined by the electron number-density. If the electron
density depends on the distance (as in the case of the sun) the
transition probabilities  between different flavor neutrinos in
matter can have the resonance character (MSW effect).

\subsection{Golden years of neutrino oscillations (1998-2004)}

In 1998, in the Super-Kamiokande atmospheric neutrino
experiment\cite{SK} (Japan)  significant up-down asymmetry of the
high-energy   muon events was observed. Neutrinos produced in the
earth atmosphere and coming from above pass distances from about 20
km to 500 km. Neutrinos coming to the detector from below pass the
earth and travel distances from  500 km to about 12 000 km. It was
discovered in the Super-Kamiokande experiment that the number of
up-going high-energy muon neutrinos was about two times smaller than
the number of the down-going high-energy muon neutrinos. Thus, it was
proved that the number of observed muon neutrinos depends on the
distance which neutrinos passed from a production point in the
atmosphere to the detector.

{\em The Super-Kamiokande atmospheric neutrino result was the first
model independent evidence of neutrino oscillations.} This result
marked a new era in the investigation of neutrino oscillations - an
era of experiments with neutrinos from different sources which
provide model independent evidence of neutrino oscillations.

{\em In 2002 in the SNO solar neutrino experiment \cite{SNO} (Canada)
model independent evidence of the disappearance of solar $\nu_{e}$
was obtained.} In this experiment high-energy solar neutrinos from
$^{8}B$-decay were detected through the observation of CC and NC
reactions. The detection of solar neutrinos through the observation
of the CC reaction allows one to determine the flux of solar
$\nu_{e}$ on the earth, while the detection of solar neutrinos
through the observation of the NC reaction allows one to determine
the {\em flux of all flavor neutrinos ($\nu_{e}$, $\nu_{\mu}$ and
$\nu_{\tau}$)}. It was shown in the SNO experiment that the flux of
the solar $\nu_{e}$ was approximately three times smaller than the
flux of $\nu_{e}$, $\nu_{\mu}$ and $\nu_{\tau}$. Thus, it was proved
that solar $\nu_{e}$'s on the way from the sun to the earth were
transferred to $\nu_{\mu}$ and $\nu_{\tau}$.

{\em In 2002-2004 model independent evidence of oscillations of
reactor $\nu_{e}$ was obtained in the KamLAND reactor experiment
\cite{Kamland} (Japan)}. In this experiment $\nu_{e}$'s from 55
reactors at an average distance of about 170 km from the large
KamLAND detector were recorded. It was found that the total number of
$\bar\nu_{e}$ events was about 0.6 of the number of the expected
events. A significant distortion of the $\bar\nu_{e}$ spectrum with
respect to the expected spectrum was observed in the experiment.

Neutrino oscillations were observed also in the long-baseline
accelerator K2K experiment \cite{K2K} (the distance $L$ between the
source and the detector was about 250 km) and in the MINOS
accelerator neutrino experiment \cite{Minos} (with a distance $L$ of
about 730 km). These experiments fully confirmed the results obtained
in  the atmospheric Super-Kamiokande experiment.

Thus, neutrino oscillations were discovered. It was proven that
neutrinos had small masses and that the flavor neutrinos $\nu_{e},
\nu_{\mu}, \nu_{\tau}$ were "mixed particles". The analysis of
existing data, which we will briefly discuss in the next subsection,
shows that existing neutrino oscillation data are well described if
we assume the three-neutrino mixing.

\subsection{Present status of neutrino oscillations}

\begin{center}
{\bf Introduction }
\end{center}
The discovery of neutrino oscillations was a result of efforts of
many physicists for many years. It required to build very large
neutrino detectors (like Super-Kamiokande, SNO, KamLAND and others)
and to overcome severe background problems. Nevertheless, there were
several "lucky circumstances'' which made it possible to discover and
investigate this phenomenon  in some detail.

In the case of tree-neutrino mixing there are two independent
mass-squared differences $\Delta m_{23}^{2}$ and $\Delta m_{12}^{2}$.
It was a "lucky circumstance" that both mass-squared differences
could be reached in neutrino experiments: the first one in the
atmospheric Super-Kamiokande experiment\footnote{Long-baseline
accelerator experiments started at the time when indications (MINOS)
and evidence (K2K, T2K)  of neutrino oscillations were obtained in
the atmospheric neutrino experiments.} and the second one in the long
baseline KamLAND reactor experiment. The second "lucky circumstance"
was the fact that the neutrino mixing angles  $\theta_{23}$ and
$\theta_{12}$ are large. As a result,  {\em effects of neutrino
oscillations  in the Super-Kamiokande and KamLAND experiments were
large.} This, of course, "simplified" the observation of neutrino
oscillations in these experiments.

\vskip0.5cm

In this subsection we will briefly discuss the present status of the
neutrino mixing and oscillations. We will consider the case of the
three-neutrino mixing.  "Mixed" flavor fields $\nu_{lL}(x)$, which
enter into CC and NC, are given in this case by the relations
\begin{equation}\label{3mixing}
\nu_{l L}(x)=\sum^{3}_{i=1} U_{l i}~\nu_{i L}(x).\quad l-e,\mu,\tau
\end{equation}
Here $U$ is the $3\times 3$ unitary PMNS mixing matrix and $\nu_{i }(x)$ is the field of neutrinos with mass $m_{i}$.

There is a lot of discussions in the literature on the methods of
derivation from (\ref{3mixing}) of the observable probability for the
transition between different types of neutrinos. Practically all
methods give the same expression for the transition probability. We
will  stress here the physical principles on which oscillations are
based and we will use the formalism which is similar to the formalism
of the $K^{0}\rightleftarrows \bar K^{0}$ oscillations.

 We know from neutrino oscillation experiments
(see later) that the neutrino mass squared differences $\Delta
m_{ik}^{2}=m_{k}^{2}-m_{i}^{2}$ are so small that
 the quantities $\frac{E}{\Delta m_{ik}^{2}}$ ($E$ is the neutrino energy)
are macroscopically large (about 10 km for reactor neutrinos and
about 100 km for accelerator neutrinos). As a result, differences
between momenta of neutrinos with different masses produced in weak
decays or reactions are much smaller than quantum mechanical
uncertainties of momenta (determined by the Heisenberg uncertainty
relation). Thus, production (and detection) of neutrinos with
different masses can not be resolved and in CC weak processes
together with a lepton $l^{+}$ a {\em flavor neutrino  $\nu_{l}$,
which is described by  the coherent superposition}
 \begin{equation}\label{3mixing1}
 |\nu_{l}\rangle=\sum^{3}_{i=1}U^{*}_{li}~|\nu_{i}\rangle,
 \end{equation}
 is produced (and detected).

We are interested in neutrino beams. Thus, the states
$|\nu_{i}\rangle$ in (\ref{3mixing1}) are states of neutrinos
$\nu_{i}$ with mass $m_{i}$, helicity -1, momentum $\vec{p}$ and
energy $E_{i}=\sqrt{p^{2}+m^{2}_{i}}\simeq p+\frac{m^{2}_{i}}{2p}$.

If at $t=0$ a flavor neutrino $\nu_{l}$ is produced, we have for the
neutrino state in vacuum at $t>0$
\begin{equation}\label{3mixing2}
|\nu_{l}\rangle_{t}= e^{-iH_{0}t}
\sum_{i}|\nu_{i}\rangle~U^{*}_{li}=
\sum_{i}|\nu_{i}\rangle e^{-iE_{i}t}~U^{*}_{li},
\end{equation}
where $H_{0}$ is the free Hamiltonian. Neutrinos are detected via the
observation of weak processes in which flavor neutrinos  take part
($\nu_{l'}+N\to l'+X$ etc). Developing (\ref{3mixing2}) over states
$|\nu_{l'}\rangle$ we find
\begin{equation}\label{3mixing3}
|\nu_{l}\rangle_{t}=\sum_{l'} |\nu_{l'}\rangle \sum_{i}U_{l'i}~e^{-iE_{i}~t}~U^{*}_{li}.
\end{equation}
If $m_{i}=m$, in this case $E_{i}=E$, $\sum_{i}U_{l'i}~U^{*}_{li}=\delta_{l'l}$ and $|\nu_{l}\rangle_{t}=e^{-iE~t}|\nu_{l}\rangle$. Thus, if all neutrino masses are equal, the produced $\nu_{l}$ will always remain $\nu_{l}$. If neutrino masses are different, in this case the initial $\nu_{l}$ can be transferred into another flavor neutrino $\nu_{l'}$. The probability of the transition $\nu_{l} \to \nu_{l'}$ is given by the expression
\begin{equation}\label{3mixing4}
P(\nu_{l}\to\nu_{l'})=|\sum_{i}U_{l'i}~e^{-iE_{i}~t}~U^{*}_{li}|^{2}.
\end{equation}
Taking into account that for ultrarelativistic neutrinos $t\simeq L$,
where $L$ is the distance between the neutrino source and the
neutrino detector, and $E_{i}-E_{2}=\frac{\Delta m_{2i}^{2}L}{2E}$ we
can rewrite  expression (\ref{3mixing4}) in the form
\begin{equation}\label{3mixing4}
P(\nu_{l}\to\nu_{l'})=|\delta_{l'l}+
\sum_{i\neq 2}U_{l'i}~(e^{-i\frac{\Delta m_{2i}^{2}L}{2E}}-1)~U^{*}_{li}|^{2}.
\end{equation}
It follows from this expression that the probability of the
transition depends periodically on the parameter $\frac{L}{E}$.
expression (\ref{3mixing4}) describes  neutrino oscillations in
vacuum. It is clear from (\ref{3mixing2}) and (\ref{3mixing3}) that
neutrino oscillations happen if the states of neutrinos with
different masses {\em  gain different phases after the evolution of
the neutrino beam during the time $t$ (at the distance $L$).}

The unitary $3\times3$ matrix $U$ is characterized by four parameters:
three mixing angles $\theta_{12}$, $\theta_{23}$,
$\theta_{13}$ and one phase $\delta$. In the case of three neutrino
masses there are two independent mass-squared differences
$\Delta m^{2}_{23}$ and $\Delta m^{2}_{12}$. Thus, in the general case
the transition probability $P(\nu_{l}\to\nu_{l'})$ depends on six parameters.

It follows from the analysis of experimental data that two parameters
are small:
\begin{equation}\label{masshierar}
\frac{\Delta m^{2}_{12}}{\Delta m^{2}_{23}}\simeq  \frac{1}{30},\quad
\sin^{2}2\theta_{13}=0.092\pm0.016\pm0.005.
\end{equation}
In the first (leading) approximation we can neglect contributions of
these parameters to neutrino transition probabilities. In this
approximation, a rather simple picture of neutrino oscillations
emerges.

In the leading approximation in the atmospheric region of
$\frac{L}{E}$ (for $\frac{\Delta m^2_{23 }L }{ 2 E}\gtrsim 1$)
$\nu_{\mu} \rightleftarrows \nu_{\tau}$ oscillations take place. In
this case, the $\nu_{\mu}\to \nu_{\mu}$ survival probability  has the
simple two-neutrino form
\begin{equation}\label{2nuoscil1}
P(\nu_{\mu}\to\nu_{\mu})\simeq 1-P(\nu_{\mu}\to\nu_{\tau})
\simeq 1-\frac{1}{2}\sin^{2}2\theta_{23}~
(1-\cos\Delta m_{23}^{2}\frac{L}{2E}).
\end{equation}
Thus, in the leading approximation neutrino oscillations in the
atmospheric region  are characterized by the parameters $\Delta
m_{23}^{2}$ and $\sin^{2}2\theta_{23}$.

In the KamLAND reactor region ($\frac{\Delta m^2_{12 }L }{ 2
E}\gtrsim 1$) $\bar \nu_{e}\rightleftarrows \bar\nu_{\mu,\tau}$ take place. For the $\bar\nu_{e}\to\bar\nu_{e}$ survival probability we have
\begin{equation}\label{2nuoscil4}
P(\bar\nu_{e}\to \bar\nu_{e})=1-\frac{1}{2}\sin^{2}2\theta_{12}~
(1-\cos\Delta m_{12}^{2}\frac{L}{2E}).
\end{equation}
Thus, in the leading approximation, neutrino oscillations in the
KamLAND reactor region are characterized by the parameters $\Delta
m_{12}^{2}$ and $\sin^{2}2\theta_{12}$.

We also remark that in leading approximation the probability of the
solar neutrinos to survive is given  by the two-neutrino
$\nu_{e}\to\nu_{e} $ survival probability in  matter which depends on
the parameters  $\Delta m_{12}^{2}$ and $\sin^{2}\theta_{12}$ and the
electron number density.

The leading approximation gives the dominant contribution to the
expressions for the neutrino transition probabilities. Until
recently, in the analysis of neutrino oscillation data two-neutrino
expressions (\ref{2nuoscil1}) and (\ref{2nuoscil4}) were used. Now
with the improvement of the accuracy of the experiments
three-neutrino transition probabilities are started to be used in the
analysis of the data.

We will briefly discuss the results that were obtained in some
neutrino oscillation experiments.

{\bf The SNO solar neutrino experiment\cite{SNO}.}

The experiment was carried out in the Creighton mine (Sudbury,
Canada) at a depth of 2092 m. Solar neutrino were detected by a large
heavy-water detector (1000 tons of  D$_2$O contained in an acrylic
vessel of 12 m in diameter). The detector was equipped with 9456
photo-multipliers to detect light created by particles which are
produced in neutrino interaction.

The high-energy $^{8}B$ neutrinos were detected in the SNO experiment.
An  important feature of the SNO experiment was the observation of
solar neutrinos {\em via three different processes}.
\begin{enumerate}
  \item The  CC process
\begin{equation}\label{SNO-CC}
\nu_e + d \to e^- + p + p ~.
\end{equation}
  \item The NC process
\begin{equation}\label{SNO-NC}
\nu_x + d \to \nu_x + p + n \quad (x=e,\mu,\tau)
\end{equation}
  \item Elastic neutrino-electron scattering (ES)
\begin{equation}\label{SNO-ES}
\nu_x + e \to \nu_x + e~.
\end{equation}
\end{enumerate}
The detection of solar neutrinos through the observation of the NC
reaction (\ref{SNO-NC}) allows one to determine the total flux of
$\nu_{e}$, $\nu_{\mu}$ and $\nu_{\tau}$ on the earth. In the SNO
experiment it was found
\begin{equation}\label{totalflux}
\Phi^{NC}_{\nu_{e,\mu,\tau}}=(5.25\pm 16  {}^{+0.11}_{-0.13})\cdot 10^{6}~\mathrm{cm}^{-2}~\mathrm{s}^{-1}.
\end{equation}
The total flux of all active neutrinos on the earth must be equal to
the total flux of $\nu_{e}$ emitted by the sun (if there are no
transitions of $\nu_{e}$ into sterile neutrinos). The flux measured
by SNO is in agreement with the total flux of $\nu_{e}$ predicted by
the Standard Solar Model:
\begin{equation}\label{totalflux1}
\Phi^{SSM}_{\nu_{e}}=(4.85\pm 0.58)\cdot 10^{6}~\mathrm{cm}^{-2}~\mathrm{s}^{-1}.
\end{equation}
The detection of the solar neutrinos via reaction (\ref{SNO-CC})
allows one to determine the total flux of $\nu_{e}$ on the earth. It
was found in the SNO experiments that the total flux of $\nu_{e}$ was
about three times smaller than the total flux of all active
neutrinos.

From the ratio of the fluxes of $\nu_{e}$ and $\nu_{e}$, $\nu_{\mu}$
and $\nu_{\tau}$ the $\nu_{e}$ survival probability can be
determined. It was shown in the SNO experiment that in the
high-energy $^{8}B$ region  the $\nu_{e}$ survival probability did
not depend on the neutrino energy and was equal to
\begin{equation}\label{ratCCNC}
    \frac{\Phi^{CC}_{\nu_{e}}}{\Phi^{NC}_{\nu_{e,\mu,\tau}}}=P(\nu_{e}\to \nu_{e})=
0.317\pm 0.016 \pm 0.009.
\end{equation}
Thus, it was proved in a direct, model independent way that solar
$\nu_{e}$ on the way to the earth are
transformed into $\nu_{\mu}$ and $\nu_{\tau}$.

{\bf The KamLAND reactor neutrino experiment\cite{Kamland}}

The KamLAND detector is situated in the Kamioka mine (Japan) at a
depth of about 1 km. The neutrino target is a 1 kiloton liquid
scintillator which is  contained in a 13 m-diameter transparent nylon
balloon suspended in 1800 $m^{3}$ non-scintillating buffer oil. The
balloon and buffer oil are contained in an 18 m-diameter
stainless-steel vessel. On the inner surface of the vessel 1879
photomultipliers are mounted.

In the KamLAND experiment $\bar\nu_{e}$ from 55
reactors situated at  distances of $175\pm 35$ km
from the Kamioka mine are detected.

Reactor $\bar\nu_{e}$'s are detected in
the KamLAND experiment through the observation of the process
\begin{equation}\label{inversebeta}
\bar\nu _{e}+p \to e^{+}+n.
\end{equation}
The signature of the neutrino event
is a coincidence between two $\gamma$-quanta produced in
the annihilation of a positron (prompt signal) and
a $\simeq 2.2$ MeV $\gamma$-quantum produced by a neutron capture in the process $n+p\to d+\gamma$ (delayed
signal).

The average energy of the reactor antineutrinos
is 3.6 MeV. For such energies, distances of about 100~km are appropriate
 to study neutrino oscillations driven by the solar
neutrino mass-squared difference $\Delta m_{12}^{2}$.

From March 2002 to May 2007 in the KamLAND experiment 1609 neutrino
events were observed. The expected number of neutrino events (if
there are no neutrino oscillations) is $2179 \pm 89 $. Thus, it was
observed in the experiment that $\bar\nu _{e}$ disappeared on the way
from the reactors to the detector.

As the $\bar\nu _{e}$ survival probability depends on the neutrino
energy we must expect that the detected spectrum of $\bar\nu _{e}$ is
different from the spectrum produced by a reactor. In fact in the
KamLAND experiment a significant distortion of the initial
antineutrino spectrum is observed.

The data of the experiment are well described if we assume that
two-neutrino oscillations take place. For the neutrino oscillation
parameters it was found
\begin{equation}\label{KLparameters}
 \Delta m_{12}^{2}=(7.66~ {}_{-0.22}^{+0.20})\cdot 10^{-5}~\mathrm{eV}^{2},\quad \tan^{2}2\theta_{12}=
0.52~{}^{+0.16}_{-0.10} .
\end{equation}
{\em From the three-neutrino analysis of all solar neutrino data and
the data of the KamLand reactor experiment} for the neutrino
oscillation parameters the following values were obtained
\begin{equation}\label{solkam}
\Delta m_{12}^{2}=(7.41 {}_{-0.19}^{+0.21})\cdot 10^{-5}~\mathrm{eV}^{2},\quad \tan^{2}\theta_{12}=
0.446{}^{+0.030}_{0.029},\quad sin^{2}\theta_{13}<0.053.
\end{equation}
{\bf  Super-Kamiokande atmospheric neutrino experiment\cite{SK}}

In the Super-Kamiokande atmospheric neutrino experiment the first
model-independent evidence in favor of neutrino oscillations was
obtained (1998). The Super-Kamiokande detector is situated in the
same Kamioka mine as the KamLAND detector. It consists of two
optically separated water-Cherenkov cylindrical detectors with a
total mass of 50 kilotons of water. The inner detector with 11146
photomultipliers  has a radius of 16.9 m and a height of 36.2 m. The
outer detector is a veto detector. It allows to reject cosmic ray
muons. The fiducial mass of the detector is 22.5 kilotons.

In the Super-Kamiokande  experiment atmospheric neutrinos
in a wide range of energies from about 100 MeV to about 10 TeV are detected .
Atmospheric $\nu_{\mu}$ ($\bar\nu_{\mu}$) and $\nu_{e}$ ($\bar\nu_{e}$)
are detected through the observation of
$\mu^{-}$ ($\mu^{+}$) and $e^{-}$ ($e^{+}$) produced in the processes
\begin{equation}\label{SKatm}
 \nu_{\mu}(\bar \nu_{\mu}) +N\to \mu^{-}(\mu^{+}) +X, \quad
\nu_{e}(\bar \nu_{e}) +N\to e^{-}(e^{+}) +X.
\end{equation}
For the study of neutrino oscillations it is important to
distinguish electrons and muons produced in the processes (\ref{SKatm}).
In the Super-Kamiokande experiment leptons are observed through the
detection of the Cherenkov radiation. The shapes of the Cherenkov rings
of electrons and muons are completely different (in the case of
electrons  the Cherenkov rings exhibit a more diffuse light  than in
the muon case). The probability of a misidentification of electrons
and muons is below  2\%.

A model-independent evidence of neutrino oscillations was
obtained by the Super-Kamiokande Collaboration through the investigation
of the zenith-angle dependence of the  electron and muon
events. The zenith angle $\theta$ is determined in such a way that
neutrinos going vertically downward have $\theta = 0$ and neutrinos
coming vertically upward through the earth have $\theta = \pi$.
 At neutrino energies $E \gtrsim 1$ GeV the fluxes
of muon and electron neutrinos are symmetric under the change
$\theta \to \pi -\theta$. Thus, if there are no neutrino
oscillations in this energy region the numbers of electron and muon events
must satisfy the relation
\begin{equation}\label{azsym}
    N_{l}(\cos\theta)=N_{l}(-\cos\theta)\quad l=e,\mu.
\end{equation}
In the Super-Kamikande experiment a large violation of this relation
for high energy muon events was established (a significant deficit of
upward-going muons was observed). The number of electron events
satisfies the relation (\ref{azsym}).

This result can naturally  be explained by the disappearance of muon
neutrinos due to neutrino oscillations.  The probability for
$\nu_{\mu}$ to survive depends on the distance between the neutrino
source and the neutrino detector. Downward going neutrinos ($\theta
\simeq 0$) pass a distance of about 20 km. On the other side upward
going neutrinos ($\theta \simeq \pi$) pass a distance of about 13000
km (earth diameter). The measurement of the dependence of the numbers
of the electron and muon events on the zenith angle $\theta$ allows
one to span the whole region of distances from about 20 km to about
13000 km.

From the data of the Super-Kamiokande experiment for high-energy electron events
was found
\begin{equation}\label{updown1}
\left(\frac{U}{D}\right)_{e}=0.961{}^{+0.086}_{-0.079}\pm 0.016.
\end{equation}
For high-energy muon  events  the value
\begin{equation}\label{updown2}
\left(\frac{U}{D}\right)_{\mu}=0.551{}^{+0.035}_{-0.033}\pm 0.004.
\end{equation}
was obtained. Here $U$ is the total number of upward going leptons
($-1< \cos\theta<-0.2$ ) and $D$ is the total number of downward
going leptons ($0.2< \cos\theta< 1$).

The data of the Super-Kamiokande atmospheric neutrino experiment are
perfectly described if we assume that $\nu_{\mu}$'s disappear mainly
due to $\nu_{\mu}\rightleftarrows  \nu_{\tau} $ oscillations. From
the three-neutrino analysis of the data for neutrino oscillation
parameters in the case  of normal (inverted) neutrino mass spectrum
it was found
\begin{eqnarray}\label{SKanalysis}
1.9(1.7)\cdot 10^{-3}~\mathrm{eV}^{2} \leq\Delta m_{23}^{2} \leq
2.6(2.7)\cdot 10^{-3}~\mathrm{eV}^{2},\nonumber\\
0.407 \leq \sin^2 \theta_{23} \leq 0.583,\quad
\sin^2 \theta_{13} < 0.04 (0.09).
\end{eqnarray}
The result of the Super-Kamiokande atmospheric neutrino experiment was
confirmed by

{\bf the long-baseline accelerator neutrino experiments K2K and MINOS
\cite{K2K,Minos}}

In the MINOS experiment, muon neutrinos produced at the Fermilab Main
Injector facility are detected. The MINOS data were obtained  with
neutrinos mostly with energies in the range $1\leq E\leq 5$ GeV.

There are two identical neutrino detectors in the  experiment. The
near detector with a mass of 1 kiloton is at a distance about 1 km
from the target of the accelerator and about 100~ m underground. The
far detector with a  mass of 5.4 kilotons is at a distance of 735 km
from the target in the Sudan mine (about 700 m underground).

Muon neutrinos (antineutrinos) are detected in the  experiment via the
observation of the process
\begin{equation}\label{Minos}
\nu_{\mu}(\bar\nu_{\mu}) +\rm{Fe}\to \mu^{-}(\mu^{+})+X
\end{equation}
The neutrino energy is given by the sum of the muon energy and the
energy of the hadronic shower.

In the near detector the initial neutrino spectrum  is measured. This
measurement allows to predict the expected spectrum of the muon
neutrinos in the far detector in the case if there were no neutrino
oscillations. A strong distortion of the spectrum of
$\nu_{\mu}(\bar\nu_{\mu})$ in the far detector was observed in the
MINOS experiment.

From the two-neutrino analysis of the $\nu_{\mu}$ data for the neutrino oscillations parameters the following values were obtained
\begin{equation}\label{Minos1}
\Delta m_{23}^{2}=(2.32~{}_{-0.08}^{+0.12})\cdot 10^{-3}~\mathrm{eV}^{2},\quad \sin^{2}2\theta_{23}>0.90.
\end{equation}

{\bf Indications in favor of nonzero $\theta_{13}$}

The value of the mixing angle $\theta_{13}$ is extremely important
for the future of the neutrino physics. If this angle is not equal to
zero (and relatively large) in this case it will be possible to
observe such a fundamental effect of the three-neutrino mixing as
$CP$ violation in the lepton sector. Another problem, the solution of
which requires nonzero $\theta_{13}$, is the problem of the neutrino
mass spectrum. In the case of the three massive neutrinos with two
mass-squared differences $\Delta m_{23}^{2}$ and $\Delta m_{12}^{2}$
two neutrino mass spectra are possible
\begin{enumerate}
  \item Normal spectrum
\begin{equation}\label{norspec}
m_{1}<  m_{2}    <  m_{3} ;\quad \Delta m^{2}_{12}  \ll    \Delta
m^{2}_{23}
\end{equation}

\item Inverted spectrum
\begin{equation}\label{invspec}
m_{3}<  m_{1} <  m_{2} ;\quad \Delta m^{2}_{12}  \ll
 | \Delta m^{2}_{13}|
\end{equation}
\end{enumerate}
Let us notice that in order to have the same notation $\Delta
m^{2}_{12}$ for the solar mass-squared difference for both spectra
the neutrino masses are usually labeled differently in the cases of
the normal and inverted neutrino mass spectra.
 In the case of the normal
spectrum  $\Delta m^{2}_{23}>0$ and in the case of  the inverted
spectrum $\Delta m^{2}_{13}<0$.

For many years only an upper bound on the parameter
$\sin^{2}\theta_{13}$ existed. This bound was obtained from the
analysis of the data of the CHOOZ reactor experiment \cite{Chooz}.

In the CHOOZ experiment the detector (5 tons of Gd-loaded liquid
scintillator) was at a distance of about 1 km from each of the two
reactors of  the CHOOZ power station (8.5 GWth). The detector had 300
m  water equivalent of rock overburden which reduced the cosmic muon
flux. The antineutrinos were detected through the observation of the
classical reaction
\begin{equation}\label{Chooz1}
\bar\nu_e + p \to e^+ + n.
\end{equation}
For the ratio $R$ of the total number of
detected $\bar\nu_{e}$ events to the number of the expected events it was found the value
\begin{equation} \label{Chooz2a}
R =1.01 \pm 2.8 \%\,(\rm{stat})\pm\pm 2.7 \%\,(\rm{syst}).
\end{equation}
The data of the experiment was analyzed  in the framework of
two-neutrino oscillations with the $\bar\nu_{e}$-survival probability
given by the expression
\begin{equation}\label{Chooz}
P(\bar\nu_{e}\to \bar\nu_{e})= 1 -\frac{1}{2}\sin^{2}2
\theta_{13}~(1-\cos\frac{\Delta m_{23}^{2}L}{2E})
\end{equation}

From the data of the CHOOZ experiment  the following upper bound
\begin{equation}\label{Chooz2}
\sin^{2}2\,\theta_{13} \leq 0.16.
\end{equation}
was obtained.

In a new long baseline  T2K neutrino experiment \cite{T2K} an
indication in favor of nonzero $\theta_{13}$ was obtained. In this
experiment muon neutrinos produced at the J-PARC accelerator  in
Japan are detected at a distance of 295 km in the water-Cherenkov
Super-Kamiokande detector. The T2K experiment is the first off-axis
neutrino experiment: the angle between the direction to the detector
and the flight direction of the parent $\pi^{+}$'s is equal to
 $2^\circ$. This allows one to obtain a narrow-band neutrino beam
with a maximal intensity at the energy  $E\simeq 0.6 $ GeV  which corresponds at
the distance of $L=295$ km to the first oscillation maximum
($E_{0}=\frac{2.54}{\pi}\Delta m_{23}^{2}L$).

At a distance of about 280 m from the target there are several
near detectors which are
used for the measurement of the neutrino spectrum and flux  and  for
the measurement of cross sections of different CC and NC processes.

The initial beam (from decays of pions and kaons) is a beam of
$\nu_{\mu}$'s with a small (about 0.4 \%) admixture of $\nu_{e}$'s.
The search for electrons in the Super-Kamiokande detector due to
$\nu_{\mu}\to \nu_{e}$ transitions was performed.  Six $\nu_{e}$
events were observed in the experiment. The expected number of
electron events (without neutrino oscillations) is equal $1.5\pm
0.3$. From the analysis of the data for the normal neutrino mass
spectrum it was found:
\begin{equation}\label{t2k}
 0.03<   \sin^{2}2\,\theta_{13}<0.28~(90\%\mathrm{CL})\quad \mathrm{best~fit:}
~~\sin^{2}2\,\theta_{13}=0.11.
\end{equation}
For the inverted neutrino mass spectrum it was found
\begin{equation}\label{t2k1}
 0.04<   \sin^{2}2\,\theta_{13}<0.34~(90\%\mathrm{CL})\quad \mathrm{best~fit:}
~~\sin^{2}2\,\theta_{13}=0.14.
\end{equation}
A similar experiment was performed by the MINOS collaboration. In this experiment for the normal (inverted) neutrino mass spectrum the following best fit value was found
\begin{equation}\label{eminos}
2\sin^2(\theta_{23})\sin^2(2\theta_{13})=0.041~{}^{+0.047}_{-0.031}~(0.079    ~{}^{+0.071}_{-0.053})
\end{equation}

The Double Chooz collaboration presented first indication in favor of
reactor $\bar\nu_{e}$'s disappearence \cite{Doublechooz}. For the
ratio of the observed and predicted $\bar\nu_{e}$ events the value
$0.944 \pm 0.016 \pm 0.040$ was found. At 90\% $CL$ it was obtained
$0.015  \sin^{2}2\theta_{13}  < 0.16$.

{\bf Measurement of nonzero $\theta_{13}$; the Daya Bay and RENO experiments}

Recently results of the Daya Bay reactor neutrino experiment were
published \cite{Dayabay}. In this experiment reactor antineutrinos
from six reactors (the thermal power of each reactor is 2.9 Gw) were
detected by three near detectors (distances 470 m and 570 m) and
three far detectors (1648 m). Antineutrinos are detected via
observation of the classical reaction
$$\bar\nu_{e}+p\to e^{+}+n.$$
Each detector contains the 20-ton $\mathrm{Gd}$-loaded liquid
scintillator. During 55 days of the data taking 10416 (80376)
$\bar\nu_{e}$-events were observed in far (near) detectors. The
number of $\bar\nu_{e}$ events in the far detectors can be predicted
(assuming that there are no neutrino oscillations) on the basis of
measurements performed in the near detectors. For the ratio $R$ of
the total numbers of the observed and predicted events the following
value was obtained
\begin{equation}\label{Daya}
  R=0.940\pm 0.011\pm 0.004.
\end{equation}
The probability of $\bar\nu_{e}$ to survive is given by the following
expression
\begin{equation}\label{Daya1}
  P(\bar\nu_{e}\to \bar\nu_{e})= 1-\sin^{2}2\theta_{13}\sin^{2}
1.267\frac{\Delta m_{23}^{2}L}{E},
\end{equation}
where $\Delta m_{23}^{2}$ is the neutrino mass-squared difference in
$\mathrm{eV}^{2}$, $L$ is the source-detector distance in m and $E$
is the antineutrino energy in MeV. From $\chi^{2}$ analysis of the
data it was found that
\begin{equation}\label{Daya2}
\sin^{2}2\,\theta_{13}=0.092\pm 0.016\pm 0.005.
\end{equation}
Thus, zero value of the parameter $\sin^{2}2\,\theta_{13}$ is
excluded at the level 5.2 $\sigma$.

The value of the parameter
$\sin^{2}2\,\theta_{13}$ obtained in the similar two-detectors reactor
RENO experiment \cite{Reno} is in agreemment with (\ref{Daya2}). In this experiment
it was found
\begin{equation}\label{Reno}
\sin^{2}2\,\theta_{13}=0.113 + 0.013 + 0.019.
\end{equation}

\section{Conclusion}

We followed here some basic facts of the history of neutrinos, unique
particles which brought three Nobel Prizes to elementary particle
physics. The neutrino history is very interesting, instructive,
sometimes dramatic. There were many wrong experiments in the history
of the neutrino (like $\beta$-decay experiments on electron-neutrino
correlation which favored $S,T$ couplings in the fifties, first
experiment on the search for $\pi\to e\nu$-decay, experiments from
which the existence of a heavy neutrino with a mass of 17 keV
followed at the beginning of the nineties, etc.) and wrong common
opinions lasting for many years (like the general opinion that the
neutrino is an undetectable particle in the thirties and forties, the
general opinion that the neutrino is a massless particle in the
fifties and sixties, etc.).

The neutrino hypothesis was born in 1930 in an attempt  to save the
law of conservation of energy and momentum ("I have hit upon a
desperate remedy to save the "exchange theorem" of statistics and the
law of conservation of energy. Namely, the possibility that in the
nuclei there could exist electrically neutral particles, which I will
call neutrons, that have spin 1/2 and obey the exclusion principle
and that further differ from light quanta in that they do not travel
with the velocity of light." Pauli's letter). The assumption of the
existence of the neutrino allowed Fermi to build a phenomenological
theory of the $\beta$-decay of nuclei and other weak processes which
could describe a lot of experimental data. However, it took more that
twenty years to prove  by a direct experiment that neutrinos exist.

 There are two unique properties of neutrinos which determine their importance
and their problems:
\begin{enumerate}
  \item Neutrinos have only weak interaction.
  \item Neutrinos have very small masses.
\end{enumerate}
Since neutrinos have only weak interaction, cross sections of
interaction of neutrinos with nucleons are extremely small. This
means that it is necessary to develop special methods of neutrino
detection (large detectors which often are situated in underground
laboratories in order to prevent cosmic ray background etc). However,
when methods of neutrino detection were developed,  neutrinos became
a unique instrument in the study of the sun (solar neutrino
experiments allow us to obtain information about the central
invisible region of the sun in which solar energy is produced in
thermonuclear reactions), in the investigation of a mechanism of the
Supernova explosion\footnote{On February 23, 1987 for the first time
antineutrinos from Supernova SN1987A in the Large Magellanic Cloud
were detected by Kamiokande, IMB and Baksan detectors. In 2002 The
Nobel Prize was awarded  to R. Davis (solar neutrinos) and M. Koshiba
(supernova neutrinos) "for pioneering contributions to astrophysics,
in particular for the detection of cosmic neutrinos".}(99\% of the
energy produced in a Supernova explosion is emitted in neutrinos), in
establishing the  quark structure of a nucleon (through the study of
the deep inelastic processes $\nu_{\mu}(\bar\nu_{\mu})+N\to
\mu^{-}(\mu^{+}+X$), etc.

In the fifties the majority of physicists believed that the neutrino
was a massless particle. This was an important, constructive
assumption (in spite of that it was wrong). The theory of the
two-component neutrino, which was based on this assumption, inspired
the creation of the phenomenological $V-A$ theory and later became
part of the Standard Model of the electroweak interaction.

Neutrino masses are very small and it is very difficult  to observe
{\em effects of neutrino masses in the $\beta$-decay and in other
weak processes.} However, small neutrino masses and, correspondingly,
mass-squared differences make it possible the production (and
detection) of the {\em coherent flavor neutrino states}(states of
$\nu_{e}$, $\nu_{\mu}$, $\nu_{\tau}$) and quantum-mechanical
periodical transitions between different flavor neutrino states
(neutrino oscillations). The observation of neutrino oscillations at
large (macroscopic) distances allowed one to resolve small neutrino
mass-squared differences.

The discovery of neutrino oscillations signifies a new era in
neutrino physics, the era of investigation of neutrino properties.
From the analysis of the existing neutrino oscillation data two
mass-squared differences $\Delta m^{2}_{23}$ and $\Delta m^{2}_{12}$
and two mixing parameters $\sin^2\theta_{23}$ and
$\tan^{2}\theta_{12}$ are determined with accuracies in the range
(3-12)\%.  The results of the first measurement of the parameter
$\sin^{2}2\,\theta_{13}$ was recently announced by the Daya Bay
collaboration.

One of the most urgent problems  which will be addressed in the next
neutrino oscillation experiments are

\begin{enumerate}
  \item {\bf CP violation  in the lepton sector.}
  \item {\bf Character of the neutrino mass spectrum (normal or inverted?)}
\end{enumerate}

Relatively "large" value of  the angle $\theta_{13}$ obtained in the
Daya Bay and other experiments open the way for the investigation of
these problems in the near years.

One of the most important problems of the physics of massive and
mixed neutrinos is the problem of the nature of neutrinos with
definite masses $\nu_{i}$.

{\bf Are neutrinos with define masses Dirac particles possessing
conserved lepton number or
truly neutral Majorana particles?}

The answer to this fundamental question can be obtained in
experiments on the search for neutrinoless double $\beta$-decay
($0\nu\beta\beta$-decay) of some even-even nuclei
\begin{equation}\label{bb}
(A,Z)\to (A,Z+2)+e^{-}+e^{-}.
\end{equation}
This process is allowed only if the total lepton number is violated.
If massive neutrinos are Majorana particles, $0\nu\beta\beta$-decay
(\ref{bb}) is the second order process in the Fermi constant with the
exchange of the virtual neutrinos between neutron-proton-electron
vertices. The matrix element of the process is proportional to the
effective Majorana mass
\begin{equation}\label{bb1}
m_{\beta\beta}=\sum_{i}U^{2}_{ei}m_{i}.
\end{equation}
Many experiments on the search for the $0\nu\beta\beta$-decay of
different nuclei were performed. No evidence in favor of the process
was obtained. The most stringent lower bound for the half-life of the
process was obtained in the experiment \cite{Hei-Mos} on the search
for the decay
$$^{76}\rm{Ge}\to ^{76}\rm{Se}+e^{-}+e^{-}$$
In this experiment the following lower bound was obtained
$$T^{0\nu}_{1/2}(^{76}\rm{Ge})>1.9\cdot 10^{25}~\rm{years}$$
Taking into account different calculations of the nuclear matrix element from this bound it can be found
$$|m_{\beta\beta}|<(0.20-0.32    )~\rm{eV}.$$
Future experiments on the search for the $0\nu\beta\beta$-decay  will be sensitive to the value
$$|m_{\beta\beta}|\simeq \rm(a~few)~10^{-2}~\rm{eV}$$
and can probe the Majorana nature of $\nu_{i}$ in the case
of the inverted hierarchy of the neutrino masses
\begin{equation}\label{Inhierar}
m_{3}\ll m_{1}<m_{2}.
\end{equation}
Another fundamental problem of the physics of massive and mixed
neutrinos is

{\bf What are the absolute value of the neutrino masse?}

From the data of neutrino oscillation experiments only the
mass-squared differences can be determined. The absolute value of the
"average" neutrino mass $m_{\beta}$ can be inferred from the
investigation of $\beta$-spectra. From the data of the latest MAINZ
and Troitsk tritium experiments the following bound was obtained
$$m_{\beta}< 2.3 ~\rm{eV},$$
where $ m_{\beta}=\sqrt{\sum_{i}|U_{e1}|^{2}m^{2}_{i}}$ is the "average" neutrino
mass. The future tritium experiment KATRIN will be sensitive to
$$m_{\beta}< 0.2 ~\rm{eV}$$
Precision modern cosmology became an important source of information
about absolute values of neutrino masses. Different cosmological
observables (Large Scale Structure of the Universe, Gravitational
Lensing of Galaxies, Primordial Cosmic Microwave Background, etc.)
are sensitive to the sum of the neutrino masses $\sum_{i}m_{i}$. From
the existing data the following bounds were obtained
\begin{equation}\label{cosmo}
\sum_{i}m_{i}<(0.2-1.3)~\mathrm{eV}.
\end{equation}
It is expected that future cosmological observables will be sensitive to the sum of neutrino masses in the range
\begin{equation}\label{cosmo1}
\sum_{i}m_{i}\simeq (0.05-0.6)~\mathrm{eV}.
\end{equation}
These future measurements, apparently,  will probe the inverted neutrino mass hierarchy
(\ref{Inhierar}) ($\sum_{i}m_{i}\simeq 0.1$ eV) and even the normal
neutrino mass hierarchy
\begin{equation}\label{Norhierar}
m_{1}\ll m_{2}\ll m_{3},\quad \sum_{i}m_{i}\simeq 0.05~\mathrm{eV}.
\end{equation}

The next question which needs to be answered is

{\bf How many neutrinos with definite masses exist in nature?}

We considered the minimal scheme with three flavor neutrinos
($\nu_{e}, \nu_{\mu},  \nu_{\tau}$) and, correspondingly, three
massive neutrinos ($\nu_{1}, \nu_{2},  \nu_{3}$). However, the number
of massive light neutrinos can be more than three. In this case
flavor neutrinos could oscillate into sterile states $\nu_{s}$, which
do not have the standard weak interaction.

For many years there was an indication in favor  of more than three
light neutrinos with definite masses obtained in a short-baseline
LSND experiment \cite{LSND}. In this experiment the $\bar\nu_{\mu}\to
\bar\nu_{e}$ transition driven by $\Delta m^{2}\simeq
1~\mathrm{eV}^{2}$, which is much larger than the atmospheric
mass-squared difference, was observed. Some indications in favor of
more than three massive neutrinos were also obtained in the MiniBooNE
and reactor experiments. New short-baseline accelerator and reactor
experiments are urgently needed. Such experiments are now at
preparation.

There are other questions connected with neutrinos
which now are
being actively discussed in the literature (neutrino magnetic
moments, nonstandard interaction of neutrinos, etc.).

An explanation of small neutrino masses requires a new, beyond the
Standard Model (Higgs) mechanism of neutrino mass generation. But
what mechanism, what kind of new physics is required to explain small
neutrino masses and peculiar neutrino mixing? This is at the moment
an open question. Several new mechanisms of neutrino mass generation
were proposed in the literature. Apparently, the most plausible
mechanism is the seesaw mechanism of the neutrino mass generation
\cite{seesaw}.

The seesaw mechanism is based on the assumption that the total lepton
number $L$ is violated at  a large scale $M$. From the seesaw
mechanism the following general consequences follow
\begin{enumerate}
  \item Neutrinos with definite masses $\nu_{i}$ are Majorana particles.
  \item Neutrino masses are given by the seesaw formula
 $$m_{i}\simeq y_{i}\frac{v^{2}}{M},$$
 where $y_{i}$ is a dimensionless Yukawa constant and the parameter $v\simeq
250$ GeV  characterizes the scale of the violation of the electroweak
symmetry.
\end{enumerate}
The scale of the violation of the lepton number $M$ depends on the
Yukawa constants $y_{i}$ which are unknown. Different options are
discussed in the literature. If $y_{i}v\simeq m^{f}_{i}$, where
$m^{f}_{i}$ is the mass of a quark or a lepton, in this case $M\simeq
(10^{14}-10^{15})$ GeV and the only implication of the violation of
the lepton number in the region of the electroweak energies are
Majorana neutrino masses.\footnote{Let us notice that  CP-violating
decays of heavy Majorana particles, the seesaw partners of Majorana
neutrinos, in the early Universe are commonly considered as a
plausible source of the baryon asymmetry of the Universe \cite{Nir}.}

If $M\simeq 1$ TeV in this case $y_{i}v\ll m^{f}_{i}$. Existence of
Majorana particles with masses $\simeq 1$ TeV could be revealed
through (see \cite{Senjanovic})
\begin{enumerate}
  \item an additional contribution to the matrix elements of
the neutrinoless double $\beta$-decay;
  \item observation of the lepton number violating processes
of production of pairs of the same sign leptons in proton-proton
collisions at LHC.
\end{enumerate}

In order to reveal the true nature of neutrino masses and mixing many
new investigations must be performed. The history of neutrinos,
unique particles, is continuing. There are no doubts that new
surprises and discoveries (and, possibly, Nobel Prizes) are ahead.

I am thankful to the theory group of TRIUMF for the hospitality.
 It is a pleasure for me to thank W. Potzel
for careful reading of the paper and numerous remarks and suggestions.


\begin{thebibliography}{99}

\bibitem{EllisWooster} C. D. Ellis and W.A. Wooster,
Proc. Roy. Soc. {\bf A117} (1927) 109.


\bibitem{Chedwick} J. Chadwick,  Nature {\bf 193} (1932) 312.


\bibitem{Heisenberg} W. Heisenberg, Z.Phys.{\bf 77} (1932)1;
Z.Phys.{\bf 78} (1932)156.

\bibitem{Majorana} E. Majorana, Z.Phys.{\bf 82} (1933)137.

\bibitem{Ivanenko} D. Ivanenko, Nature {\bf 129} (1932)798.


\bibitem{Fermi} E. Fermi,  Zeitschr. f. Phys.  {\bf 88} (1934) 161.

\bibitem{Perrin}  F. Perrin, Comptes Rendus
{\bf 197} (1933) 1625.

\bibitem{GamovTeller}
G. Gamow and E. Teller,  Phys. Rev. {\bf 49} (1936) 895.



\bibitem{BethePeierls} H.Bethe and R.Peierls, Nature {\bf 133} (1934)  532.

\bibitem{BPonte46} B. Pontecorvo, Report PD-205, Chalk River Laboratory, 1946.

\bibitem{Davis}B.T. Cleveland {\it et al.}   Astrophys. J. {\bf 496} (1998) 505.



\bibitem{BPontmue} B. Pontecorvo, Phys. Rev.{\bf 72} (1947) 246.

\bibitem{Puppi}G. Puppi, Nuovo Cimento {\bf 5} (1948) 587.

\bibitem{Klein}O. Klein, Nature {\bf 161} (1948) 897.

\bibitem{YangTiomno} C. N. Yang and J. Tiomno, Phys. Rev.{\bf 79} (1950) 495.



\bibitem{LeeYang56} T. D. Lee and C. N. Yang,  Phys. Rev. {\bf 104} (1956) 254.


\bibitem{Wu} C. S. Wu {\it et al.}, Phys. Rev. {\bf 105} (1957) 1413.

\bibitem{Lederman} R. L. Garwin, L. M. Lederman and W.
Weinrich, Phys. Rev. {\bf 105} (1957) 1415.








\bibitem{Landau}L. D. Landau,  Nucl. Phys. {\bf 3 } (1957) 127.


\bibitem{LeeYang}T. D. Lee and C. N. Yang, Phys. Rev. {\bf 105} (1957) 1671.

\bibitem{Salam} A. Salam, Nuovo Cim. {\bf 5 } (1957) 299.


\bibitem{Weil}H. Weil, Z. Physik  {\bf 56 } (1929) 330.

\bibitem{Pauli33}W. Pauli, Handbuch der  Physik, Springer Verlag, Berlin   {\bf v.24 } (1933) 226-227.


\bibitem{Majorana} E. Majorana, Nuovo Cimento \textbf{5} (1937) 171.





\bibitem{Goldhaber} M. Goldhaber, L. Grodzins and A. W. Sunyar,
 Phys. Rev. {\bf 109} (1958) 1015.




\bibitem{FeyGel} R. P. Feynman and M. Gell-Mann, Phys.~Rev. {\bf 109} (1958) 193.

\bibitem{MarSud}  E. C. G. Sudarshan and R. E. Marshak,  Phys.~Rev.
{\bf 109} (1958) 1860.

\bibitem{GerZeld} S.S.Gerstein and Ja.B Zeldovich, Sov. Phys. JETP    {\bf2}(1956)576.




\bibitem{Anderson} H.L. Anderson and C. Lattes, Nuovo Cimento {\bf 6} (1957) 1356.


\bibitem{Fidecaro} T. Fazzini, G. Fidecaro {\it et al.}, Phys. Rev. Lett. {\bf 1} (1958) 247.

\bibitem{Reinesnue} F. Reines, H.S. Gurr and H.W. Sobel, Phys. Rev. Lett. {\bf 37} (1976) 315.
\bibitem{Cabibbo} N. Cabibbo, Phys. Rev. Lett. {\bf 10} (1963) 531.

\bibitem{Klein} O. Klein, Proc. Symp. on Les Nouvelles
Theories de la Physique, Warsaw, 1938 (Institut
International de Coopï¿-ration Intellectuelle, Paris, 1939), p.6.



\bibitem{Reines}
F. Reines and C. L. Cowan,  Phys. Rev. {\bf 92} (1953) 830;
F. Reines and C. L.  Cowan,  Nature {\bf 178} (1956) 446;
F. Reines and C. L.  Cowan,  Phys. Rev. {\bf 113} (1959) 273.

\bibitem{Davis1} R. Davis, Bull. Am. Phys. Soc., Washington meeting, 1959.

\bibitem{Ponte62} B. Pontecorvo, Journal de Physique {\bf 43} N. 12 (1959) p.C8-221.

\bibitem{Brookhaven} G. Danby, J.-M. Gaillard, K. Goulianos, L.M.
Lederman, N. Mistry, M. Schwartz and J. Steinberger,  Phys. Rev. Lett. {\bf 9} (1962) 36.

\bibitem{Feinberg} G. Feinberg, Phys. Rev. {\bf 110} (1958) 1482.


\bibitem{Ponte59}
B. Pontecorvo, Sov. Phys. JETP {\bf10} (1960) 1236.


\bibitem{PDG}K. Nakamura {\em et al.} (Particle Data Group), J. Phys. {\bf G37}G 37 (2010) 075021.


\bibitem{GIM} S. L. Glashow, J. Iliopoulos and L. Maiani,  Phys.~Rev. {\bf D2} (1970) 1258.
\bibitem{BilPonte78} S. M. Bilenky, B. Pontecorvo,  Phys. Rep. {\bf 41} (1978) 225.

\bibitem{Perl}M. L. Perl {\em et al.}, Phys. Rev. Lett. {\bf35}, 1489 (1975).

\bibitem{Donut}K. Kodama {\em et al.} (DONUT Collaboration),  Physics Letters {\bf B 504} (2001) 218.


\bibitem{KobMas} M. Kobayashi and T. Maskawa,  Progress of Theoretical Physics {\bf49} (2) (1973) 652.

\bibitem{PDG}K. Nakamura {\em et al.} (Particle Data Group), J. Phys. {\bf G37}G 37 (2010) 075021.


\bibitem{Weinberg67}S. Weinberg,  Phys. Rev. Lett. {\bf 19} (1967) 1264.


\bibitem{Salam68}A. Salam, Proc. of the Eighth Nobel Symposium
(ed. N. Svartholm, Wiley-Interscience, New York (1968).

\bibitem{CMS}S. Chatrchyan {\em et al.}, Phys. Lett. {\bf B716} (2012) 30.


\bibitem{Atlas}G. Aad {\em et al.}, Phys. Lett. {\bf B 716} (2012) 1.

\bibitem{Glashow61}S. L. Glashow, Nucl. Phys. {\bf 22} (1961) 579.

\bibitem{Hooft} G. 't Hooft,  Nucl. Phys. {\bf B35} (1971) 1967.

\bibitem{YangMills} C. N. Yang and R. Mills, Phys. Rev.{\bf 96} (1954) 191

\bibitem{Gargamelle} F. J. Hasert {\it et al.}, Phys. Lett. {\bf B46} (1973) 138.

\bibitem{NuTeV}G.P.Zeller {\it et al.}, Phys. Rev. Lett. {\bf 88} (2002) 091802.

\bibitem{GMPais}M. Gell-Mann and A. Pais, Phys. Rev. {\bf 97} (1955) 1387.

\bibitem{BPonte57}B. Pontecorvo,
J.Exptl. Theoret. Phys. {\bf 33} (1957) 549
 [Sov. Phys. JETP {\bf 6} (1958) 429].

\bibitem{BPonte58}B. Pontecorvo, J.Exptl. Theoret.
Phys. {\bf 34} (1958) 247 [Sov. Phys. JETP {\bf 7} (1958) 172].



\bibitem{BPonte67}B. Pontecorvo, J. Exptl.
Theoret. Phys. {\bf 53} (1967) 1717.
[Sov. Phys. JETP {\bf 26} (1968) 984].

\bibitem{1Davis}R. Davis {\em et al.},Proc. Conf. "Neutrino 72", Hungary, vol.I p. 29,
(1972).

\bibitem{GPonte69}V. Gribov and B. Pontecorvo, Phys. Lett. {\bf B28} (1969) 493.


\bibitem{BF69}J. Bahcall and S. Frautschi, Phys. Lett. {\bf 29} (1969) 623.

\bibitem{BilPonte76}S.M. Bilenky and B. Pontecorvo,
 Phys. Lett. {\bf B61} (1976) 248; Yad.
Fiz. {\bf 3} (1976) 603 .

\bibitem{FM76}H. Fritzsch and P. Minkowski, Phys. Lett.  {\bf B62} (1976) 72.

\bibitem{ES76}S. Eliezer and A. Swift,~ Nucl.Phys. {\bf B105} (1976) 45.

\bibitem{MNS}Z. Maki, M. Nakagava and S. Sakata, Prog. Theor.
Phys. {\bf 28} (1962) 870.



\bibitem{BilPonte77} S.M. Bilenky and B. Pontecorvo, Phys.
Rep. {\bf 41} (1978) 225 .

\bibitem{BilPonteNC}S.M. Bilenky and B. Pontecorvo, Lett.
Nuovo Cim. {\bf17} (1976) 569.

\bibitem{Kamiokande}K.S. Hirata  {\em et al.}, Phys. Rev. Lett.  {\bf63} (1989) 16.
\bibitem{Kamatmospheric}K. Hirata  {\em et al.}, Phys. Rev. Lett. {\bf58} (1987) 1490.


\bibitem{seesaw}
P. Minkowski, Phys. Lett. {\bf B 67} (1977) 421; M.~Gell-Mann, P.~Ramond and R.~Slansky,
in \textit{Supergravity}, p.~315, edited by F. van Nieuwenhuizen and D. Freedman, North Holland, Amsterdam, (1979); T.~Yanagida, Proc. of the \textit{Workshop on Unified Theory and the Baryon Number of the
  Universe}, KEK, Japan, (1979); S.L. Glashow, NATO Adv.Study Inst. Ser. B Phys. {\bf 59} (1979) 687; R.N. Mohapatra and G.~Senjanovi{\'c}, Phys. Rev. {\bf D23} (1981) 165.

\bibitem{Gallex} P. Anselmann   {\em et al.} (GALLEX Collaboration), Phys. Lett.  {\bf B327} (1994) 377.

 \bibitem{Sage}J.N. Abdurashitov  {\em et al.} (SAGE Collaboration), Phys. Lett. {\bf B328} (1994) 234.


\bibitem{MSW} L. Wolfenstein, Phys. Rev. {\bf D17} (1978) 2369;
S. P. Mikheev and A. Yu. Smirnov,  Nuovo Cim. {\bf C9} (1986) 17.


\bibitem{SK}Y. Fukuda {\em et al.} (Super-Kamiokande Collaboration), Phys. Rev. Lett. {\bf 81} (1998) 1562. R. Wendell   {\em et al.} (Super-Kamiokande Collaboration), Phys.Rev. {\bf D81} (2010) 092004.




\bibitem{SNO} Q.R.   Ahmad  {\em et al.} (SNO Collaboration), Phys. Rev. Lett. {\bf89} (2002) 011301: B. Aharmim et al (SNO Collaboration) arXiv:1109.0763.


\bibitem{Kamland} K. Eguchi  {\em et al.} (KamLAND Collaboration), Phys. Rev. Lett. {\bf90} (2003) 021802;T. Araki  {\em et al.} (KamLAND Collaboration),    Phys.Rev.Lett.{\bf94} (2005) 081801. Abe, S. {\em et al.} (KamLAND Collaboration), Phys. Rev. Lett. {\bf100} (2008) 221803.


\bibitem{K2K}M.H. Ahn  {\em et al.} (K2K Collaboration), Phys. Rev. Lett. {\bf90} (2003) 041801.


\bibitem{Minos} D.G. Michael  {\em et al.} (MINOS Collaboration), Phys. Rev. Lett.  {\bf97} (2006) 191801. P. Adamson  {\em et al.} (MINOS Collaboration), Phys. Rev. Lett. {\bf106} (2011) 181801.



\bibitem{Chooz} M. Apollonio {\it et al.} (CHOOZ Collaboration)
 Eur. Phys. J. {\bf C27} (2003) 331.




\bibitem{T2K} K.Abe  {\em et al.} (T2K Collaboration),
Phys. Rev. Lett. {\bf107} (2011) 041801.

\bibitem{Hei-Mos}
M. Gunther {\em et al.} (Heidelberg - Moscow Collaboration),
Phys. Rev.{\bf D55} (1997) 54.

\bibitem{LSND} A. Aguilar {\em et al.} (LSND Collaboration),
Phys. Rev. {\bf D64} (2001) 112007.

\bibitem{Doublechooz}Y. Abe {\em et al.} (Double Chooz Collaboration),
arXiv:1112.6353v3.

\bibitem{Dayabay}F. P. An {\em et al.} (Daya Bay Collaboration),
Phys. Rev. Lett. {\bf 108} (2012) 171803; arXiv:1203.1669v1.

\bibitem{Reno}J. Ahn {\em et al.} (Reno Collaboration),
Phys. Rev. Lett. {\bf 108} (2012) 191802.





\bibitem{Nir} S. Davidson, E. Nardi, Y. Nir, Phys. Rept. {\bf 466}, 105 (2008).


\bibitem{Senjanovic} G. Senjanovic,    arXiv:1012.4104.















\end{thebibliography}
\end{document}